\documentclass[longauth]{aa}

\usepackage{graphicx}
\usepackage{float}
\usepackage{array}
\usepackage{xspace}

\newcolumntype{C}[1]{>{\centering\arraybackslash}p{#1}}
\usepackage{multirow}

\newcommand{\duets}{\texttt{Cosmic Duets}\xspace}
\usepackage[dvipsnames]{xcolor}

\usepackage[breaklinks=true,colorlinks=true,citecolor=blue,linkcolor=blue,urlcolor=blue]{hyperref}
\usepackage{tablefootnote}
\usepackage{natbib}
\usepackage{amsmath} 
\usepackage{amssymb}
\usepackage{mathtools}
\usepackage{txfonts}
\usepackage{microtype}
\usepackage{array}
\usepackage{makecell}
\usepackage{multirow}
\usepackage{siunitx}
\usepackage{tabularx}
\usepackage{booktabs}
\usepackage{color}
\definecolor{myorange}{rgb}{0.8, 0.3, 0.0}

\usepackage[caption=false]{subfig}
\usepackage{verbatim}

\newcommand{\civ}{C\,{\sc{iv}}\xspace}
\newcommand{\mgii}{Mg\,{\sc{ii}}\xspace}

\begin{document}

\title{Cosmic Duets II. }
\subtitle{Growth and accretion in Gaia Multi-Peak dual AGN at Cosmic Noon  }

\titlerunning{Cosmic Duets II}
\authorrunning{Scialpi et al.}
\author{
M.~Scialpi\inst{\ref{iUNITN},\ref{iUNIFI},\ref{Inaf-oaa}}
\and F.~Mannucci\inst{\ref{Inaf-oaa}}
\and D.~Izquierdo-Villalba \inst{\ref{ICE}}
\and A.~Marconi\inst{\ref{iUNIFI},\ref{Inaf-oaa}}
\and J.~Singh\inst{\ref{Inaf-brera}}
\and M.~Volonteri\inst{\ref{Paris}}
\and \\C.~Marconcini\inst{\ref{iUNIFI},\ref{Inaf-oaa}}
\and Q.~D'Amato\inst{\ref{Inaf-roma}}
\and B.~Moreschini\inst{\ref{iUNIFI},\ref{Inaf-oaa}}
\and M.V.~Zanchettin\inst{\ref{Inaf-roma}}
\and G.~Tozzi\inst{\ref{MPE}}
\and M.~Ceci\inst{\ref{iUNIFI},\ref{Inaf-oaa}}
\and L.~Ulivi\inst{\ref{Madrid}}
\and \\E.~Bertola\inst{\ref{Inaf-oaa}}
\and S.~Bonoli\inst{\ref{DIPC},\ref{IKERBASQUE}}
\and C.~Bracci\inst{\ref{iUNIFI},\ref{Inaf-oaa}}
\and E.~Cataldi\inst{\ref{iUNIFI},\ref{Inaf-oaa}}
\and C. Cicone \inst{\ref{Oslo}}
\and A. Ciurlo \inst{\ref{UCLA}}
\and M.~Colpi\inst{\ref{bicocca},\ref{Inaf-brera},\ref{Infn-unimib}}
\and \\G.~Cresci\inst{\ref{Inaf-oaa}}
\and A.~De Rosa\inst{\ref{Inaf-roma}}
\and M.~Dotti\inst{\ref{bicocca},\ref{Inaf-brera}}
\and M.~Fumagalli\inst{\ref{bicocca},\ref{inaf_trieste}}
\and M.~Ginolfi\inst{\ref{iUNIFI},\ref{Inaf-oaa}}
\and I.~Lamperti\inst{\ref{Madrid}}
\and E.~Nardini\inst{\ref{Inaf-oaa}}
\and \\M.~Parvatikar\inst{\ref{Inaf-roma},\ref{torvergata}}
\and M.~Perna\inst{\ref{Madrid}}
\and F.~Rigamonti\inst{\ref{Inaf-brera},\ref{Infn-unimib},\ref{CLAP}}
\and K.~Rubinur\inst{\ref{Oslo}}
\and B.~Sala\inst{\ref{iUNITN}, \ref{Inaf-brera}, \ref{Infn-unimib}}
\and A.~Sesana\inst{\ref{bicocca},\ref{Inaf-brera},\ref{Infn-unimib}}
\and \\P.~Severgnini\inst{\ref{Inaf-brera}}
\and C.~Spingola\inst{\ref{Inaf-bologna}}
\and D.~Spinoso\inst{\ref{bicocca}}
\and G.~Venturi\inst{\ref{Inaf-oaa}}
\and C. Vignali\inst{\ref{iUNIBO},\ref{OASBO}}
\and S. Yeh\inst{\ref{keck}}
}
\institute{
University of Trento, Via Sommarive 14, I-38123 Trento, Italy
\label{iUNITN}
\and
Università di Firenze, Dipartimento di Fisica e Astronomia, via G. Sansone 1, 50019 Sesto F.no, Firenze, Italy\\
\email{martina.scialpi@unifi.it}
\label{iUNIFI}
\and
INAF -- Osservatorio Astrofisico di Arcetri, Largo E. Fermi 5, 50125 Firenze, Italy
\label{Inaf-oaa}
\and
Institute of Space Sciences (ICE, CSIC), Campus UAB, Carrer de Magrans, E-08193 Barcelona, Spain
\label{ICE}
\and
INAF -- Osservatorio Astronomico di Brera,
Milano: via Brera 28, I-20121 Milano, Italy,
Merate: Via Bianchi 46, I-23807 Merate (LC), Italy
\label{Inaf-brera}
\and
Institut d'Astrophysique de Paris, 98bis Bd Arago, 75014 Paris, France
\label{Paris}
\and
INAF - Istituto di Astrofisica e Planetologia Spaziali, Via Fosso del Cavaliere 100, 00133 Rome, Italy
\label{Inaf-roma}
\and
Max Planck Institute for extraterrestrial Physics, Giessenbachstraße 1, 85748 Garching, Germany
\label{MPE}
\and
Centro de Astrobiología (CAB), CSIC--INTA, Ctra. de Ajalvir Km 4, 28850 Torrejón de Ardoz, Madrid, Spain
\label{Madrid}
\and
Donostia International Physics Centre (DIPC), Paseo Manuel de Lardizabal 4, 20018 Donostia-San Sebastian, Spain
\label{DIPC}
\and
IKERBASQUE, Basque Foundation for Science, E-48013, Bilbao, Spain
\label{IKERBASQUE}
\and
Institute of Theoretical Astrophysics, University of Oslo, P.O. Box 1029, Blindern, 0315 Oslo, Norway
\label{Oslo}
\and
Department of Physics and Astronomy, University of California Los Angeles, 430 Portola Plaza, Los Angeles, CA 90095, USA
\label{UCLA}
\and
Dipartimento di Fisica "G. Occhialini", Universit\`a degli Studi di Milano-Bicocca, Piazza della Scienza 3, 20126 Milano, Italy
\label{bicocca}
\and
INFN, Sezione di Milano-Bicocca, Piazza della Scienza 3, I-20126 Milano, Italy
\label{Infn-unimib}
\and
INAF -- Osservatorio Astronomico di Trieste, via G.B. Tiepolo 11, I-34143 Trieste, Italy
\label{inaf_trieste}
\and
Dipartimento di Fisica, Università di Roma Tor Vergata, Via della Ricerca Scientifica, I-00133, Roma, Italy
\label{torvergata}
\and
Como Lake centre for AstroPhysics (CLAP), DiSAT, Università dell’Insubria, via Valleggio 11, 22100 Como, Italy
\label{CLAP}
\and
INAF -- Istituto di Radioastronomia, Via Piero Gobetti 101, 40129 Bologna, Italy
\label{Inaf-bologna}
\and
Physics and Astronomy Department ``Augusto Righi'', Università di Bologna, Via Gobetti 93/2, 40129 Bologna, Italy
\label{iUNIBO}
\and
INAF -- Osservatorio di Astrofisica e Scienza dello Spazio di Bologna, Via Gobetti 93/3, I-40129 Bologna, Italy
\label{OASBO}
\and
W. M. Keck Observatory, 65-1120 Mamalahoa Highway, Kamuela, HI 96743, USA
\label{keck}
}

   \date{}
    

\abstract{
We present the first statistical characterization of the physical properties of \duets: dual AGN in the redshift range $0.5 \leq z \leq 3.0$ with projected separations of 1--7~kpc, corresponding to systems in which both AGN are expected to reside within the same merger remnant.

We analyze the largest homogeneous sample currently available, consisting of 30 spectroscopically confirmed \duets selected via the Gaia Multi-Peak technique.
Using rest-frame UV-to-optical integral-field spectroscopy for all targets, primarily with VLT/MUSE, VLT/ERIS, and Keck/OSIRIS, we spatially resolve the two AGN components and extract separate spectra for each system. We decompose the continuum, emission lines, and Fe\,\textsc{ii} complexes through spectral fitting.
We derive black hole masses of $\log(M_{\rm BH}/M_\odot)\sim7.4$--10.1 and bolometric luminosities of $\log(L_{\rm bol}/{\rm erg\ s^{-1}})\sim44.8$--47.2, consistent with those of single AGN at similar redshifts.

We finally carry out the first comparison of dual AGN properties with semi-analytical models and hydrodynamical simulations, including \textsc{L-GalaxiesBH} and \textsc{Astrid}. 
\textsc{Astrid} provides a closer match to the observed black-hole mass-ratio distribution, but its selected systems are predominantly at lower redshifts, resulting in lower bolometric luminosities at fixed $G$-band magnitude. 
\textsc{L-GalaxiesBH} perfectly reproduce the observed parameter space, but it predicts more heterogeneous pairs, with higher primary-bolometric luminosity, lower-mass secondary SMBHs, and higher Eddington ratios, resulting in an excess of near-Eddington accreting AGN due to its adopted accretion prescription.

Our results suggest that, in GMP-selected systems, secondary SMBH growth  is more efficient than predicted by current models, leading to more comparable black-hole masses and accretion power between the two components. 
}

   \maketitle
%

\section{Introduction}

During the hierarchical assembly of galaxies, the supermassive black holes (SMBHs) hosted at the centers of galaxies are brought together through galaxy mergers \citep[e.g.,][]{White+91}. 
Through the accretion of surrounding gas, SMBHs can power active galactic nuclei (AGN) and drive feedback processes that influence the evolution of their host galaxies and regulate star formation \citep{Dimatteo05,Harrison_almeida24}. 
When both SMBHs involved in a merger are simultaneously active, the system is observed as a dual AGN. At this stage, the two SMBHs remain embedded in their respective host galaxies and are typically observed at projected separations of tens of kpc \citep[e.g.,][]{derosa20,pfeifle23_dual}. 
As the galaxy merger proceeds, dynamical friction drives the SMBHs toward the center of the common gravitational potential formed by the merging system. 
When galaxies form a single remnant, the two AGN are typically separated by few kpc, where each of the two SMBHs may retain a surrounding reservoir of gas and an accretion disk capable of sustaining AGN activity \citep{Li2023,Johnstone25}.
At these late merger stages, both observations and simulations indicate enhanced gas inflows toward the nuclear regions, leading to elevated star formation, increased AGN luminosities, and higher levels of nuclear obscuration compared to earlier merger phases \citep{Ricci17,Koss18,Troncoso25}. 
These systems therefore trace an advanced stage of interaction where efficient orbital decay and merger-driven accretion are expected to occur. To distinguish them from dual AGN at larger separations, we refer to these systems as \duets.
At smaller separations, of the order of parsecs, the two SMBHs become gravitationally bound, forming a binary system. The subsequent evolution is driven by gravitational interactions with surrounding stars and gas \citep[see][for a review]{Colpi14}, which extract energy and angular momentum from the binary and further shrink its orbit. At milliparsec scales, gravitational wave (GW) emission dominates the removal of energy and angular momentum, driving the inspiral and eventual coalescence of the SMBHs \citep{Amaro-Seoane2023}.

The \duets\ systems targeted in this work represent a specific subset of dual AGN at $0.5 \leq z \leq 3.0$ with projected separations of $\lesssim 7$ kpc, for which both AGN are expected to reside within the same merger remnant \citep{Volonteri22,Chen23b}.
Characterizing their observed projected separations, accretion rates, black-hole masses, and host-galaxy properties provides key constraints on the efficiency of orbital decay and the physical mechanisms driving SMBH pairing.
In particular, the combination of SMBH masses and separations determines the timescales for dynamical friction, stellar hardening, and gas-driven migration, ultimately setting how efficiently systems can reach the GW-dominated regime. These quantities directly impact both the amplitude and the spectral shape of the stochastic GW background, as well as the event rates of individually resolvable mergers \citep{Sesana08}. Future missions such as Laser Interferometer Space Antenna (LISA) will be sensitive to coalescing SMBHs across a broad mass range, from $\sim10^{4}$ to $10^{8}\,M_{\odot}$ out to redshifts of $z\sim30$, with peak sensitivity to systems hosting SMBHs of $\sim10^{5-6}\,M_{\odot}$ \citep{Colpi2024arxiv,AmaroSeoane2023}.
In contrast, Pulsar Timing Array (PTA) experiments probe the low-frequency GW spectrum ($10^{-9}$–$10^{-7}$ Hz), providing evidence of a stochastic GW background generated by slowly inspiraling, more massive binary SMBHs ($\gtrsim 10^{8}\,M_{\odot}$) at redshifts $z \lesssim 2$ \citep{Agazie2023a,EPTACollaboration2023,2023ApJ...951L...6R,2023RAA....23g5024X}.
Therefore, measuring masses, luminosities, and accretion properties of dual AGN provides a fundamental observational anchor for theoretical predictions of SMBH merger rates and the resulting GW signal. Quantifying their abundance and demographics is essential to link galaxy evolution with the emerging GW landscape \citep{derosa20, puertosanchez24,villalba_lisa26}.

In the local Universe, studies of dual AGN have provided detailed insights into the physical processes occurring during galaxy mergers, including SMBH accretion, gas inflows, star formation, and feedback (e.g. \cite{Treister18,Treister20,Koss23,Ceci26,Lamperti26,desi_dual,mezcua26,Kouroumpatzakis26}). However, it is still unclear whether these nearby systems are representative of the dual AGN population at higher redshift. Large samples of distant dual AGN are therefore essential to study their evolution across cosmic time.

\begin{figure*}[ht!]
        \centering
        \includegraphics[width=\linewidth]{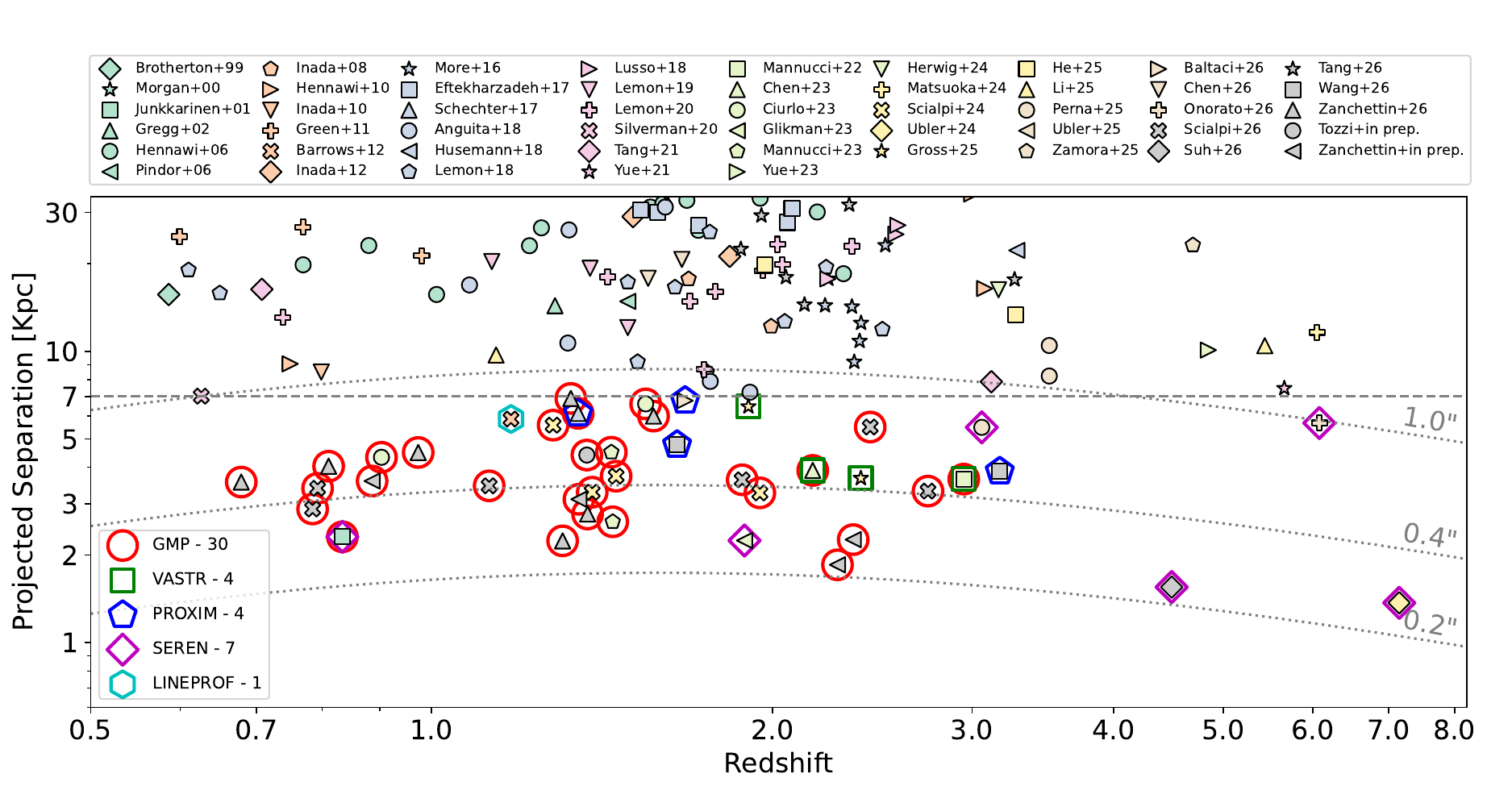}
    \hfill
        \caption{Distribution of spectroscopically confirmed dual AGN systems in the redshift--projected separation plane. Literature sources are shown as filled symbols, with references annotated. Systems with projected separations below 7 kpc are additionally marked with colored symbols indicating the selection technique. The GMP systems presented in this work are highlighted by red circles.}
        \label{fig:distrib_dualAGN}
\end{figure*}

The direct identification of \duets remains challenging because of their intrinsic rarity, which requires wide-area surveys to build statistically significant samples, as well as the high spatial resolution needed to resolve the two nuclei.
A number of dual AGN systems have been discovered serendipitously (\texttt{SEREN}) in targets observed for other reasons \citep{Junkkarinen01,Inada08,Inada10,Inada12,Silverman20,Glikman23,Husemann18}, and more recently with the \textit{James Webb Space Telescope} (JWST) at high redshift \citep{Ubler2024dual,Ubler2025triple,Maiolino2024dual,Matsuoka2024,Li+25,Perna2025dual,Zamora25,suh26,Onorato26}. Several dedicated observational techniques have also been developed to identify dual AGN candidates, each probing different regions of parameter space and subject to specific selection biases in luminosity, redshift, and projected separation. Early approaches include the selection of double-peaked emission lines (\texttt{LINEPROF}; \citealt{Ge12_doublepeaked,Barrows12}) and spectroscopic searches for close galaxy and quasar pairs (\texttt{PROXIM}; \citealt{Morgan00,Hennawi06,Hennawi10,Anguita18,Green11,Liu11_sdsspair,Scheckter17,Lemon18,Lemon19,Lemon20,yue+21,Chen22a,yue+23,Shen23b,Jing25_desidual,Tang+26,wang+26,Chen+26}). 
Complementary multi-wavelength approaches include the identification of multiple compact nuclei in high-resolution radio surveys \citep{Fu15_a,Fu15_b,Baltaci+26}, mid-infrared selections \citep{Satyapal17,Pfeifle19_triple}, and X-ray-selected AGN pairs identified through the combination of X-ray, optical, and infrared diagnostics \citep{DeRosa2023,Battistini26}. More recent techniques exploit variability \citep{Kochanek06,KroneMartins19} and astrometric jitter (“vastrometry”; \citealt{Shen19,Hwang20,Gross23,gross25_vodka,Chen23a}), enabling the discovery of systems at sub-arcsec separations. Machine-learning techniques applied to \textit{Euclid} imaging data have recently emerged as a promising avenue for identifying dual AGN candidates \citep{Ulivi2025ArXiv}.
The outcomes of the different observational techniques used to identify dual AGN systems at $z>0.5$ and the corresponding confirmed samples are reported in Fig.~\ref{fig:distrib_dualAGN}, where systems with $<7$\,kpc separations  are marked according to their selection method. However, these methods often suffer from heterogeneous selection functions and limited completeness, especially at sub-arcsec separations.

The Gaia multi-peak technique \citep[\texttt{GMP};][]{Mannucci22} provided a major step forward, enabling systematic all-sky searches for close dual AGN using the \textit{Gaia} mission. The method identifies candidates by detecting multiple peaks in the one-dimensional \textit{Gaia} scan profiles.
Thanks to the combination of high angular resolution (0.11" along the scan direction) and full-sky coverage, GMP provides a relatively homogeneous and well-characterized selection of dual AGN candidates at sub-arcsec separations. A hundred of these systems have been spectroscopically confirmed through spatially resolved observations from both space- and ground-based facilities \citep{Junkkarinen01,Scheckter17,Mannucci22,Mannucci23,Ciurlo23,Chen23a,Scialpi24,Scialpi+25_arxive}. This effort is still ongoing but has already established one of the largest homogeneous samples of 30 confirmed dual AGN at kpc-scale separations.
As shown in Fig.~\ref{fig:distrib_dualAGN}, the GMP selection probes a previously underexplored region of the parameter space, corresponding to close nuclear separations at cosmic noon (i.e. $r_{\rm p}<$~7 kpc), and provides a unique opportunity to study dual AGN in the late stages of galaxy mergers. For the first time, a statistically significant sample of dual AGN can be used to test predictions for the formation and evolution of SMBH pairs across cosmic time.

A wide range of modelling approaches has been developed, which can be broadly organised following the scheme proposed by \citet{villalba_lisa26}.
Large-volume cosmological hydrodynamical simulations form the backbone of this effort, enabling statistically representative galaxy populations. This class includes \textsc{Illustris}\ \citep{Genel14, Vogelsberger14}, its successor suite \textsc{IllustrisTNG}\ (TNG50, TNG100, TNG300) \citep{Pillepich18, Nelson19}, \textsc{EAGLE} \citep{Schaye15, Rosas-Guevara19}, \textsc{Horizon-AGN} \citep{Dubois16, Volonteri16}, SIMBA \citep{Dave19, Thomas19}, \textsc{BlueTides} \citep{Feng16}, \textsc{Astrid} \citep{Ni22_astrid, Bird22, Chen23b}, and \textsc{FABLE} \citep{Buttigieg25_fable}.
These are complemented by high-resolution zoom-in simulations, which target individual halos or merger systems and provide a more detailed treatment of SMBH pairing and accretion physics. Representative examples include \textsc{NewHorizon} \citep{Volonteri22} and \textsc{Romulus} \citep{Saeedzadeh24}, which achieve significantly improved spatial resolution in selected environments.
In parallel, semi-analytic models and dark-matter-based structure formation frameworks provide a complementary and computationally efficient approach to SMBH evolution. These include semi-analytic models such as \textsc{L-GalaxiesBH} \citep{Izquierdo-Villalba22_Lgalaxies, Spinoso23,Bonoli25_Lgalaxies}, \textsc{GALFORM} \citep{Cole00_galform,Benson10_galform}, and \textsc{SHARK} \citep{Lagos18_shark}, as well as particle-based or merger-tree frameworks such as \textsc{Pinocchio} \citep{Monaco02,Munari17,singh26}.

These models predict a wide range of properties for dual and binary SMBH systems, including the distribution of separations, mass ratios, accretion rates, luminosities, and their redshift evolution \citep{Steinborn2016,Capelo17,Rosas-Guevara19,Volonteri22,Chen22a,2023MNRAS.519.2083I}. In particular, they consistently show that SMBH growth and orbital decay are most efficient during the late stages of galaxy mergers, where gas inflows and dynamical interactions enhance SMBH accretion and drive the inward black-hole migration. They also predict significantly different number densities of these systems across models, reflecting the strong dependence on the underlying physical assumptions and subgrid prescriptions \citep{puertosanchez24}.
Simultaneously capturing cosmological volumes and resolving kpc-to-pc scale physics remains computationally challenging. Large-volume simulations are required to sample statistically significant populations of mergers, while high spatial resolution is essential to trace the detailed dynamics of merger remnants and properly model orbital decay, feedback, and variability. As a result, different simulation approaches are complementary: large-volume runs provide statistical predictions for SMBH populations, whereas zoom-in and high-resolution simulations are crucial to connect these predictions to systems comparable to the observed \duets.

In this paper, we present the first statistical study of the physical properties of an homogeneously-selected sample of \duets. We investigate the distributions of black-hole masses and Eddington ratios for both components of each system as a function of redshift and luminosity, exploiting the well-characterized selection function of the GMP sample to obtain robust constraints on the population of \duets. We then compare these observational results with predictions from state-of-the-art cosmological simulations and semi-analytical models, in order to probe the accretion physics of both AGN components during galaxy mergers. 
This comparison also allow us to refine predictions for the SMBH merger population relevant to GW experiments, including PTA and LISA.

The paper is structured as follows. 
In Sect.~\ref{sec:sample} we describe the sample and data used in this work. In Sect.~\ref{sec:specfit} we present the spectral fitting procedure and the measurement of AGN properties. In Sects.~\ref{sec:properties} and \ref{sec:sdsscomparison} we derive and present the distributions of bolometric luminosities, Eddington ratios, and black-hole masses. Finally, in Sect.~\ref{sec:models} and Sect.~\ref{sec:comparison} we introduce the simulations and compare them with the observational results.
We adopted a flat $\Lambda$CDM cosmology with parameters from \cite{Planck2020}, with $H_{\rm 0}=67.7\ {\rm km\ s^{-1}\ Mpc^{-1}}$ and $\Omega_{\rm m,0}=0.31$.

\section{Sample selection}
\label{sec:sample}

To date, the literature reports 39 spectroscopically confirmed dual AGN with proper projected separations below 7 kpc and redshifts in the range $0.5 \le z \le 3.0$ (see Fig.~\ref{fig:distrib_dualAGN}). 
Among these, 30 systems are identified via the GMP technique presented in \cite{Mannucci22}, and constitute the sample analysed in this paper.
In particular, 27 systems were selected directly within the GMP framework \citep[Zanchettin et al.\ (subm.); Zanchettin, Ceci et al.\ in prep.; Tozzi et al.\ in prep.]{Mannucci22, Mannucci23, Ciurlo23, Scialpi24, Scialpi+25_arxive}, while 3 additional systems were initially discovered through alternative techniques: serendipitous identification (\citealt{Junkkarinen01}), photometric color selection (\citealt{Scheckter17}), and astrometric methods (\citealt{Chen23a}), and were later found to also satisfy the GMP selection criteria, leading to their inclusion within the same framework.
In this work, we focus on the 30 systems that satisfy the GMP selection criteria to enable a statistical analysis based on a well-defined and quantifiable selection function.
Of these 30 GMP systems, J1837--6711 has only the H$\delta$ emission line available in the observed wavelength range, which cannot be used to estimate $M_{\rm BH}$ with the single-epoch (SE) virial method, while J0103--2753 exhibits strong broad absorption in the brighter component, preventing a reliable $M_{\rm BH}$ estimate. We therefore exclude these two systems from the $M_{\rm BH}$ measurements and related analyses. 

As shown by \citet{Mannucci23}, the GMP method efficiently selects type 1 AGN with angular separations ranging from $\sim 0.15''$ to $0.8''$, and magnitude limits of $G_{\rm brighter} \leq 20.5$ and $G_{\rm fainter} \leq 21.0$ for the brighter and fainter AGN of the pair, respectively. The method identifies dual systems in Gaia DR3 \citep{Gaia_EDR3_1} at $z \gtrsim 0.5$, corresponding to projected separations below $\sim 7$ kpc at these redshifts. At these scales, contamination from the diffuse host galaxy emission in the Gaia detections is negligible \citep{Mannucci22}, maximizing the completeness of the dual AGN selection.

All these systems have been spectroscopically confirmed using high spatial resolution facilities, including the Multi Unit Spectroscopic Explorer (MUSE, \citealt{Bacon10}), the Enhanced Resolution Imager and Spectrograph (ERIS, \citealt{Davies2023_eris}) at the Very Large Telescope (VLT); the Space Telescope Imaging Spectrograph (STIS, \cite{STIS98}) on board the Hubble Space Telescope (HST); the OH-Suppressing Infrared Imaging Spectrograph (OSIRIS, \citealt{osiris_Larkin2006}) at the Keck Observatory; the Gemini Multi-Object Spectrograph (GMOS, \citealt{Gemini_north04}) at the Gemini Observatory; and the Inamori Magellan Areal Camera and Spectrometer (IMACS, \citealt{IMACS_baade}) on the Magellan Baade telescope \citep{Magellan_telescope03}.
In terms of wavelength coverage, MUSE, GMOS, STIS, and IMACS operate in the optical, typically probing rest-frame UV-optical emission lines (depending on redshift), such as \mgii\ $\lambda2799$, \civ\ $\lambda1549$, H$\beta$ $\lambda4861$, and [O\textsc{iii}] $\lambda\lambda4959,5007$. In contrast, ERIS and OSIRIS probe the near-infrared, accessing rest-frame optical lines such as H$\alpha$ and H$\beta$ at the redshifts considered.
Details of the instrumental configurations, spectral and spatial resolutions, and data-reduction procedures can be found in the original confirmation studies.
All observations have sufficient spatial resolution to resolve the two AGN components ($\lesssim0.1''$) and adequate spectral resolution to characterize their emission-line profiles ($\sim 30{-}150\ \mathrm{km s^{-1}}$).


\section{Spectral fitting}
\label{sec:specfit}

To characterize each individual Type 1 AGN in the pair and derive the physical quantities used throughout this work, we perform a detailed spectral fitting of each extracted spectrum after correcting for Galactic extinction. In particular, the fitting procedure provides accurate measurements of the systemic redshift, the emission-line fluxes and widths, and the continuum luminosity. For each AGN, a single spectrum was extracted (see \cite{Mannucci23, Ciurlo23, Scialpi+25_arxive} for details).
We apply a custom interactive fitting procedure designed to simultaneously reproduce each spectrum with a power-law continuum, emission-line components arising from the broad-line region (BLR) and narrow-line region (NLR), Fe\textsc{ii} emission complexes, and, when present, interstellar medium (ISM) absorption features. The continuum emission was modeled with a power-law function defined as:
\begin{equation}
F(\lambda) = A \left( \frac{\lambda}{\lambda_{\rm ref}} \right)^{\alpha},
\label{eq:cont_powerlaw}
\end{equation}
where $\lambda_{\rm ref}$ is the reference wavelength, $\alpha$ is the spectral index, and $A$ is the luminosity density at $\lambda_{\rm ref}$.

Broad emission lines associated with the BLR were modeled using different analytical profiles, including asymmetric exponential profiles convolved with a Gaussian, Lorentzian profiles, Gaussian profiles convolved with a broken power law, and single Gaussian functions. The adopted profile was selected on a case-by-case basis through $\chi^2$ minimization to provide the best representation of the observed line shape, following common practice in AGN spectral analyses (e.g., \citealt{Mejia-Restrepo16, Kollatschny13, Maiolino2024dual}). We verified that the choice of functional form does not significantly affect the derived line widths or virial SMBH mass estimates. In all cases, the differences in FWHM between models are within the measurement uncertainties, with a negligible impact on $M_{\mathrm{BH}}$ compared to the typical SE systematic uncertainty of $\sim0.4$ dex.

Narrow emission lines from the NLR were modeled as single Gaussian component grouped into line systems with tied velocity shifts and dispersions, with flux ratios tied when required by atomic physics. In particular, we fixed the following ratios:
{\small{
[O\textsc{iii}] $\lambda5007/\lambda4959 = 3$,
[N\textsc{ii}] $\lambda6583/\lambda6548 = 3$, and
[O\textsc{i}] $\lambda6300/\lambda6363 = 3$.
}}
The Fe\textsc{ii} emission complexes were modeled using linear combinations of \texttt{Cloudy}-based templates \citep{Ferland98}, each with an independent normalization. The templates were generated with \texttt{CLOUDY} v23.01 \citep{Gunasekera23}, adopting the AGN ionizing continuum table of \cite{Mathews_Ferland87}. We used the full Fe,\textsc{ii} atom model implemented in \texttt{CLOUDY}, based on the Stout database \citep{Ferland13} with updated atomic data from \cite{Smyth19}, ensuring that all relevant atomic levels and transitions were included.

Our grid spans a wide range of physical conditions, with hydrogen densities in the range $\log(n_\mathrm{H}/\mathrm{cm}^{-3}) = 8$--14 and ionizing photon fluxes $\log(\Phi/\mathrm{cm^{-2}s^{-1}}) = 17$--23, sampled in steps of 2 dex. We considered both Solar and super-Solar metallicities (up to $20\ Z_\odot$), as well as sub-Solar values down to $0.1\ Z_\odot$. However, given the typical SMBH masses of the systems in our sample at Cosmic Noon, Solar metallicity is expected to provide a more motivated description of the BLR gas, which is generally found to be metal-rich compared to the host ISM in star-forming galaxies. Since the Fe \textsc{ii} templates are applied with an independent normalization, the impact of moderate metallicity variations (e.g. {\small {$Z \sim 0.5$--$1\ Z_\odot$}}) on the resulting fits is expected to be limited.

We also considered models with either no turbulence or a microturbulent velocity of $300~\mathrm{km\ s^{-1}}$, adopted as an extreme value based on Fig.~7 of \cite{Temple21}, where such velocities are shown to significantly affect the strength and shape of UV resonance lines such as \ion{C}{iv} and \ion{N}{v}. This choice is intended to bracket the upper limit of plausible microturbulent broadening in BLR clouds and to assess its impact on the Fe\textsc{ii} emission.

When present, narrow absorption features were modeled as negative Gaussian component, allowing for both intrinsic absorbers associated with the AGN and therefore placed at the source redshift, and intervening systems along the line of sight.

The model parameters were optimized through $\chi^2$ minimization using the Levenberg-Marquardt algorithm implemented in the Python package \texttt{Lmfit} \citep{lmfit}, with the $\chi^2$ statistic computed from the flux uncertainties of each spectral pixel. Parameters entering linearly in the model, primarily component normalizations and template weights, were solved at each iteration using a non-negative least-squares approach to ensure physically meaningful positive fluxes. The remaining non-linear parameters, including continuum slope, velocity shifts, velocity dispersions, and higher-order kinematic coefficients, were varied freely during the minimization. This hybrid strategy reduces parameter degeneracies and promotes robust convergence toward the best-fitting solution.

Uncertainties on the fitted parameters and derived quantities were estimated through Monte Carlo simulations. The best-fit model was first obtained from the observed spectrum, and the residuals, defined as data minus model, were used to estimate the noise level. Multiple realizations were then generated by adding Gaussian noise drawn from the standard deviation of the residual distribution, and each realization was re-fitted using the same procedure as the original spectrum. The uncertainties were finally derived from the 16th and 84th percentiles of the resulting distributions, accounting for observational noise, parameter degeneracies, and line-profile asymmetries.

Representative examples of the resulting spectral fits are presented in Appendix~\ref{app:fit}.

\subsection{Line Measurements and Continuum Luminosity}
\label{sec:line_measurements}
\label{sec:continuumlum}
We measured fluxes and FWHM of the lines of the BLR from the modeled line profiles. Equivalent widths (EW) of the broad emission lines were computed as the ratio between the integrated line flux and the continuum flux density at the line center ($\lambda_0$). We consider fluxes, FWHM and EW as complementary diagnostics of the broad-line region properties in dual AGN systems: while FWHM traces the kinematics of the emitting gas, EW provides a measure of the line strength relative to the continuum and is therefore sensitive to differences in accretion rate and continuum emission between the two components.
We focus on FWHM for SE virial black hole mass estimates, while both FWHM and EW are used in the comparative analysis of the two AGN within each system. 
Uncertainties were estimated following the Monte Carlo procedure described in Sect.~\ref{sec:specfit}. 

The continuum emission is modeled using the best-fit AGN power-law continuum (Eq.~\ref{eq:cont_powerlaw}). Monochromatic flux densities are evaluated from the best-fit model at standard rest-frame wavelengths (1350\AA, 3000\AA, 4400\AA, and 5100\AA), selected according to available spectral coverage and the emission line used for the virial mass estimate. Specifically, we adopt~$\lambda_0\simeq~3000$\AA\ for \ion{Mg}{II},~$\lambda_0 \simeq 5100$ \AA\ for \ion{H}{$\beta$} and~H$\alpha$,~$\lambda_0~\simeq~1350$\AA\ for~\ion{C}{IV}, and the additional $\lambda_0 \simeq 1700$ \AA\ for \ion{C}{III}]. When the required $\lambda_0$ falls outside the observed spectral range, the continuum power-law is extrapolated. We also evaluate the continuum flux density at 4400 \AA, which provides a common reference for deriving homogeneous bolometric luminosities across the full sample \citep{duras20}, independently of the emission line adopted for the virial mass estimate. Monochromatic luminosities, {\small $L_{\lambda}(\lambda_0)$}, are then computed from the continuum model using the luminosity distance.


\section{Estimate of dual AGN physical properties}
\label{sec:properties}
In this section, we derive the fundamental AGN properties for the dual AGN sample, including the relative velocity ($\Delta v$), bolometric luminosities ($L_{\rm bol}$), and Eddington ratios ($\lambda_{\rm Edd}$). 
These quantities are used to characterize each AGN in the dual system and to compare their accretion properties with those of single AGN and with model predictions.

\subsection{Relative velocity}

\begin{figure}[h!]
\centering
\includegraphics[width=1\linewidth]{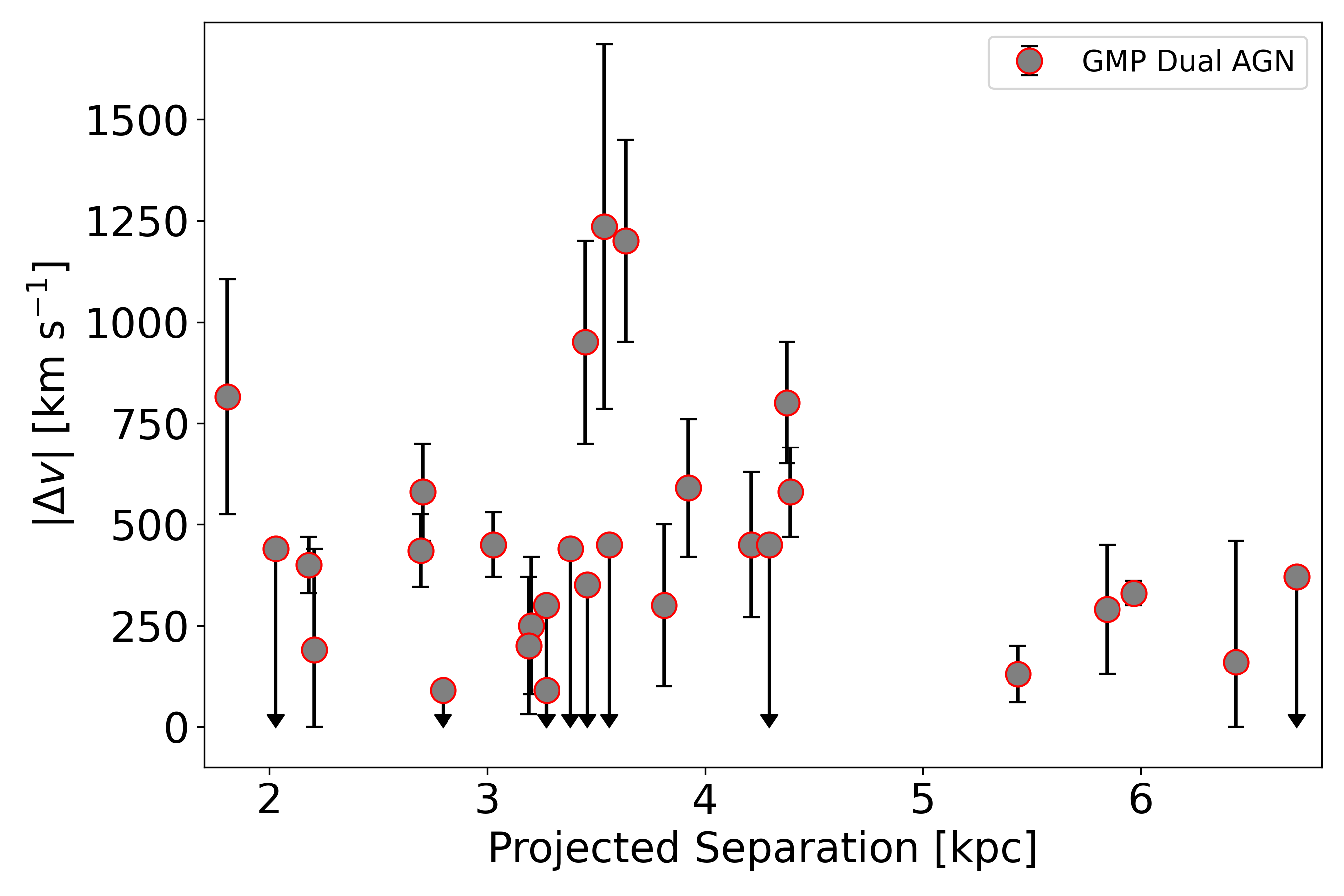}
\caption{
Line-of-sight velocity differences as a function of projected separation for the \duets sample.}
\label{fig:velocity}
\end{figure}

The combination of projected separation and relative line-of-sight velocity is used to assess whether galaxy pairs are physically associated and are undergoing interaction. 
Figure~\ref{fig:velocity} shows velocity offset as a function of projected separation for the 30 GMP dual AGN sample.
$\Delta v$ is defined as the difference between the line-of-sight velocities of the two nuclei, derived from redshifts measured independently through spectral fitting of the narrow emission lines, when present, or otherwise of the strongest available broad emission line over the full extracted spectral range. The redshift difference is converted into a velocity offset following the prescription of \citet{hogg99}. The effective velocity resolution is evaluated at the observed wavelength of the spectral features used to determine the redshift.
For systems with velocity separations smaller than the effective velocity resolution, we report $\Delta v$ as an upper limit. When the uncertainty on the systemic redshift exceeds the instrumental velocity resolution (e.g., when only broad emission lines are available), the upper limit is instead determined by the redshift uncertainty.

The measured line-of-sight velocity offsets span a broad range, with most systems exhibiting $|\Delta v| \lesssim 600$ km s$^{-1}$ over projected separations of $\sim1$--7 kpc. These values are consistent with expectations for interacting galaxy pairs. Using reconstructed orbits of galaxy pairs in the IllustrisTNG100 simulation \citep{Patton24_illustristng}, \citet{Bigmac25} showed that 95\% of systems that experienced a close passage within the previous $\sim1$ Gyr occupy the region defined by projected separations $r_{\rm p}\lesssim110$ kpc and line-of-sight velocity differences $|\Delta v|\lesssim600$ km s$^{-1}$.
Most of our sources fall within this expected velocity range. The few exceptions are associated with large uncertainties in the systemic redshift determination and therefore do not alter this overall picture. In J2204--6530 \citep{Scialpi+25_arxive}, absorption features affect the emission lines used for the redshift estimate, resulting in a large uncertainty in the systemic redshift. Similarly, J0103--2753 falls outside the range shown in Fig.~\ref{fig:velocity}, with a reported $\Delta v \sim 3900$ km s$^{-1}$ based on the C{\sc iv} emission-line centroids, which are affected by strong broad absorption and possible C{\sc iv} blueshifts \citep{Junkkarinen01}. In J0300--0532 (Zanchettin et al., subm.), the H$\alpha$ emission line lies close to the edge of the spectral coverage, reducing the accuracy of the centroid measurement. Finally, in J0052--3045 \citep{Scialpi24}, narrow emission lines are detected only in the brighter component of the pair, while the systemic velocity of the fainter nucleus is derived from broader spectral features due to the lower signal-to-noise ratio.


\subsection{Bolometric luminosity}
\label{sec:lbol}

We derive bolometric luminosities from the continuum measurements described in Sect.~\ref{sec:continuumlum}. Specifically, we compute $L_{\rm bol}$ from monochromatic continuum luminosities evaluated at 3000\ \AA, 4400\ \AA, and 5100\ \AA, allowing us to assess the consistency of different bolometric corrections and to construct a homogeneous luminosity scale across the full sample.

Bolometric luminosities are computed as
$
L_{\mathrm{bol}} = k_{\mathrm{bol}}\,(\lambda L_\lambda),
$
where the bolometric correction factors are $k_{\rm bol}=5.15$ at 3000\ \AA\ \citep{Richards06}, $k_{\rm bol}=5.13$ at 4400\ \AA\ \citep{duras20}, and $k_{\rm bol}=9.26$ at 5100\ \AA\ \citep{Richards06}. The statistical uncertainties are propagated from the continuum measurements, and the intrinsic scatter of the adopted bolometric correction is added in quadrature to obtain the final uncertainty.
We adopt these wavelength-dependent corrections from the literature to remain consistent with the original empirical calibrations. They provide conversions between monochromatic and bolometric luminosities and do not imply different assumptions on the intrinsic AGN spectral energy distribution.

To define a uniform bolometric luminosity estimator, we adopt 4400$\AA$~as the common reference wavelength. This wavelength lies within the optical continuum and provides a suitable anchor for the different spectral configurations in our sample, minimizing the extrapolation required for objects observed around~H$\alpha$,~H$\beta$,~or~\ion{Mg}{II}. 
For each source, the continuum luminosity is measured directly at~4400$\AA$~when covered by the spectrum; otherwise, it is inferred from the nearest available continuum measurement using the empirical inter-wavelength relations described in Appendix~\ref{app:lbol}. This procedure places all objects on a homogeneous luminosity scale before applying the corresponding bolometric correction.

We find that the bolometric luminosities derived at different wavelengths are tightly correlated, with the smallest scatter between the 4400 and 5100\ \AA\ estimates. The adopted calibration is robust against continuum extrapolation and does not introduce systematic differences between the various spectral subsamples. As an additional validation, the resulting bolometric luminosities are compared with independent line-based estimates (Appendix~\ref{app:lbol}), showing no significant systematic offsets for either H$\alpha$- or \ion{Mg}{II}-based measurements (Fig.~\ref{fig:lbol_lines}).


\subsection{Single-Epoch Black-Hole Mass Estimates}
\label{sec:masses}
Black-hole masses were estimated using the SE virial method, which assumes that the BLR gas is virialized and that its size scales with the AGN luminosity through the empirical radius--luminosity relation derived from reverberation mapping studies. Following \citet{WuShen22}, we derived black-hole masses from continuum and broad emission-line measurements using the calibrations of \citet{VestergaardPeterson06} (hereafter VP06) for H$\beta$ and C\,\textsc{iv}, and \citet{Shen+11} for Mg\,\textsc{ii}. For these estimators, the black-hole mass is computed as:
\begin{equation}
\log\left(\frac{M_{\mathrm{BH}}}{M_\odot}\right)
=
a
+
b\,\log\left(
\frac{\lambda L_\lambda}
{10^{44}\ \mathrm{erg\ s^{-1}}}
\right)
+
A\,\log\left(
\frac{\mathrm{FWHM}}
{10^3\ \mathrm{km\ s^{-1}}}
\right), \label{eq:MBH_SE}
\end{equation}
where $\lambda L_\lambda$ is the monochromatic continuum luminosity at 5100, 3000, and 1350\ \AA\ for H$\beta$, Mg\,\textsc{ii}, and C\,\textsc{iv}, respectively. The adopted coefficients $(a,b,A)$ are reported in Table~\ref{tab:virial}.
Several studies discussed the unreliability of C\,\textsc{iv}-based mass estimates, due to gas outflows strongly affecting the profile of the line \citep{Netzer07,Assef11,Marziani12,Runnoe13}. However, C\,\textsc{iv}-based estimates are used for less than 10\% of our sample, where no alternative tracers are available.

In addition, we include the C\,\textsc{iii}] $\lambda1909$ emission line as a virial proxy \citep[e.g.,][]{Trevese14}. The C\,\textsc{iii}] feature is a semi-forbidden doublet that is blended with nearby emission lines (primarily Si\,\textsc{iii}] $\lambda1892$ and Al\,\textsc{iii} $\lambda1857$) and may also include contributions from non-virial gas components outside the BLR. Given these complexities and the possible presence of line asymmetries, we adopt the calibration of \citet{Marziani22} and apply a multiplicative correction factor to its measured FWHM ($\xi_{{\rm C\,\textsc{iii}]}} \approx 1.25$). The C\,\textsc{iii}]-based estimate is used for only one object in our sample, and therefore it does not affect our results.

For H$\alpha$, we adopt the line-luminosity-based virial calibration of \cite{Shen+11}, originally derived by \cite{GreeneeHo05}:

\begin{equation}
\log\left(\frac{M_{\mathrm{BH}}}{M_\odot}\right)
=
a
+
b\,\log\left(
\frac{L_{\mathrm{H}\alpha}}
{10^{42}\ \mathrm{erg\ s^{-1}}}
\right)
+
A\,\log\left(
\frac{\mathrm{FWHM}_{\mathrm{H}\alpha}}
{\mathrm{km\ s^{-1}}}
\right). \label{eq:MBH_SE_Ha}
\end{equation}

Unlike the VP06 H$\beta$ calibration, which uses the continuum luminosity to estimate $R_{\rm BLR}$, this relation relies on the H$\alpha$ line luminosity. The two calibrations are broadly consistent, with a mean offset of $\sim 0.08$ dex and a scatter of $\sim 0.18$ dex for broad-line quasars \citep{Shen+11}, consistent with the dispersion between H$\alpha$- and H$\beta$-based virial mass estimates.

These calibrations combine luminosity and broad-line width in a framework widely adopted in the literature. Uncertainties on $\log M_{\mathrm{BH}}$ are obtained by propagating the independent errors on luminosity and FWHM, adding in quadrature the intrinsic scatter of the adopted virial relations (Table~\ref{tab:virialmass}).

We also note that SE black-hole mass estimates are subject to significant systematic uncertainties, which are further amplified by the use of different broad emission lines across the sample. 
Since only two objects have multiple emission lines available in their observed spectra, we do not perform a formal statistical comparison between the different virial estimators. However, for J2100+0612 and J0144$-$5745, for which $M_{BH}$ can be independently estimated from both H$\beta$ and Mg\textsc{ii}, we find differences of $\Delta \log(M_{\rm BH})<0.1$ dex, indicating excellent consistency between the estimators and well within the typical uncertainties of SE virial mass estimates ($\sim 0.4$ dex).

\begin{table*}[ht]
\centering
\caption{Coefficients adopted for the single-epoch virial black-hole mass estimators used in this work (Eq.~\ref{eq:MBH_SE} and Eq.~\ref{eq:MBH_SE_Ha}). Scatter values represent the intrinsic dispersion of the corresponding calibrations.}
\label{tab:virial}
\renewcommand{\arraystretch}{1.1}

\begin{tabular}{l|cccccl}
Line & $a$ & $b$ & $A$ & Scatter (dex) & Method & Reference \\
\hline
C\,\textsc{iv} $\lambda1549$   & 6.66 & 0.53 & 2.0 & 0.50 & $L_{1350}$, FWHM & \cite{VestergaardPeterson06} \\
Mg\,\textsc{ii} $\lambda2798$  & 6.74 & 0.62 & 2.0 & 0.40 & $L_{3000}$, FWHM & \cite{Shen+11} \\
H$\beta$ $\lambda4861$ & 6.91 & 0.50 & 2.0 & 0.43 & $L_{5100}$, FWHM & \cite{VestergaardPeterson06} \\
H$\alpha$ $\lambda6563$ & 0.379 & 0.43 & 2.1 & 0.43 & $L_{\mathrm{H}\alpha}$, FWHM & \cite{Shen+11}
\\
C\,\textsc{iii}] $\lambda1909$ & 6.8 & 0.676 & 2.0 & 0.35 & $L_{1350}$, FWHM & \cite{Marziani22} \\
\bottomrule
\end{tabular}
\label{tab:virialmass}
\end{table*}


\subsection{Eddington Ratios}
\label{sec:eddingtonratio}
Using the SE black-hole mass estimates, we compute the Eddington luminosities for each AGN component as:
$$
L_{\rm Edd} = 1.26 \times 10^{38} \left(\frac{M_{\rm BH}}{M_\odot}\right)\ {\rm erg\ s^{-1}},
$$
which follows from the standard definition of the Eddington limit under the assumption of pure hydrogen, spherical accretion, and electron scattering opacity. The Eddington ratio is then defined as $\lambda_{\rm Edd} = L_{\rm bol}/L_{\rm Edd}$.

We note that SE mass estimates rely on the radius--luminosity relation calibrated from reverberation-mapped AGN, which is assumed to be linear in luminosity. However, at high accretion rates ($\lambda_{\rm Edd} \gtrsim 0.1$), deviations from this relation have been reported, with systematically smaller broad-line region radii compared to the standard scaling \citep[e.g.,][]{Wang14,Du14,Du16,Du18, Lupi24}. In such cases, virial SMBH masses may be overestimated, leading to a systematic underestimation of $\lambda_{\rm Edd}$. We caution that this effect may be relevant for the highest-accretion sources in our sample, including objects with $\lambda_{\rm Edd} \sim 1$.

\section{Observational Results}
\label{sec:sdsscomparison}
\begin{figure}[ht!]
        \centering
        \includegraphics[width=\linewidth]{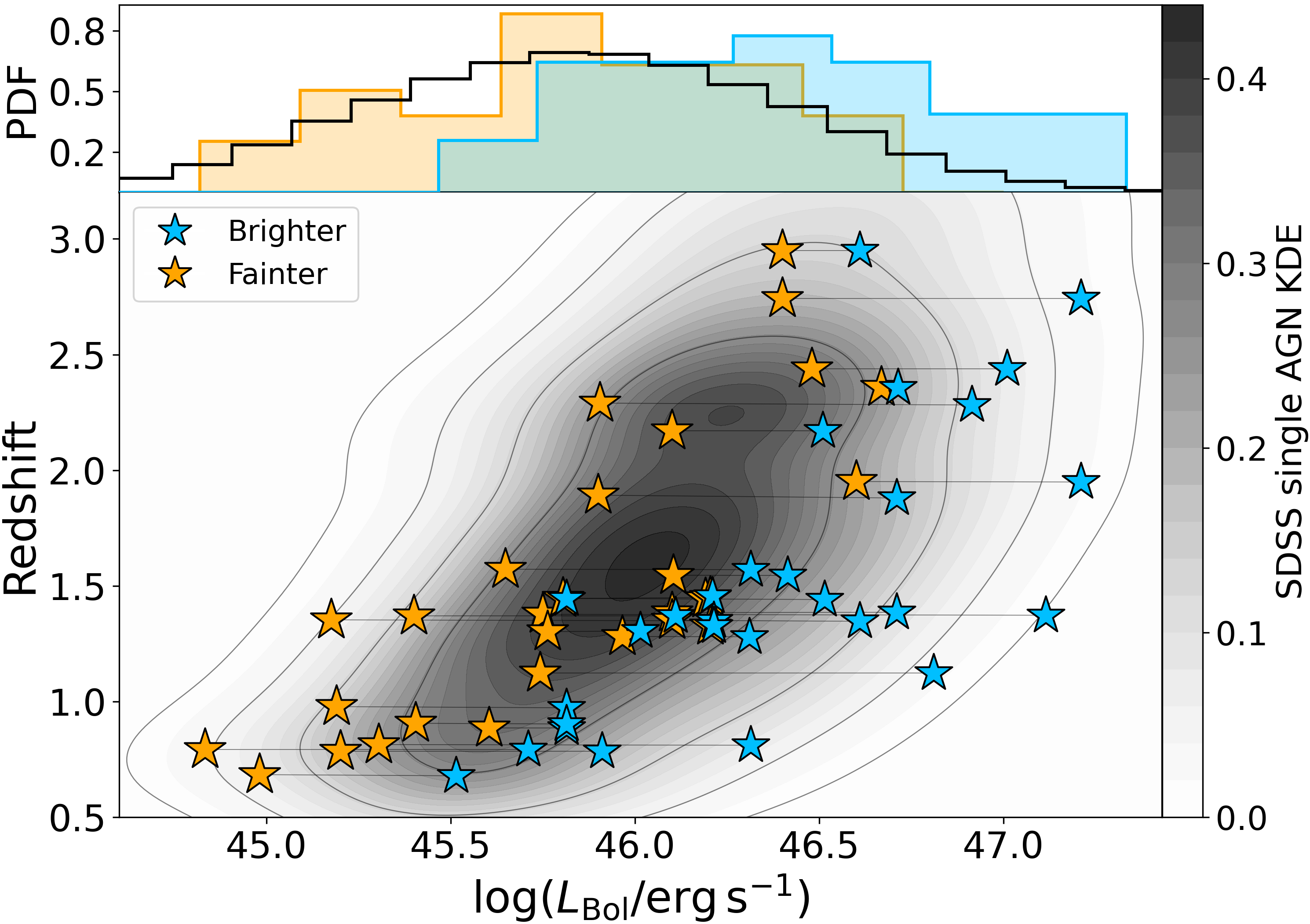}
        \hfill
        \includegraphics[width=\linewidth]{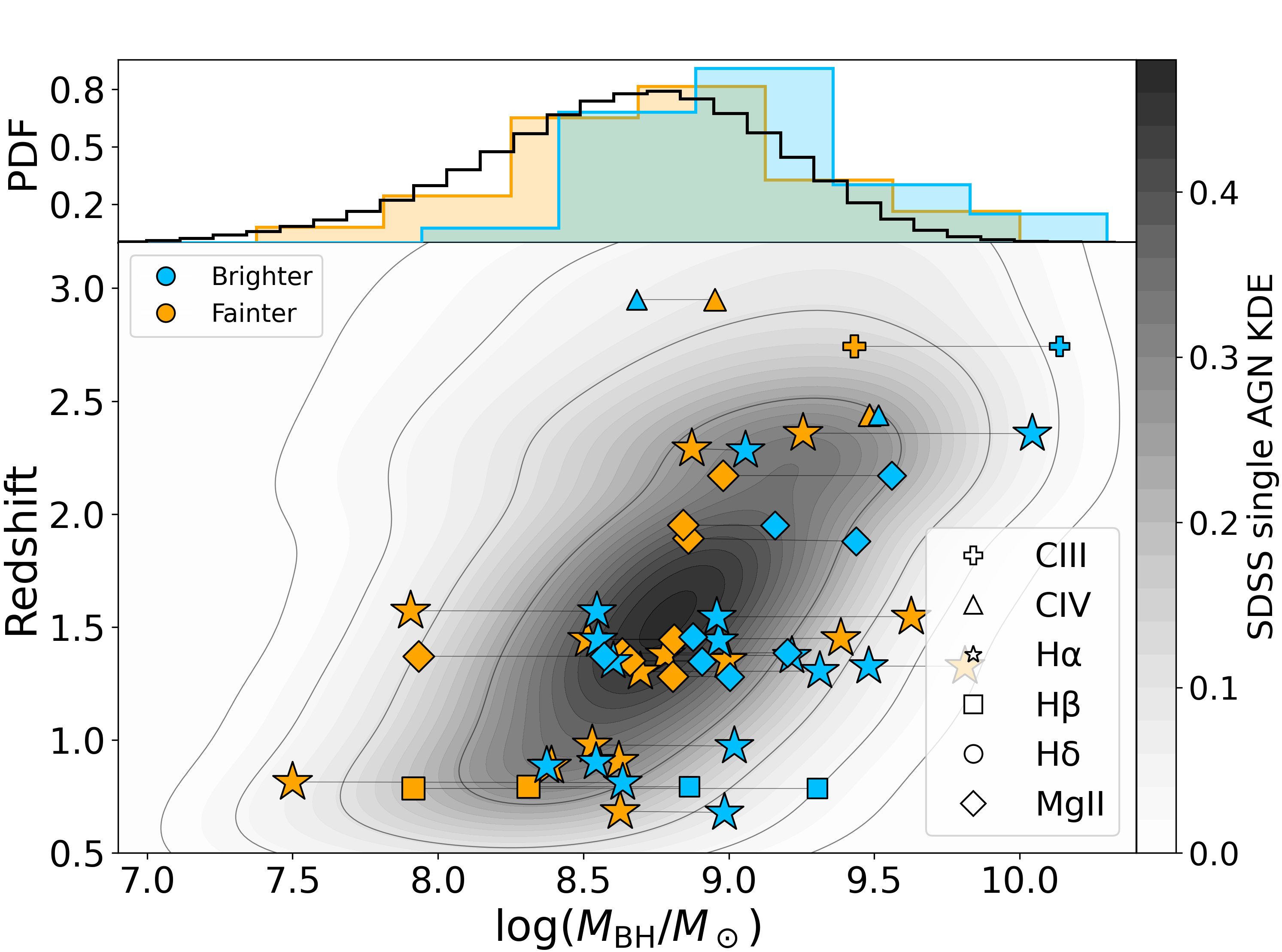}
        \caption{Bolometric luminosity (top) and black hole mass (bottom) as a function of redshift for \duets. Light-blue and orange symbols represent the brighter and fainter components of each pair, respectively. Components belonging to the same dual system are connected by grey line segments. In the right panel, marker shapes identify the broad emission line used for the SE mass estimate. Grey contours show the {\small KDE} of single SDSS AGN \citep{WuShen22} after the GMP selection cuts\protect\footnotemark. The marginal histograms show the corresponding distributions.}
        \label{fig:mass_lbol_shen}
\end{figure}

In this section, we present the observational results for our sample of 30 \duets. 
In Fig.~\ref{fig:mass_lbol_shen}, we compare the bolometric luminosity (top panel) and black hole mass (bottom panel) of the dual AGN as a function of redshift with the distribution of single SDSS AGN from \cite{WuShen22}. The brighter and fainter components of each dual system are shown in light blue and orange, respectively, while the grey contours represent the Kernel Density Estimation (KDE) distribution of the SDSS comparison sample after applying the same redshift and selection cuts as for the GMP sample (Sect.~\ref{sec:sample}).

\footnotetext{For SDSS objects at $z>2$, whose $M_{BH}$ are estimated from the C\,IV emission line, we applied a correction of $+0.4$ dex to $\log(M_{\rm BH})$ to approximately account for known systematic uncertainties affecting C\,IV-based virial mass estimates.}

The bolometric luminosities cover $\log(L_{\rm Bol}/{\rm erg\ s^{-1}})\sim44.8$--47.2, fully consistent with the SDSS single AGN population after applying the same redshift and selection cuts. As expected from the sample definition, a clear luminosity stratification is observed between the two components: the fainter AGN is, on average, $0.5$ dex less luminous than the brighter one, with a scatter of $0.4$ dex. Brighter systems preferentially populate the upper envelope of the single AGN luminosity distribution, whereas the fainter components are generally consistent with the bulk of the SDSS comparison sample. 
Since the fainter component must still be sufficiently luminous to be detected by Gaia and included in the sample, its luminosity distribution closely resembles that of the SDSS AGN population. Conversely, the brighter components are, by construction, biased toward higher luminosities and preferentially populate the upper envelope of the single-AGN luminosity distribution. 
To test whether this contrast can arise from selection effects alone, we constructed 1000 random GMP-like pairs from the SDSS AGN population, requiring $|\Delta z|<0.1$, and repeated this procedure over 100 realizations. The median luminosity difference is $0.35\pm0.07$ dex, suggesting that the observed luminosity contrast is at least partly driven by the selection function and the underlying AGN luminosity distribution, rather than necessarily reflecting an intrinsic difference in dual AGN.\\

The SMBH masses span the range $\log(M_{\rm BH}/M_\odot)\sim7.4$--10.1, consistent with the SDSS AGN population. In contrast to the luminosity distribution, although the brighter components reach the highest black hole masses, the two populations do not show evidence of a systematic mass offset, instead exhibiting a strong overlap in parameter space.
Moreover, we find that the most luminous AGN in each pair is not necessarily the most massive black hole nor the one accreting at the highest Eddington ratio.

The ratio of the Eddington ratios of the two AGN ($\lambda_{\rm Edd,brighter}/\lambda_{\rm Edd,fainter}$) show significant deviations from unity across the merger sequence, supporting a scenario in which gas is supplied asymmetrically to the two black holes during the merger process. The brighter AGN span $\lambda_{\rm Edd} \simeq 0.02$--$0.9$, with a median value of $\lambda_{\rm Edd}=0.17$, while the fainter AGN span $\lambda_{\rm Edd} \simeq 0.01$--$0.05$, with a median value of $\lambda_{\rm Edd}=0.10$. 

We then tested whether either the Eddington ratio contrast ($\lambda_{\rm Edd,brighter}/\lambda_{\rm Edd,fainter}$) or the black hole mass ratio ($M_{\rm BH,brighter}/M_{\rm BH,fainter}$) depends on the projected separation between the two AGN, finding no significant correlation. These results suggest that, within the separation range probed by our sample (\duets; $r_{\rm p}$ $<7$ kpc), the relative contributions of black hole mass and accretion to the luminosity contrast do not vary systematically with projected separation. Although accretion activity may evolve with merger stage on larger scales \citep{Battistini26}, no such dependence is evident in our data.

\section{Cosmological models}
\label{sec:models}
This section summarizes the main characteristics of the cosmological models considered for comparison with our observational results. We explore five state-of-the-art models: \textsc{L-GalaxiesBH}, \textsc{Astrid}, \textsc{Fable}, \textsc{Horizon-AGN}, and \textsc{Romulus}. \\From each simulation, we select actively accreting AGN with $L_{\rm bol}>10^{43}~\mathrm{erg,s^{-1}}$, where the bolometric luminosity is derived from the SMBH accretion rate according to the accretion and radiative-efficiency prescriptions adopted by each model. We then compute the Gaia $G$-band apparent magnitudes ($G_{\mathrm{SIM}}$) for each simulated AGN following \cite{Truant26}. Since our observational sample is selected from observed photometry, we also include dust attenuation by the ISM following \cite{Henriques15} (see Appendix~\ref{app:simulations}).

\subsection{Selection of GMP-like objects}
\label{sec:models_gmplike}
In order to reproduce our observational pairs as closely as possible, we identify dual AGN systems in each simulation and apply the GMP selection criteria. 
The two AGN in each dual system are ordered according to their apparent magnitudes, defining the brighter and fainter components. 
We then keep only systems in the redshift range $0.5\leq z\leq3.0$, with apparent magnitudes $G_{\mathrm{brighter}}\leq20.5$ and $G_{\mathrm{fainter}}\leq21.0$, and projected separations $1<r_{\rm p}\leq7$ kpc. \footnote{For the L-GalaxiesBH sample, we do not impose a sharp lower cut at $r_{\rm p}=1$ kpc, as the separation is derived from the simulation post-processing and is not an exact quantity. All the systems considered probe the kpc-scale physical regime relevant to our observations, and we do not expect small differences in $r_{\rm p}$ to significantly affect the black hole masses or luminosities \citep{Volonteri16}.}

After applying these cuts, only a subset of the simulations yields systems directly comparable to the observed GMP sample. We identify~185~systems in \textsc{L-GalaxiesBH},~31~in \textsc{Astrid},~4~in \textsc{Fable},~1~in \textsc{Horizon-AGN}, and none in \textsc{Romulus}, depending on the simulation volume, numerical resolution, and gravitational softening (see Table~\ref{tab:simulations}).
The physical-property distributions of all simulations are shown in Appendix~\ref{app:simulations}.

\textsc{L-GalaxiesBH} provides a large and diverse population of dual systems spanning a broad range of redshift, luminosity, separation, and $M_{BH}$, enabling a statistically meaningful comparison with the observed sample. 
\textsc{Astrid}, despite providing a sample size comparable to the observed one, has selected dual systems predominantly concentrated at $z\simeq0.5$, whereas our observed sample spans a broader redshift range. As a result, at fixed $G$ magnitude, these systems correspond to lower bolometric luminosities than the higher-redshift systems probed by our observations, limiting the extent to which \textsc{Astrid} can reproduce the full observed luminosity range (see Fig.~\ref{fig:lbol_primary_secondary}).
By contrast, \textsc{Fable}, \textsc{Horizon-AGN}, and \textsc{Romulus} are significantly more affected by the combined redshift and magnitude cuts, resulting in a very limited or negligible overlap with the observational sample ($<5$ targets). 
For this reason, the following subsections provide a detailed description of the \textsc{L-GalaxiesBH} and \textsc{Astrid} simulations, since they are the two most suitable simulations for comparison with our observations and are therefore used in the subsequent analysis. Details on the remaining simulations are provided in Appendix \ref{app:simulations}.

\subsection{\textsc{L-GalaxiesBH}}
The \textsc{L-GalaxiesBH} catalogs used in this work are based on the most recent super-Eddington version of the semi-analytic model presented by \citet{Bonoli25_Lgalaxies}, which builds on previous developments of the \textsc{L-Galaxies} model \citep{Spinoso23}. This new version includes several improvements over previous implementations in the modeling of massive black hole formation, growth, and dynamics, and is designed to simultaneously reproduce the nanoHz stochastic GW background observed by PTAs and the high-redshift AGN population revealed by JWST. The model of \citet{Bonoli25_Lgalaxies} is built on dark matter (DM) merger trees extracted from the Millennium (MS, $M_\mathrm{DM} = 8.6 \times 10^8~h^{-1}~M_\odot$) and Millennium-II (MSII, $M_\mathrm{DM} = 6.8 \times 10^6~h^{-1}~M_\odot$) simulations, re-scaled to match the Planck 2015 cosmology, which are combined through a grafting technique to simultaneously achieve large cosmological volume and high mass resolution. This approach allows the evolution of galaxies and SMBHs to be traced across a wide dynamical range, from dwarf systems to massive clusters, while accurately accounting for merger histories and unresolved progenitors.

The model reproduces a broad range of observed properties of galaxies and massive black holes, including the stellar and black hole mass functions and the AGN luminosity function \citep{Bonoli25_Lgalaxies}. In particular, the latest implementation is designed to simultaneously reproduce the high-redshift AGN population revealed by JWST and the nanoHz stochastic GW background inferred from PTA observations \citep{Bonoli25_Lgalaxies}.
In contrast to previous works where dual AGN were identified in discrete simulation snapshots, here they are extracted from a lightcone catalogue, providing a continuous galaxy distribution in right ascension, declination, and redshift. The generated lightcone covers an area of $260~\mathrm{deg}^2$, enabling a more direct comparison with observational surveys.

Within \textsc{L-GalaxiesBH}, SMBHs of $10-10^4 M_{\odot}$ initial mass are seeded in DM halos depending on the environmental conditions. Afterwards, these seeds are allowed to grow through both merger-driven and secular gas accretion channels, with accretion rates capped at the Eddington limit, or allowed to exceed it during rare episodes of exceptionally large gas inflows.
The bolometric luminosity is related to the Eddington ratio through $L_{\rm bol}=f_{\rm Edd}L_{\rm Edd}$, where $f_{\rm Edd}$ is determined by the accretion light curve \citep{Bonoli25_Lgalaxies}.

SMBH mergers occur following galaxy mergers, with coalescence timescales ($t_f$) governed primarily by dynamical friction, which gradually removes orbital energy and angular momentum from the black holes via gravitational interactions with the surrounding stars, and by the subsequent orbital evolution of the host galaxies, without explicitly resolving sub-kiloparsec black hole dynamics. As a semi-analytic model, \textsc{L-GalaxiesBH} does not directly resolve the small-scale gas physics around SMBHs, and instead relies on sub-grid prescriptions calibrated on observations and higher-resolution simulations.

The underlying DM simulation on which L-GalaxiesBH is built allows the spatial separation between dual AGN to be tracked accurately before the merger of their host galaxies. However, this information is no longer available once the host galaxies merge (typically when their nuclei are at a few kpc separation) and the two active SMBHs become associated with the same remnant galaxy. To model their subsequent orbital evolution, L-GalaxiesBH assumes that the AGN associated with the less massive progenitor is placed at the radius where its host galaxy has lost 50\% of its initial mass through tidal stripping. The separation between these two AGN is then evolved according to a simple orbital-decay prescription, decreasing as $(1-\Delta t/t_f)$, where $\Delta t$ is the time elapsed since the galaxy merger and $t_f$ is the corresponding dynamical-friction timescale.
These modeling assumptions are particularly relevant for the predicted dual AGN population, as they determine both the triggering of nuclear activity and the timescales over which SMBH pairs remain observable before coalescence, making this framework well suited for comparison with observational results.

\subsection{\textsc{Astrid}}
The \textsc{Astrid} simulation is a large-scale cosmological hydrodynamic simulation designed to study galaxy formation and dual AGN evolution \citep{Chen23b}. In this work, we use the \citet{Ni25} outputs, extended to $z=0$, to compare directly with our observational sample at $z=0.5$--$3.0$. The model reproduces a broad range of observed properties of SMBHs and their host galaxies, including the black hole mass function, AGN luminosity functions, and the $M_\mathrm{BH}$--$M_*$ relation \citep{Ni22_astrid, Ni25} \footnote{Note that the \textsc{Astrid} values do notaccount for dust attenuation.}.

\textsc{Astrid} evolves $5500^3$ DM particles and an equal number of smoothed particle hydrodynamics (SPH) elements in a $250~h^{-1}~\mathrm{Mpc^{3}}$ box, with initial conditions set at $z = 99$. The adopted cosmological parameters follow \cite{Planck2020}.
The simulation has a mass resolution of $M_\mathrm{DM} = 6.74 \times 10^6~h^{-1}~M_\odot$ 
for DM particles and $M_\mathrm{gas} = 1.27 \times 10^6~h^{-1}~M_\odot$ for gas particles, 
with a gravitational softening length of $\epsilon_g = 1.5~h^{-1}~\mathrm{kpc}/(1+z)$. 
Star particles have a median mass of $10^{5.5}~M_\odot$.
In \textsc{Astrid}, SMBHs are seeded in sufficiently massive halos with total mass $M_\mathrm{halo} > 5 \times 10^9~h^{-1}~M_\odot$ and stellar mass $M_* > 2 \times 10^6~h^{-1}~M_\odot$, ensuring formation in halos with enough cold gas and collisionless star particles to experience dynamical friction \citep{Ni22_astrid}. BH seed masses are drawn from a power-law distribution between $3 \times 10^4$ and $3 \times 10^5~h^{-1}~M_\odot$.
BH accretion follows a Bondi–Hoyle–Lyttleton prescription, corrected by a dimensionless factor $\alpha = 100$ to account for unresolved ISM phases, and is capped at twice the Eddington rate. The bolometric luminosity is estimated from the black hole accretion rate as $L_{\rm bol}\propto\epsilon_r\dot{M}_{\rm BH}$, assuming a radiative efficiency of $\epsilon_r=0.1$ \citep{Ni25}. The dynamical mass $M_\mathrm{dyn}$ tracks the physical $M_\mathrm{BH}$, and is used to compute gravitational forces and dynamical friction \citep{chen22_astrid_dynfriction}.

As shown by \cite{Ni25}, \textsc{Astrid} avoids repositioning SMBHs and instead uses a sub-grid model to track their dynamical evolution down to the spatial resolution, merging bound pairs when their separation falls below $2\epsilon_g$. In reality, black hole binaries can spend a long time spiraling inward at separations below $\sim1~h^{-1}$~kpc/(1+z) before merging. However, this simplification is unlikely to affect our results, as most such close pairs would not be detectable with our observational method.
AGN feedback operates in radiatively efficient (quasar) and inefficient (jet) modes. The former injects thermal energy into the surrounding gas, while the latter provides kinetic feedback and is activated for SMBHs with $M_\mathrm{BH} \gtrsim 5\times10^8~h^{-1}~M_\odot$ and low Eddington ratios, typically at $z\lesssim2.3$. This dual-mode feedback regulates star formation and SMBH growth, reproducing observed correlations between AGN activity and host galaxy properties \citep{Ni22_astrid}.


\section{Comparing dual AGN properties with models} 
\label{sec:comparison}

Having identified the subset of simulated systems that closely matches the observational selection, we compare their physical properties with those of the observed GMP sample to investigate the physical mechanisms that regulate SMBH growth and AGN activity during galaxy mergers.
The observational sample consists of 30 dual AGN systems, while the simulation samples include 185 systems from \textsc{L-GalaxiesBH} and 31 systems from \textsc{Astrid} after applying the same observational selection criteria.
For the analysis, we define the primary and secondary AGN as the most and least massive black holes, respectively, in both simulations and observations, ensuring a consistent comparison between equivalent components. We also tested an alternative ordering based on $G$-band magnitude and found qualitatively similar trends.

When comparing observational driven results and predicted properties, we focus on the bolometric luminosity, black hole mass, Eddington ratio, and the mass ratio between the two components.
To account for measurement uncertainties and ensure a realistic, consistent, and quantitative model-to-data comparison, we introduce a statistical scatter in the simulated $L_{\rm bol}$ to capture the variability associated with the accretion and radiative-efficiency prescriptions adopted by the simulations.

\begin{figure}[h!]
    \centering
    \includegraphics[width=0.968\linewidth]{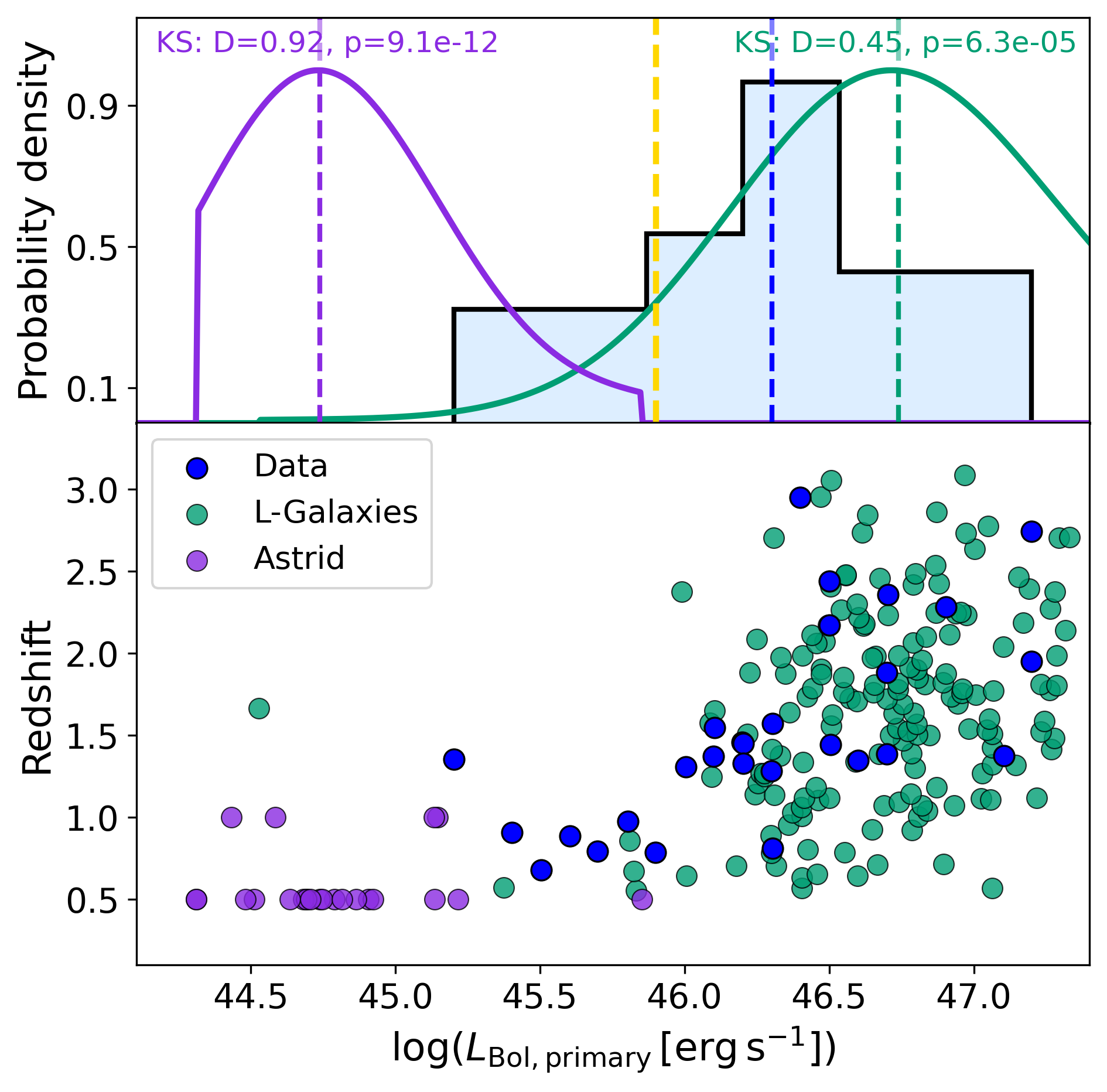}
    \hfill
            \hfill

    \includegraphics[width=0.968\linewidth]{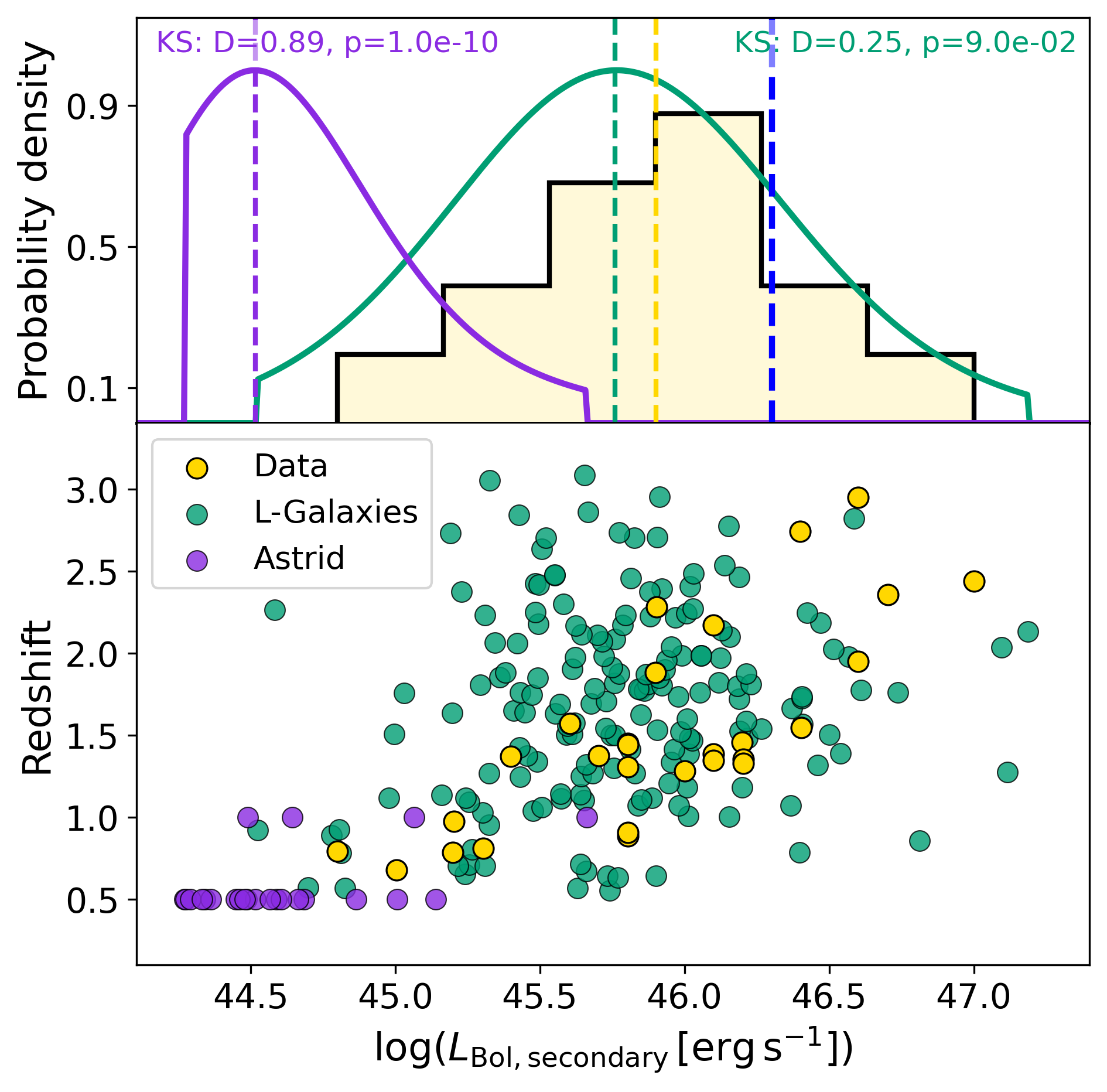}
    \caption{Bolometric luminosity versus redshift for the primary (top) and secondary (bottom) AGN, comparing the observational samples (blue/yellow) with the \textsc{L-GalaxiesBH} (teal) and \textsc{Astrid} (purple) simulations. The upper panels show the normalized bolometric-luminosity distributions, represented by histograms for the observational samples and KDE for the simulations, with dashed vertical lines indicating the corresponding median values.}
    \label{fig:lbol_primary_secondary}
\end{figure}

\begin{figure}[h!]
        \caption{Joint distribution of black hole mass and Eddington ratio in dual AGN systems. The observational samples are shown in blue (primary, top panel) and yellow (secondary, bottom panel), while the \textsc{L-GalaxiesBH} and \textsc{Astrid} predictions are shown in teal and purple, respectively. The corresponding marginal distributions are shown as histograms for the observational samples and as KDE for the models, with dashed lines indicating the corresponding medians. The KS statistics and p-values for the comparison between the observational and simulated distributions are reported.\\}
        \label{fig:MBH_Ledd}
    \hfill
        \centering
        \includegraphics[width=1\linewidth]{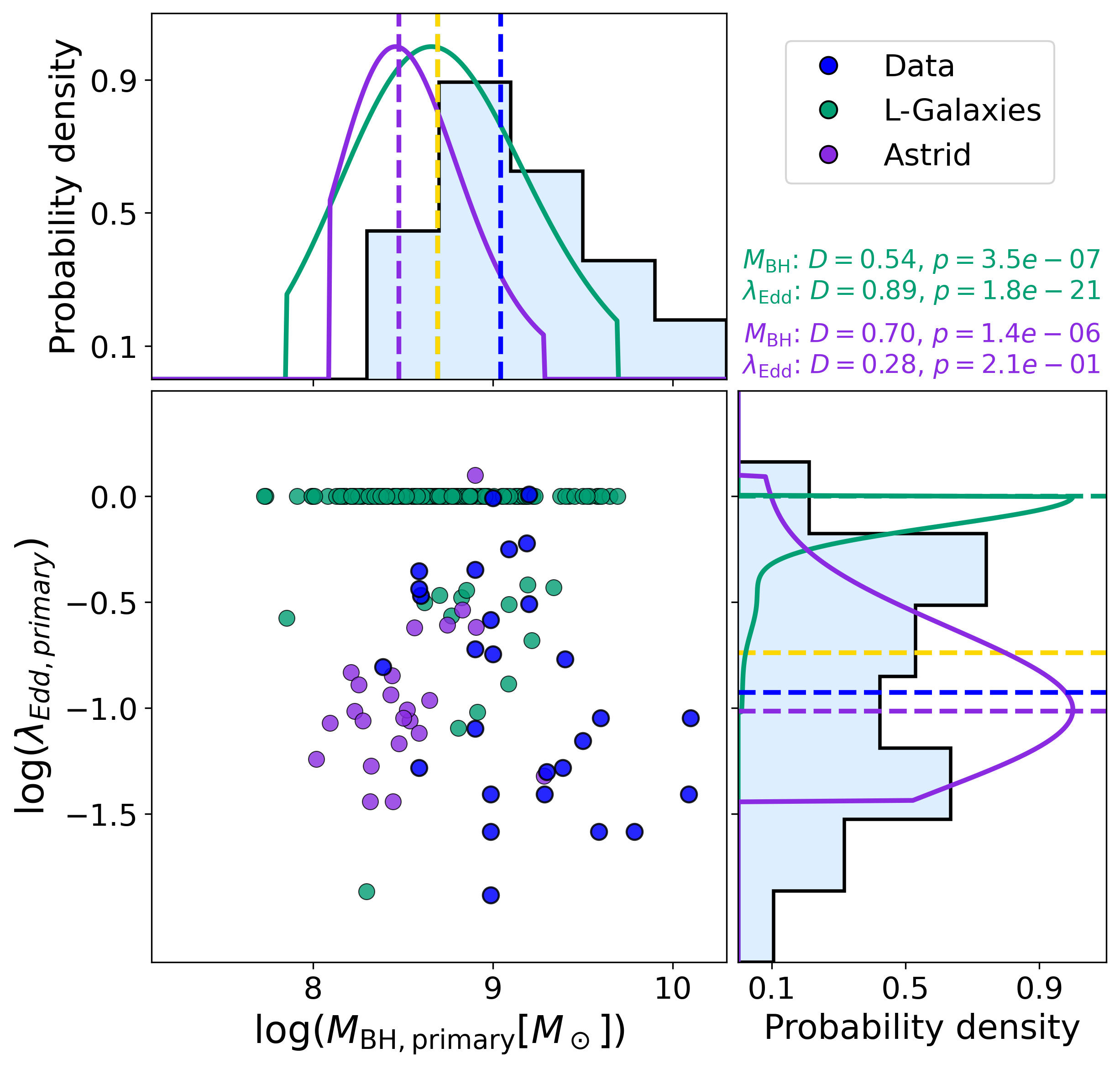}
        \hfill
                \hfill

        \includegraphics[width=1\linewidth]{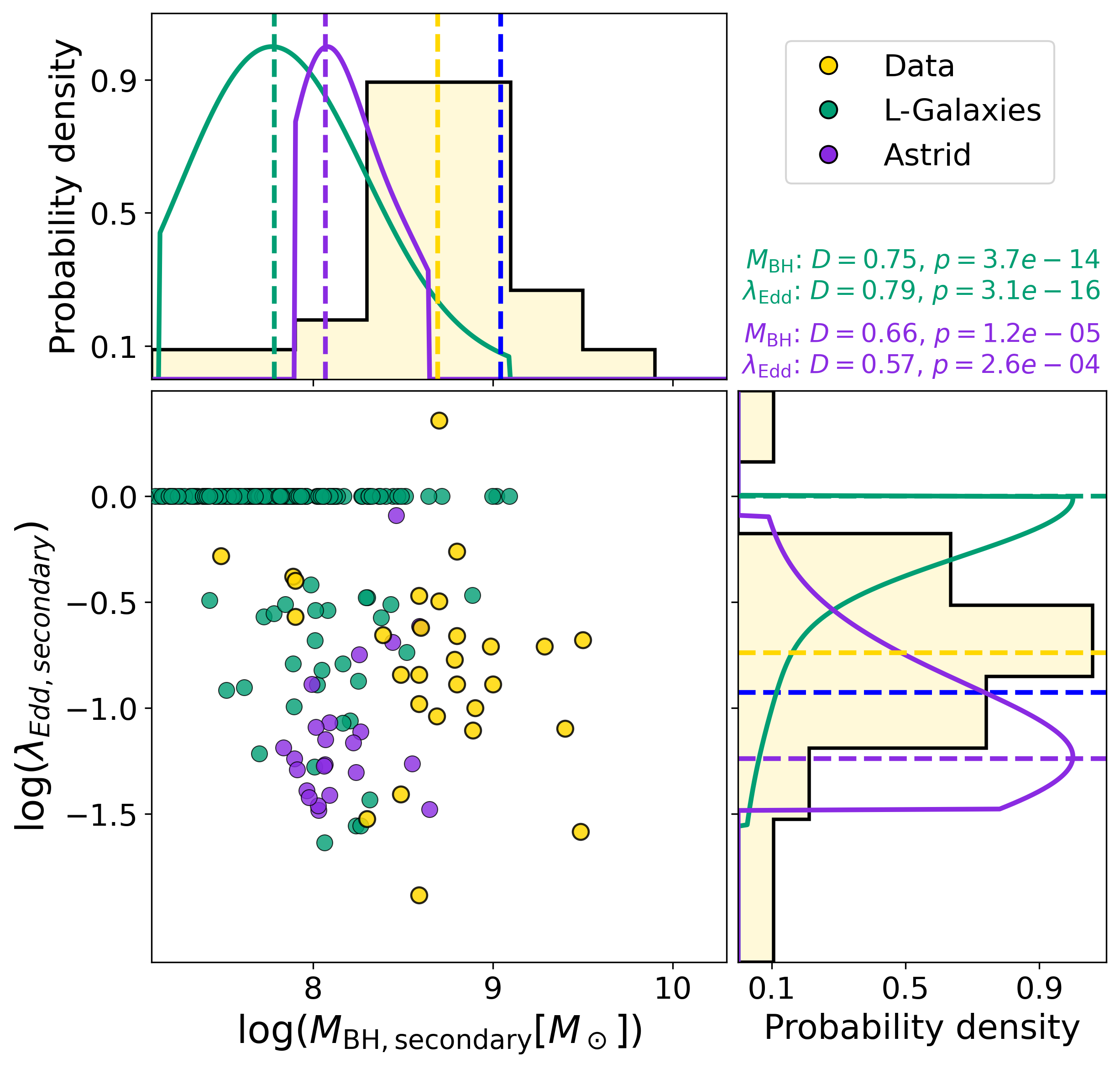}
\end{figure}

\begin{figure}[h!]
    \centering
    \includegraphics[width=0.94\linewidth]{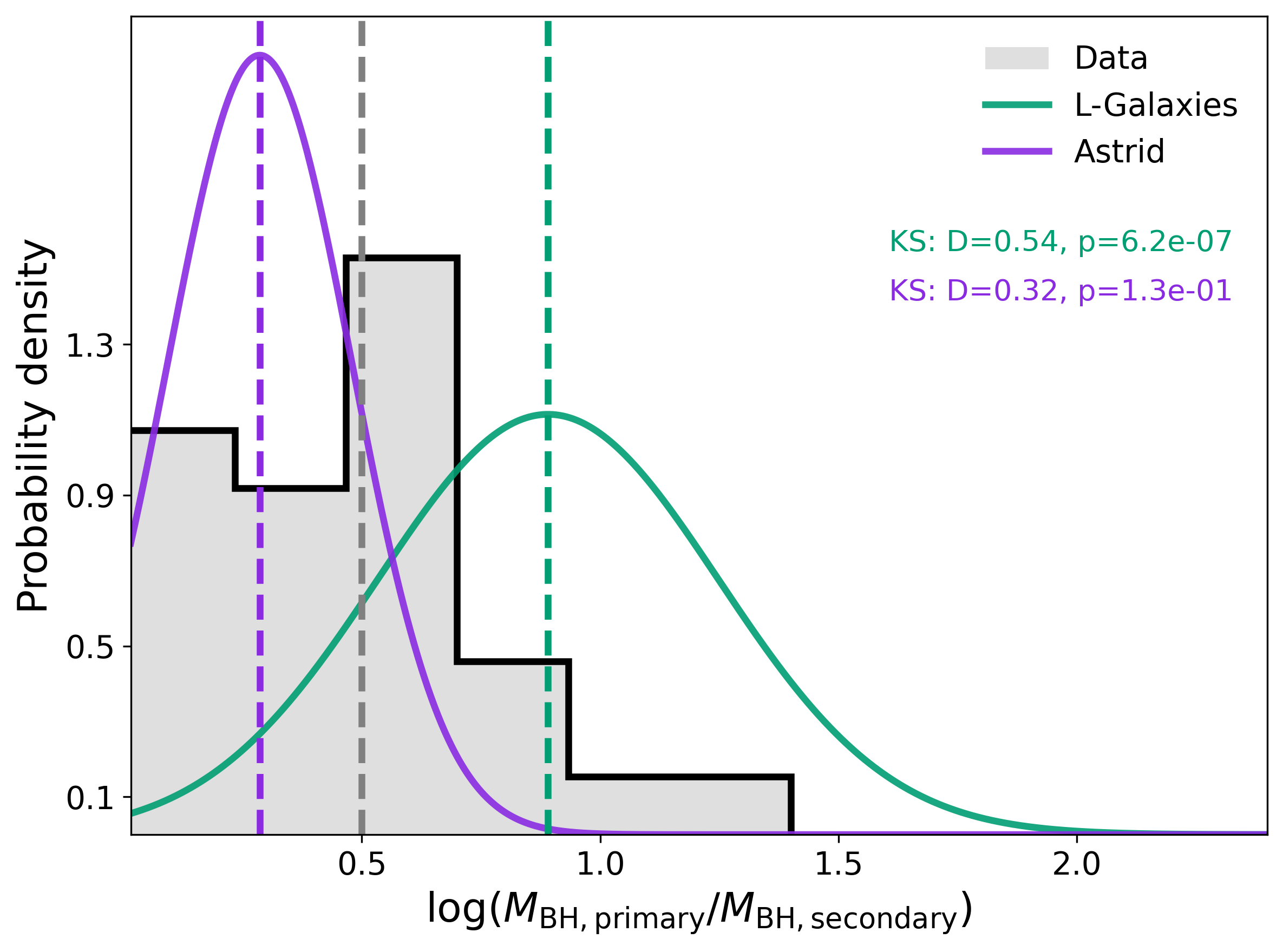}
    \caption{Distribution of the $M_{BH}$ ratio for observed and simulated \duets. 
    Observations are shown in grey, while \textsc{L-GalaxiesBH} and \textsc{Astrid} in teal and purple, respectively. he KS statistics and p-values are reported with the same color coding. }
    \label{fig:massratio}
\end{figure}

We derive probability density functions (PDFs) and cumulative distribution functions (CDFs) for each physical property, treating the two AGN components separately. The uncertainty on the observed CDFs is assessed through bootstrap resampling with replacement, generating multiple realizations of the observed dataset and their corresponding empirical CDFs.
The agreement between models and observations is quantified using two-sample Kolmogorov--Smirnov (KS) tests \citep{Massey_KS}, reporting the KS statistic $D_{\rm KS}$ and associated $p$-value. 
The results are shown alongside the PDFs in Figs.~\ref{fig:lbol_primary_secondary}, \ref{fig:MBH_Ledd}, and \ref{fig:massratio}, where observational data are shown as histograms, with \textsc{L-GalaxiesBH} in teal and \textsc{Astrid} in purple.

\paragraph{Bolometric luminosity.}
Fig.~\ref{fig:lbol_primary_secondary} shows the bolometric luminosity distributions for the primary (top panel) and secondary (bottom panel) AGN. The bolometric luminosity distributions reveal distinct behaviors for the two AGN components and the two models.

In \textsc{Astrid}, both primary and secondary AGN exhibit systematically lower bolometric luminosities compared to the observations ($D_{\rm KS}=0.92$, $p=9.1\times10^{-12}$ for secondaries; $D_{\rm KS}=0.89$, $p=1.0\times10^{-10}$ for primaries). This offset is driven by the fact that the \textsc{Astrid} sample is dominated by AGN at $z\sim0.5$ (see Fig.~\ref{fig:lbol_primary_secondary}), which, at fixed $G$-band magnitude, have lower bolometric luminosities than higher-redshift sources.
To check whether this difference is related to the different redshift distributions, we repeated the comparison using only the observed systems at $z<1$. The observed $L_{\rm bol}$ distribution shifts to lower values, as expected, but remains systematically higher than that of \textsc{Astrid}, with $D_{\rm KS}=0.968$ and $p=1.27\times10^{-6}$ for primary AGN, and $D_{\rm KS}=0.806$ and $p=3.0\times10^{-4}$ for secondary AGN. We note that this subsample contains fewer than 8 observed systems, so the comparison is limited by the small number of sources. We also repeated the comparison restricting the observed sample to $L_{\rm bol}<45.8$, finding consistent results, but only with five observed pairs. Moreover, since $L_{\rm bol}\propto\epsilon_{\rm r}\dot{M}_{\rm BH}$ in \textsc{Astrid}, even allowing for a factor of three variation in the assumed radiative efficiency, corresponding to $\sim0.48$ dex in $L_{\rm bol}$, the systematic offset remains. These tests suggest that the difference may not be driven solely by the higher-redshift ($z>1$) systems, but may also reflect differences in the simulated AGN population (see \cite{Ni25}), potentially related to an underestimation of the black hole accretion efficiency and/or duty cycle arising from the adopted sub-grid accretion and feedback prescriptions.

By contrast, \textsc{L-GalaxiesBH} shows better agreement with the observed luminosity scale of the secondary AGN ($D_{\rm KS}=0.25$, $p=9.0\times10^{-2}$), while it performs less well for the primary AGN ($D_{\rm KS}=0.45$, $p=6.3\times10^{-5}$). In this case, the model predicts primary AGN with systematically higher bolometric luminosities compared to the observed sample. 
However, this difference is relatively small when accounting for the scatter in the accretion prescription, for which a factor-of-two variation would correspond to $\sim0.3$ dex in $L_{\rm bol}$.
This discrepancy may reflect the impact of the modeling assumptions on gas accretion onto SMBHs. In particular, the higher predicted luminosities could be associated with an enhanced accretion efficiency required to reproduce the amplitude of the stochastic GW background recently reported by PTA experiments. As a consequence of this increased accretion efficiency, the model predicts a somewhat higher number density of both luminous AGN and massive SMBH binaries \citep{Izquierdo-Villalba24}.
This difference between the primary and secondary populations indicates that the model reproduces a different luminosity scaling between the two components, suggesting possible differences in the modeled accretion history or interaction-driven fueling during the dual AGN phase.

\paragraph{Black hole mass.}
Fig.~\ref{fig:MBH_Ledd} presents the joint distributions of $\log_{10}(M_{\rm BH}/M_\odot)$ and $\log_{10}(\lambda_{\rm Edd})$, highlighting the connection between black hole mass and growth rate. 
The two models predict broadly consistent black hole mass distributions for both the primary and secondary components, with no significant differences between \textsc{L-GalaxiesBH} and \textsc{Astrid}. 

For the primary black hole, we find only limited agreement between observations and both models, with KS statistics of
$D_{\rm KS} = {0.54}$ for \textsc{L-GalaxiesBH} and
$D_{\rm KS} = {0.70}$ for \textsc{Astrid}$ $
(with $p = 3.5\times10^{-7}$ and $p = 1.4\times10^{-6}$, respectively), indicating the presence of more BHs at lower masses and a consequent change in the overall distribution shape. The difference in Eddington-ratio distributions between \textsc{Astrid} and \textsc{L-GalaxiesBH} is expected given their similar SMBH mass distributions but different redshift distributions. Since the sample is selected to be G-band bright, SMBHs are effectively selected by mass, accretion rate, and redshift. As a result, at fixed SMBH mass, \textsc{L-GalaxiesBH} SMBHs require higher Eddington ratios to satisfy the luminosity threshold, while \textsc{Astrid} SMBHs can meet the same apparent-magnitude limit at lower Eddington ratios because they reside at lower redshifts.

In contrast, the secondary black hole masses in the observational sample are systematically higher than those predicted by both models, suggesting a potential mismatch between model assumptions and the observed population. 
To assess whether this discrepancy could arise from the adopted mass estimator, we tested multiple independent SE black hole mass calibrations from the literature. Although these introduce systematic differences at the object level, the overall trend remains unchanged, with the observed secondary black hole masses remaining systematically higher than predicted by both models.

Part of this tension may nevertheless be attributable to known systematic uncertainties affecting SE virial masses. In particular, the standard BLR radius--luminosity relation is known to break down at high Eddington ratios, potentially leading to overestimated black hole masses in the highest-accretion sources \citep[e.g.,][]{Wang14,Du14,Du16,Du18, Lupi24}. In addition, recent interferometric observations with \textsc{GRAVITY} suggest that SE masses of luminous AGN may be systematically biased high owing to uncertainties in the virial factor and BLR orientation \citep{gravity_L_R}. 
Accounting for these effects would reduce, but not eliminate, the discrepancy with the model predictions.

The persistently higher masses of the observed secondary SMBHs are instead qualitatively consistent with the high-resolution hydrodynamical simulations of \citet{Callegari11}, which show that tidal torques and orbital circularization during galaxy mergers can trigger enhanced accretion onto the secondary SMBH, driving the SMBH mass ratio toward more equal values during the merger. Independent observational evidence for enhanced secondary SMBH accretion is provided by \citet{Battistini26}, who find that in nearby X-ray-selected AGN pairs the less massive SMBH exhibits higher bolometric luminosities and Eddington ratios toward smaller projected separations. While physically consistent, their sample explores a different region of parameter space, including wider-separation systems ($r_{\rm p}=1$--$100$ kpc), lower redshifts ($z<0.1$), and a broader luminosity range compared to the \duets population considered here.
Within this framework, the comparison with \textsc{L-GalaxiesBH} is particularly relevant, as the model already incorporates merger-driven enhancement of secondary SMBH accretion \citep[e.g.,][]{Capelo15}. However, the remaining excess of secondary SMBH masses in the observations suggests that the amplitude or duration of this enhancement may still be underestimated in the semi-analytic prescription.

When considering systems ranked by magnitude rather than black hole mass, the brighter component is broadly consistent with the model predictions, whereas the fainter component still exhibits systematically higher masses than expected. This reinforces the conclusion that the observed mass distribution is more symmetric than predicted by the current semi-analytic models, despite the fact that the models already account for merger-driven enhancement of secondary SMBH accretion.

\vspace{-5mm}
\paragraph{Black hole mass ratio.}
Figure~\ref{fig:massratio} shows the distribution of the black hole mass ratio.
Its distribution further highlights differences between the two models. In particular, \textsc{Astrid} is in good agreement with the observed distribution, with both showing typical mass ratios of $q \sim 3$. In contrast, \textsc{L-GalaxiesBH} predicts significantly higher mass ratios, corresponding to a population skewed toward more unequal-mass systems. This behavior is consistent with the systematically lower secondary black hole masses predicted by \textsc{L-GalaxiesBH}, discussed in the previous section.

A consistent behavior is found when ranking systems by luminosity rather than mass, i.e. considering the brighter and fainter components instead of the most and least massive black holes. Observations again favor more equal-brightness (and hence more equal-mass) systems than those predicted by \textsc{L-GalaxiesBH}, while remaining broadly consistent with \textsc{Astrid}.

Overall, these results reinforce the conclusion that the merger-driven growth of secondary SMBHs may be more efficient than currently captured by the semi-analytic prescriptions, leading to SMBH pairs with more comparable masses.

\vspace{-8mm}
\paragraph{Eddington ratio.}
The Eddington ratio distributions show trends opposite to the black hole mass distributions when compared with \textsc{Astrid} (Fig.~\ref{fig:MBH_Ledd}). This is expected since $\lambda_{\rm Edd}\propto L_{\rm bol}/M_{\rm BH}$, making its evolution dependent on both the bolometric luminosity and black hole mass distributions.
In the case of the primary AGN, the observations tend to favor slightly higher accretion rates than predicted by \textsc{Astrid}, indicating a negligible discrepancy. In contrast, the observations suggest higher Eddington ratios for the secondary AGN compared to \textsc{Astrid}.

In \textsc{L-GalaxiesBH}, the accretion prescription is explicitly linked to merger- or disc-instability-driven gas inflows, where black holes are assumed to accrete at or near the Eddington limit for a fixed fraction of the available gas reservoir (see Section 2.5.2 in \citealt{Bonoli25_Lgalaxies}). This naturally produces a narrower distribution of $\lambda_{\rm Edd}$ clustered around unity, partially explaining the differences relative to the observational sample. 
Future developments should therefore explore more flexible accretion prescriptions, for example by introducing stochasticity in the accretion rate, to reproduce the broader distribution of Eddington ratios inferred from observations.

\subsection{Physical interpretation}
Our results suggest that, in GMP-selected systems (i.e. $0.5\leq z\leq3.0$, $G\lesssim21$, and $\mathrm{sep}<7 \mathrm{kpc}$), the secondary SMBH undergoes more efficient fueling during galaxy mergers than predicted by current semi-analytic models such as \textsc{L-GalaxiesBH}.  This is reflected in dual AGN populations with more comparable black hole masses and AGN luminosities between the two members of each pair, pointing to more symmetric merger-driven fueling than typically implemented in \textsc{L-GalaxiesBH} and similar sub-grid prescriptions. 
Overall, observations favor a more balanced distribution of black hole masses and accretion power between the two members of each dual AGN system, whereas the models preferentially produce more asymmetric configurations. This suggests that current implementations of dual AGN growth and triggering may either underestimate the efficiency of secondary black hole growth during mergers or overestimate the contrast between primary and secondary activity during the dual phase.

This interpretation is consistent with hydrodynamical simulations showing that merger-driven processes can preferentially enhance accretion onto the secondary SMBH \citep[e.g.,][]{Callegari11}, as well as with recent observational evidence for enhanced secondary SMBH activity in local AGN pairs \citep{Battistini26}, although selected using a different approach. The discrepancy with \textsc{L-GalaxiesBH} therefore likely reflects differences in the efficiency or duration of the secondary SMBH fueling phase rather than the absence of such a mechanism in the model, which already includes merger-induced accretion enhancement \citep[e.g.,][]{Capelo15}.

\subsection{Selection effects} 
\label{sec:selec_effect}

Since we have applied to \textsc{L-GalaxiesBH} the same selection effects present in our observations, in principle, the differences we find should not depend on the sample selection. We have performed a number of checks to test for remaining differential effects that could bias the results.

In particular, we have verified that the magnitude-ratio distribution of the two components, as a function of redshift and projected separation, is consistent between the observations and simulations, to test whether it could drive the observed difference in their relative black hole masses. We find that both samples cover the same parameter space and show no significant differences. As an independent check of the GMP selection, following the approach of \cite{Mannucci22,Mannucci23}, we have also examined the magnitude-ratio distribution of the two components of GMP-selected double stars. The much larger observed sample provides better statistical constraints, and we still find no significant differences between the observed double-source sample and the simulated AGN sample.

A possible residual systematic effect that we must consider, however, is the prescription used to derive the $G$-band magnitudes for the \textsc{L-GalaxiesBH} systems. We have shown that this prescription correctly reproduces the $G$-band magnitude distribution of single SDSS spectra (Fig.~\ref{fig:gmag}), but whether it is valid for dual AGN systems remains uncertain, particularly regarding the obscuration of the two components. In fact, \cite{Ricci17} found that AGN obscuration increases during the late stages of galaxy mergers, suggesting that merging systems may have different nuclear environments from isolated AGN. If the two components experience different levels of obscuration, this could affect their observed $G$-band magnitudes and the inferred magnitude contrast between them, motivating dedicated models of dust obscuration in merging systems.

\section{Discussion and conclusions}
In this paper, we analyzed 30 spectroscopically confirmed dual AGN systems selected from the GMP sample in the redshift range $0.5\leq z \leq 3$, with $G_{\mathrm{brighter}} \leq20.5$, $G_{\mathrm{fainter}} \leq 21$, and projected separations between 1 and 7 $\mathrm{kpc}$ (see Figure~\ref{fig:distrib_dualAGN}). We performed spectral fitting of both AGN in each pair (Section \ref{sec:specfit}), measured continuum luminosities and emission-line properties, and derived bolometric luminosities, and SE black-hole mass (Sections~\ref{sec:properties}).

The observed dual AGN population spans a broad range of black-hole masses and luminosities, consistent with optically selected SDSS AGN populations after applying the same redshift and luminosity selection (Section~\ref{sec:models_gmplike}). The brighter AGN in the pairs are generally more luminous than the bulk of the SDSS single AGN population and, on average, $\sim0.5$ dex more luminous than the fainter components, which lie within the main distribution. This difference is driven by the sample selection rather than reflecting an intrinsic difference between the two AGN. The two components, however, often exhibit comparable black-hole masses and strongly overlapping mass distributions. The brighter AGN does not necessarily host the most massive black hole or have the highest Eddington ratio, highlighting the complex interplay between black-hole mass and accretion activity.

The significant deviations of the Eddington-ratio contrast from unity indicate asymmetric accretion onto the two black holes, while neither the Eddington-ratio contrast nor the black-hole mass ratio shows a significant dependence on projected separation within the $\lesssim7$ kpc range probed by our sample. This suggests that, on these small physical scales, the relative contributions of black-hole mass and accretion to the luminosity contrast do not evolve systematically with merger stage, pointing instead to a complex and potentially stochastic fueling process.

In Section~\ref{sec:comparison}, we compared the observed population with the semi-analytical model \textsc{L-GalaxiesBH} and the hydrodynamical simulation \textsc{Astrid}, after applying the same observational selection criteria (Section~\ref{sec:models_gmplike}). While both models reproduce part of the observed parameter space, they differ in the relative distribution of mass and accretion power between the two SMBHs (Figs.~\ref{fig:MBH_Ledd} and \ref{fig:massratio}). \textsc{Astrid} provides a closer match to the observed black-hole mass-ratio distribution, but its AGN luminosities are systematically lower, consistent with the predominance of AGN at lower redshifts. \textsc{L-GalaxiesBH}, instead, better reproduces the observed luminosity scale of the secondary AGN but predicts more asymmetric systems, with lower-mass secondary SMBHs and larger mass ratios than observed. Its Eddington-ratio distribution is also narrower and more strongly peaked toward high accretion rates than observed.

The observed secondary SMBHs are systematically more massive than predicted by both models, although systematic uncertainties in SE virial mass estimates may contribute to this discrepancy. The same qualitative trend is found when systems are ranked by magnitude rather than black-hole mass, with the fainter components also showing higher masses than predicted by \textsc{L-GalaxiesBH}. This suggests that secondary SMBH growth during galaxy mergers may be more efficient or symmetric than assumed in current semi-analytical prescriptions. This interpretation is consistent with hydrodynamical simulations showing preferential enhancement of secondary SMBH accretion during mergers \citep[e.g.,][]{Callegari11}, as well as with recent observations of enhanced secondary SMBH activity in AGN pairs \citep{Battistini26}. The remaining tension with \textsc{L-GalaxiesBH}, which already includes merger-induced accretion enhancement \citep[e.g.,][]{Capelo15}, may therefore reflect differences in the efficiency or duration of secondary SMBH fueling.

The comparison between observations and models is subject to several observational and systematic limitations. The GMP selection requires both AGN components to be sufficiently bright in the Gaia $G$ band, potentially favoring systems with more comparable component luminosities. Although analogous cuts are applied to the simulated samples, differences in the underlying $G$-band and dust prescriptions may introduce differential selection effects (see Section \ref{sec:selec_effect}). Spectroscopic confirmation may also favor systems with detectable broad emission lines and moderate-to-low obscuration. In addition, black-hole masses derived from SE virial estimators and bolometric luminosities inferred from emission-line or continuum-based proxies are affected by systematic uncertainties, particularly at high accretion rates.

The box size is the main limitation of cosmological simulations affecting the comparison with our observational sample. Due to their rarity, obtaining a statistically significant population of bright ($\gtrsim10^{45}$ erg s$^{-1}$) and massive SMBHs ($\gtrsim10^{7.5} M_\odot$) at close ($<7$ kpc projected) separations within a cosmological run would require substantially larger volumes. At the same time, the resolution needed to characterize kpc-scale AGN pairs in such volumes would require a prohibitively large number of particles. A possible way forward is represented by cosmological simulations initialized with constrained realizations \citep{1987ApJ...323L.103B,1991ApJ...380L...5H}, in which the initial density field is modified to ensure the presence of massive haloes hosting heavy galaxies and SMBHs within a limited cosmological volume \citep[see][for an example applied to $z>6$ quasars]{2019MNRAS.488.4004L}.

Semi-analytical models can explore larger volumes at a lower computational cost, providing larger samples of massive and bright GMP-like systems than cosmological simulations. However, they remain limited by their adopted physical prescriptions. In \textsc{L-GalaxiesBH}, for example, active SMBHs start accreting close to their Eddington limit, producing a narrow $\lambda_{\rm Edd}$ distribution. A more flexible prescription, possibly introducing stochastic sampling of $\lambda_{\rm Edd}$ during accretion events, may therefore be required to reproduce the broader observed distribution. Furthermore, semi-analytical models do not self-consistently follow the gravitational and hydrodynamical perturbations induced by galaxy mergers. These effects are expected to be stronger in the smaller galaxy, reducing the SMBH mass contrast in unequal-mass mergers \citep{Callegari11}. Although \textsc{L-GalaxiesBH} already enhances gas accretion onto the lighter SMBH during mergers, the difference between the modeled and observed mass-ratio distributions suggests that this enhancement may be underestimated and that alternative prescriptions should be explored.

Overall, our observations suggest that secondary SMBH growth during the dual AGN phase is more efficient than predicted by current models. 
These differences may also have implications for GW observations. Since the chirp mass ($\mathcal{M}$), defined as $\mathcal{M} = (M_1 M_2)^{3/5}(M_1 + M_2)^{-1/5}$, determines the amplitude and phase evolution of the GW signal, differences in the black-hole mass distributions between models and observations will affect the predicted gravitational-wave signals (Mannucci, in prep.). A larger abundance of nearly equal-mass systems naturally increases the amplitude of the stochastic GW background, potentially improving the agreement between theoretical predictions and current PTA measurements \citep{Agazie2023a}. The implications for space-based detectors such as \textit{LISA} are less direct, since the SMBHs in our sample are generally more massive than the binaries to which \textit{LISA} is most sensitive, and extrapolation to lower masses is required.

\vspace{5mm}
\begin{acknowledgements}

We thank P. Schechter, N. Morgan, and 
T. Chen for providing the dual AGN data published in their work. We also thank V. Saeedzadeh, L. Mayer, A. Babul, S. Buttigieg, and K. Li for providing the simulation data and for fruitful discussions on the comparison with the observations. We also thank Monika, Giacomo, and all the ESO team in Garching for their support and assistance, as well as Marco Berton and the entire ESO team on the mountain for their help during the observations.
We further thank B. Trefoloni and M. Signorini for their suggestions on the fit and estimation of SMBH properties, and T. Shimizu for comments on the analysis, including considerations of the GRAVITY results. 
We also thank the referee for constructive comments and suggestions that helped improve the manuscript.
We thank the Galaxy and SMBH Formation, Growth, and Co-Evolution Workshop in Nepal for providing valuable discussions and connections that helped to improve the comparison with the models presented in this paper.
This publication  was produced while attending the PhD program in Space Science and Technology at the University of Trento, Cycle XXXIX, with the support of a scholarship financed by the Ministerial Decree no. 118 of 2nd march 2023, based on the NRRP - funded by the European Union - NextGenerationEU - Mission 4 "Education and Research", Component 1 "Enhancement of the offer of educational services: from nurseries to universities” - Investment 4.1 “Extension of the number of research doctorates and innovative doctorates for public administration and cultural heritage”.    
\\
Based on observations collected at the European Southern Observatory (ESO) programs 109.22W5, 110.23SM, 114.27BY with VLT/MUSE, and 112.25GT, 113.26BW, 114.26ZT, 115.27WA, and 116.28MD with VLT/ERIS.
Some of the data presented herein were obtained at Keck Observatory, which is a private 501(c)3 non-profit organization operated as a scientific partnership among the California Institute of Technology, the University of California, and the National Aeronautics and Space Administration. The Observatory was made possible by the generous financial support of the W. M. Keck Foundation. The Keck Observatory date presented herein, once public, can be accessed from the Keck Observatory Archive (KOA), which is operated by the W. M. Keck Observatory and the NASA Exoplanet Science Institute (NExScI), under contract with the National Aeronautics and Space Administration.

This work has made use of data from the European Space Agency (ESA) mission Gaia (https://www.cosmos.esa. int/gaia), processed by the Gaia Data Processing and Analysis Consortium (DPAC, https://www.cosmos.esa.int/ web/gaia/dpac/consortium). Funding for the DPAC has been provided by national institutions, in particular the institutions participating in the Gaia Multilateral Agreement.

 \\
We acknowledge financial contribution from INAF Large Grant “Dual AGN and binary supermassive black holes in the multimessenger era: from galaxy mergers to gravitational waves” (Bando Ricerca Fondamentale INAF 2022), from the INAF project "VLT-MOONS" CRAM 1.05.03.07, from the French National Research Agency (grant ANR-21-CE31-0026, project MBH\_waves), from the INAF Large Grant 2022 "The metal circle: a new sharp view of the baryon cycle up to Cosmic Dawn with the latest generation IFU facilities", from INAF Large Grant "The Quest for dual and binary massive black
holes in the gravitational wave era", (Bando Ricerca Fondamentale INAF 2024), the MIRACLE INAF 2024 GO grant "A JWST/MIRI MIRACLE: Mid-IR Activity of Circumnuclear Line Emission". 
EB acknowledges the support of the “Ricerca Fondamentale 2024” INAF RSN1 mini-grant 1.05.24.07.01. 
GV acknowledges financial support from the Italian National Institute for Astrophysics (INAF) under the IAF - Astrophysics Fellowships in Italy grant CUP C59J21034720001 - ``AD MAJORA''.

We acknowledge financial support from the ASI agreements n.  2024-36-HH.1-2025 and 2025-29-HH.0.

We also acknowledge financial support by the PRIN-MUR project ``PROMETEUS''  202223XDPZM financed by the European Union -  Next Generation EU, Mission 4 Component 1, CUP B53D23004750006 and C53D2300080006. 

AS acknowledges the financial support provided under the European Union’s H2020 ERC Advanced Grant ``PINGU'' (Grant Agreement: 101142079)
This study is supported by the Italian Ministry for Research and University (MUR) under Grant 'Progetto Dipartimenti di Eccellenza 2023-2027' (BiCoQ). 

The project that gave rise to these results received the support of a fellowship from ``la Caixa'' Foundation (ID 100010434). The fellowship code is LCF/BQ/PI25/12100024. Proyecto PID2025-173874NA-100 financiado por MICIU/AEI/10.13039/501100011033 y por FEDER, UE

CS acknowledges financial support by the Italian Ministry of University and Research (grant FIS2023$-$01611, CUP C53C25000300001) and by the INAF (through \textsl{Ricerca Fondamentale 2024}, Ob. Fu. 1.05.24.07.04).

MP acknowledges support through the grants PID2021-127718NB-I00, PID2024-159902NA-I00, and RYC2023-044853-I, funded by the Spain Ministry of Science and Innovation/State Agency of Research MCIN/AEI/10.13039/501100011033 and El Fondo Social Europeo Plus FSE+.
IL is supported by the European Research Council (ERC) under the European Union’s Horizon 2020 research and innovation programme (DistantDust, Grant agreement No. 101117541).
S.B. acknowledges support from the Spanish Ministerio de Ciencia e Innovaci\'on through project PID2024-159201NB-C21
\end{acknowledgements}
\newpage
\bibliographystyle{aa}
\bibliography{references} 

\begin{thebibliography}{183}
\expandafter\ifx\csname natexlab\endcsname\relax\def\natexlab#1{#1}\fi

\bibitem[{{Agazie} {et~al.}(2023){Agazie}, {Anumarlapudi}, {Archibald}, {Arzoumanian}, {Baker}, {B{\'e}csy}, {Blecha}, {Brazier}, {Brook}, {Burke-Spolaor}, {Burnette}, {Case}, {Charisi}, {Chatterjee}, {Chatziioannou}, {Cheeseboro}, {Chen}, {Cohen}, {Cordes}, {Cornish}, {Crawford}, {Cromartie}, {Crowter}, {Cutler}, {Decesar}, {Degan}, {Demorest}, {Deng}, {Dolch}, {Drachler}, {Ellis}, {Ferrara}, {Fiore}, {Fonseca}, {Freedman}, {Garver-Daniels}, {Gentile}, {Gersbach}, {Glaser}, {Good}, {G{\"u}ltekin}, {Hazboun}, {Hourihane}, {Islo}, {Jennings}, {Johnson}, {Jones}, {Kaiser}, {Kaplan}, {Kelley}, {Kerr}, {Key}, {Klein}, {Laal}, {Lam}, {Lamb}, {Lazio}, {Lewandowska}, {Littenberg}, {Liu}, {Lommen}, {Lorimer}, {Luo}, {Lynch}, {Ma}, {Madison}, {Mattson}, {McEwen}, {McKee}, {McLaughlin}, {McMann}, {Meyers}, {Meyers}, {Mingarelli}, {Mitridate}, {Natarajan}, {Ng}, {Nice}, {Ocker}, {Olum}, {Pennucci}, {Perera}, {Petrov}, {Pol}, {Radovan}, {Ransom}, {Ray}, {Romano}, {Sardesai}, {Schmiedekamp}, {Schmiedekamp}, {Schmitz},
  {Schult}, {Shapiro-Albert}, {Siemens}, {Simon}, {Siwek}, {Stairs}, {Stinebring}, {Stovall}, {Sun}, {Susobhanan}, {Swiggum}, {Taylor}, {Taylor}, {Turner}, {Unal}, {Vallisneri}, {van Haasteren}, {Vigeland}, {Wahl}, {Wang}, {Witt}, {Young}, \& {Nanograv Collaboration}}]{Agazie2023a}
{Agazie}, G., {Anumarlapudi}, A., {Archibald}, A.~M., {et~al.} 2023, \apjl, 951, L8

\bibitem[{{Amaro-Seoane} {et~al.}(2023{\natexlab{a}}){Amaro-Seoane}, {Andrews}, {Arca Sedda}, {Askar}, {Baghi}, {Balasov}, {Bartos}, {Bavera}, {Bellovary}, {Berry}, {Berti}, {Bianchi}, {Blecha}, {Blondin}, {Bogdanovi{\'c}}, {Boissier}, {Bonetti}, {Bonoli}, {Bortolas}, {Breivik}, {Capelo}, {Caramete}, {Cattorini}, {Charisi}, {Chaty}, {Chen}, {Chru{\'s}li{\'n}ska}, {Chua}, {Church}, {Colpi}, {D'Orazio}, {Danielski}, {Davies}, {Dayal}, {De Rosa}, {Derdzinski}, {Destounis}, {Dotti}, {Du{\c{t}}an}, {Dvorkin}, {Fabj}, {Foglizzo}, {Ford}, {Fouvry}, {Franchini}, {Fragos}, {Fryer}, {Gaspari}, {Gerosa}, {Graziani}, {Groot}, {Habouzit}, {Haggard}, {Haiman}, {Han}, {Istrate}, {Johansson}, {Khan}, {Kimpson}, {Kokkotas}, {Kong}, {Korol}, {Kremer}, {Kupfer}, {Lamberts}, {Larson}, {Lau}, {Liu}, {Lloyd-Ronning}, {Lodato}, {Lupi}, {Ma}, {Maccarone}, {Mandel}, {Mangiagli}, {Mapelli}, {Mathis}, {Mayer}, {McGee}, {McKernan}, {Miller}, {Mota}, {Mumpower}, {Nasim}, {Nelemans}, {Noble}, {Pacucci}, {Panessa}, {Paschalidis},
  {Pfister}, {Porquet}, {Quenby}, {Ricarte}, {R{\"o}pke}, {Regan}, {Rosswog}, {Ruiter}, {Ruiz}, {Runnoe}, {Schneider}, {Schnittman}, {Secunda}, {Sesana}, {Seto}, {Shao}, {Shapiro}, {Sopuerta}, {Stone}, {Suvorov}, {Tamanini}, {Tamfal}, {Tauris}, {Temmink}, {Tomsick}, {Toonen}, {Torres-Orjuela}, {Toscani}, {Tsokaros}, {Unal}, {V{\'a}zquez-Aceves}, {Valiante}, {van Putten}, {van Roestel}, {Vignali}, {Volonteri}, {Wu}, {Younsi}, {Yu}, {Zane}, {Zwick}, {Antonini}, {Baibhav}, {Barausse}, {Bonilla Rivera}, {Branchesi}, {Branduardi-Raymont}, {Burdge}, {Chakraborty}, {Cuadra}, {Dage}, {Davis}, {de Mink}, {Decarli}, {Doneva}, {Escoffier}, {Gandhi}, {Haardt}, {Lousto}, {Nissanke}, {Nordhaus}, {O'Shaughnessy}, {Portegies Zwart}, {Pound}, {Schussler}, {Sergijenko}, {Spallicci}, {Vernieri}, \& {Vigna-G{\'o}mez}}]{Amaro-Seoane2023}
{Amaro-Seoane}, P., {Andrews}, J., {Arca Sedda}, M., {et~al.} 2023{\natexlab{a}}, Living Reviews in Relativity, 26, 2

\bibitem[{{Amaro-Seoane} {et~al.}(2023{\natexlab{b}}){Amaro-Seoane}, {Andrews}, {Arca Sedda}, {Askar}, {Baghi}, {Balasov}, {Bartos}, {Bavera}, {Bellovary}, {Berry}, {Berti}, {Bianchi}, {Blecha}, {Blondin}, {Bogdanovi{\'c}}, {Boissier}, {Bonetti}, {Bonoli}, {Bortolas}, {Breivik}, {Capelo}, {Caramete}, {Cattorini}, {Charisi}, {Chaty}, {Chen}, {Chru{\'s}li{\'n}ska}, {Chua}, {Church}, {Colpi}, {D'Orazio}, {Danielski}, {Davies}, {Dayal}, {De Rosa}, {Derdzinski}, {Destounis}, {Dotti}, {Du{\c{t}}an}, {Dvorkin}, {Fabj}, {Foglizzo}, {Ford}, {Fouvry}, {Franchini}, {Fragos}, {Fryer}, {Gaspari}, {Gerosa}, {Graziani}, {Groot}, {Habouzit}, {Haggard}, {Haiman}, {Han}, {Istrate}, {Johansson}, {Khan}, {Kimpson}, {Kokkotas}, {Kong}, {Korol}, {Kremer}, {Kupfer}, {Lamberts}, {Larson}, {Lau}, {Liu}, {Lloyd-Ronning}, {Lodato}, {Lupi}, {Ma}, {Maccarone}, {Mandel}, {Mangiagli}, {Mapelli}, {Mathis}, {Mayer}, {McGee}, {McKernan}, {Miller}, {Mota}, {Mumpower}, {Nasim}, {Nelemans}, {Noble}, {Pacucci}, {Panessa}, {Paschalidis},
  {Pfister}, {Porquet}, {Quenby}, {Ricarte}, {R{\"o}pke}, {Regan}, {Rosswog}, {Ruiter}, {Ruiz}, {Runnoe}, {Schneider}, {Schnittman}, {Secunda}, {Sesana}, {Seto}, {Shao}, {Shapiro}, {Sopuerta}, {Stone}, {Suvorov}, {Tamanini}, {Tamfal}, {Tauris}, {Temmink}, {Tomsick}, {Toonen}, {Torres-Orjuela}, {Toscani}, {Tsokaros}, {Unal}, {V{\'a}zquez-Aceves}, {Valiante}, {van Putten}, {van Roestel}, {Vignali}, {Volonteri}, {Wu}, {Younsi}, {Yu}, {Zane}, {Zwick}, {Antonini}, {Baibhav}, {Barausse}, {Bonilla Rivera}, {Branchesi}, {Branduardi-Raymont}, {Burdge}, {Chakraborty}, {Cuadra}, {Dage}, {Davis}, {de Mink}, {Decarli}, {Doneva}, {Escoffier}, {Gandhi}, {Haardt}, {Lousto}, {Nissanke}, {Nordhaus}, {O'Shaughnessy}, {Portegies Zwart}, {Pound}, {Schussler}, {Sergijenko}, {Spallicci}, {Vernieri}, \& {Vigna-G{\'o}mez}}]{AmaroSeoane2023}
{Amaro-Seoane}, P., {Andrews}, J., {Arca Sedda}, M., {et~al.} 2023{\natexlab{b}}, Living Reviews in Relativity, 26, 2

\bibitem[{{Anguita} {et~al.}(2018){Anguita}, {Schechter}, {Kuropatkin}, {Morgan}, {Ostrovski}, {Abramson}, {Agnello}, {Apostolovski}, {Fassnacht}, {Hsueh}, {Motta}, {Rojas}, {Rusu}, {Treu}, {Williams}, {Auger}, {Buckley-Geer}, {Lin}, {McMahon}, {Abbott}, {Allam}, {Annis}, {Bernstein}, {Bertin}, {Brooks}, {Burke}, {Carnero Rosell}, {Carrasco-Kind}, {Carretero}, {Cunha}, {D'Andrea}, {De Vicente}, {DePoy}, {Desai}, {Diehl}, {Doel}, {Flaugher}, {Garc{\'\i}a-Bellido}, {Gerdes}, {Gruen}, {Gruendl}, {Gschwend}, {Hartley}, {Hollowood}, {Honscheid}, {James}, {Kuehn}, {Lima}, {Maia}, {Miquel}, {Plazas}, {Sanchez}, {Scarpine}, {Smith}, {Soares-Santos}, {Sobreira}, {Suchyta}, {Tarle}, \& {Walker}}]{Anguita18}
{Anguita}, T., {Schechter}, P.~L., {Kuropatkin}, N., {et~al.} 2018, \mnras, 480, 5017

\bibitem[{{Assef} {et~al.}(2011){Assef}, {Denney}, {Kochanek}, {Peterson}, {Koz{\l}owski}, {Ageorges}, {Barrows}, {Buschkamp}, {Dietrich}, {Falco}, {Feiz}, {Gemperlein}, {Germeroth}, {Grier}, {Hofmann}, {Juette}, {Khan}, {Kilic}, {Knierim}, {Laun}, {Lederer}, {Lehmitz}, {Lenzen}, {Mall}, {Madsen}, {Mandel}, {Martini}, {Mathur}, {Mogren}, {Mueller}, {Naranjo}, {Pasquali}, {Polsterer}, {Pogge}, {Quirrenbach}, {Seifert}, {Stern}, {Shappee}, {Storz}, {Van Saders}, {Weiser}, \& {Zhang}}]{Assef11}
{Assef}, R.~J., {Denney}, K.~D., {Kochanek}, C.~S., {et~al.} 2011, \apj, 742, 93

\bibitem[{{Aubert} {et~al.}(2004){Aubert}, {Pichon}, \& {Colombi}}]{Aubert04}
{Aubert}, D., {Pichon}, C., \& {Colombi}, S. 2004, \mnras, 352, 376

\bibitem[{Bacon {et~al.}(2010)Bacon, Accardo, Adjali, Anwand, Bauer, Biswas, Blaizot, Boudon, Brau-Nogue, Brinchmann, Caillier, Capoani, Carollo, Contini, Couderc, Daguis{\'{e}}, Deiries, Delabre, Dreizler, Dubois, Dupieux, Dupuy, Emsellem, Fechner, Fleischmann, Fran{\c{c}}ois, Gallou, Gharsa, Glindemann, Gojak, Guiderdoni, Hansali, Hahn, Jarno, Kelz, Koehler, Kosmalski, Laurent, Floch, Lilly, Lizon, Loupias, Manescau, Monstein, Nicklas, Olaya, Pares, Pasquini, P{\'{e}}contal-Rousset, Pell{\'{o}}, Petit, Popow, Reiss, Remillieux, Renault, Roth, Rupprecht, Serre, Schaye, Soucail, Steinmetz, Streicher, Stuik, H, Vernet, Weilbacher, Wisotzki, \& Yerle}]{Bacon10}
Bacon, R., Accardo, M., Adjali, L., {et~al.} 2010, in {SPIE} Proceedings, ed. I.~S. McLean, S.~K. Ramsay, \& H.~Takami ({SPIE})

\bibitem[{{Baltaci} {et~al.}(2026){Baltaci}, {Gross}, {Liu}, {Zakamska}, {Shen}, \& {Liu}}]{Baltaci+26}
{Baltaci}, S., {Gross}, A.~C., {Liu}, X., {et~al.} 2026, arXiv e-prints, arXiv:2606.03592

\bibitem[{{Barrows} {et~al.}(2012){Barrows}, {Stern}, {Madsen}, {Harrison}, {Assef}, {Comerford}, {Cushing}, {Fassnacht}, {Gonzalez}, {Griffith}, {Hickox}, {Kirkpatrick}, \& {Lagattuta}}]{Barrows12}
{Barrows}, R.~S., {Stern}, D., {Madsen}, K., {et~al.} 2012, \apj, 744, 7

\bibitem[{{Battistini} {et~al.}(2026){Battistini}, {De Rosa}, {Severgnini}, {Vignali}, {Singh}, {Capelo}, {Bertola}, {Bianchi}, {D'Amato}, {Guainazzi}, {La Franca}, {Lamperti}, {Mannucci}, {Parvatikar}, {Piconcelli}, {Ricci}, {Rigamonti}, {Scialpi}, \& {Vittoria Zanchettin}}]{Battistini26}
{Battistini}, L., {De Rosa}, A., {Severgnini}, P., {et~al.} 2026, arXiv e-prints, arXiv:2604.14900

\bibitem[{{Benson} \& {Bower}(2010)}]{Benson10_galform}
{Benson}, A.~J. \& {Bower}, R. 2010, \mnras, 405, 1573

\bibitem[{{Bertschinger}(1987)}]{1987ApJ...323L.103B}
{Bertschinger}, E. 1987, \apjl, 323, L103

\bibitem[{{Bird} {et~al.}(2022){Bird}, {Ni}, {Di Matteo}, {Croft}, {Feng}, \& {Chen}}]{Bird22}
{Bird}, S., {Ni}, Y., {Di Matteo}, T., {et~al.} 2022, \mnras, 512, 3703

\bibitem[{{Bonoli} {et~al.}(2025){Bonoli}, {Izquierdo-Villalba}, {Spinoso}, {Colpi}, {Sesana}, {Polkas}, \& {Springel}}]{Bonoli25_Lgalaxies}
{Bonoli}, S., {Izquierdo-Villalba}, D., {Spinoso}, D., {et~al.} 2025, arXiv e-prints, arXiv:2509.12325

\bibitem[{{Buttigieg} {et~al.}(2025){Buttigieg}, {Sijacki}, {Moore}, \& {Bourne}}]{Buttigieg25_fable}
{Buttigieg}, S., {Sijacki}, D., {Moore}, C.~J., \& {Bourne}, M.~A. 2025, \mnras, 542, 2019

\bibitem[{{Callegari} {et~al.}(2011){Callegari}, {Kazantzidis}, {Mayer}, {Colpi}, {Bellovary}, {Quinn}, \& {Wadsley}}]{Callegari11}
{Callegari}, S., {Kazantzidis}, S., {Mayer}, L., {et~al.} 2011, \apj, 729, 85

\bibitem[{{Capelo} {et~al.}(2017){Capelo}, {Dotti}, {Volonteri}, {Mayer}, {Bellovary}, \& {Shen}}]{Capelo17}
{Capelo}, P.~R., {Dotti}, M., {Volonteri}, M., {et~al.} 2017, \mnras, 469, 4437

\bibitem[{{Capelo} {et~al.}(2015){Capelo}, {Volonteri}, {Dotti}, {Bellovary}, {Mayer}, \& {Governato}}]{Capelo15}
{Capelo}, P.~R., {Volonteri}, M., {Dotti}, M., {et~al.} 2015, \mnras, 447, 2123

\bibitem[{{Cardelli} {et~al.}(1989){Cardelli}, {Clayton}, \& {Mathis}}]{Cardelli1989}
{Cardelli}, J.~A., {Clayton}, G.~C., \& {Mathis}, J.~S. 1989, \apj, 345, 245

\bibitem[{{Ceci} {et~al.}(2026){Ceci}, {Marconcini}, {Marconi}, {Feltre}, {Lamperti}, {Belfiore}, {Bertola}, {Bracci}, {Carniani}, {Cataldi}, {Cresci}, {D'Amato}, {Fritz}, {Ginolfi}, {Hatziminaoglou}, {Hirschmann}, {Mingozzi}, {Moreschini}, {Mannucci}, {Sabatini}, {Salvestrini}, {Scialpi}, {Tozzi}, {Ulivi}, {Venturi}, {Vidal-Garc{\'\i}a}, {Vignali}, \& {Zanchettin}}]{Ceci26}
{Ceci}, M., {Marconcini}, C., {Marconi}, A., {et~al.} 2026, \aap, 707, A376

\bibitem[{{Chen} {et~al.}(2023{\natexlab{a}}){Chen}, {Di Matteo}, {Ni}, {Tremmel}, {DeGraf}, {Shen}, {Holgado}, {Bird}, {Croft}, \& {Feng}}]{Chen23b}
{Chen}, N., {Di Matteo}, T., {Ni}, Y., {et~al.} 2023{\natexlab{a}}, \mnras, 522, 1895

\bibitem[{{Chen} {et~al.}(2022{\natexlab{a}}){Chen}, {Ni}, {Tremmel}, {Di Matteo}, {Bird}, {DeGraf}, \& {Feng}}]{chen22_astrid_dynfriction}
{Chen}, N., {Ni}, Y., {Tremmel}, M., {et~al.} 2022{\natexlab{a}}, \mnras, 510, 531

\bibitem[{{Chen} {et~al.}(2026){Chen}, {He}, {Deng}, {Jing}, {Zhu}, \& {Wu}}]{Chen+26}
{Chen}, Q., {He}, Z., {Deng}, Z., {et~al.} 2026, arXiv e-prints, arXiv:2607.00093

\bibitem[{{Chen} {et~al.}(2022{\natexlab{b}}){Chen}, {Hwang}, {Shen}, {Liu}, {Zakamska}, {Yang}, \& {Li}}]{Chen22a}
{Chen}, Y.-C., {Hwang}, H.-C., {Shen}, Y., {et~al.} 2022{\natexlab{b}}, \apj, 925, 162

\bibitem[{{Chen} {et~al.}(2023{\natexlab{b}}){Chen}, {Liu}, {Foord}, {Shen}, {Oguri}, {Chen}, {Di Matteo}, {Holgado}, {Hwang}, \& {Zakamska}}]{Chen23a}
{Chen}, Y.-C., {Liu}, X., {Foord}, A., {et~al.} 2023{\natexlab{b}}, \nat, 616, 45

\bibitem[{{Ciurlo} {et~al.}(2023){Ciurlo}, {Mannucci}, {Yeh}, {Amiri}, {Carniani}, {Cicone}, {Cresci}, {Lusso}, {Marasco}, {Marconcini}, {Marconi}, {Nardini}, {Pancino}, {Rosati}, {Rubinur}, {Severgnini}, {Scialpi}, {Tozzi}, {Venturi}, {Vignali}, \& {Volonteri}}]{Ciurlo23}
{Ciurlo}, A., {Mannucci}, F., {Yeh}, S., {et~al.} 2023, \aap, 671, L4

\bibitem[{{Cole} {et~al.}(2000){Cole}, {Lacey}, {Baugh}, \& {Frenk}}]{Cole00_galform}
{Cole}, S., {Lacey}, C.~G., {Baugh}, C.~M., \& {Frenk}, C.~S. 2000, \mnras, 319, 168

\bibitem[{{Colpi}(2014)}]{Colpi14}
{Colpi}, M. 2014, \ssr, 183, 189

\bibitem[{{Colpi} {et~al.}(2024){Colpi}, {Danzmann}, {Hewitson}, {Holley-Bockelmann}, {Jetzer}, {Nelemans}, {Petiteau}, {Shoemaker}, {Sopuerta}, {Stebbins}, {Tanvir}, {Ward}, {Weber}, {Thorpe}, {Daurskikh}, {Deep}, {Fern{\'a}ndez N{\'u}{\~n}ez}, {Garc{\'\i}a Marirrodriga}, {Gehler}, {Halain}, {Jennrich}, {Lammers}, {Larra{\~n}aga}, {Lieser}, {L{\"u}tzgendorf}, {Martens}, {Mondin}, {Piris Ni{\~n}o}, {Amaro-Seoane}, {Arca Sedda}, {Auclair}, {Babak}, {Baghi}, {Baibhav}, {Baker}, {Bayle}, {Berry}, {Berti}, {Boileau}, {Bonetti}, {Brito}, {Buscicchio}, {Calcagni}, {Capelo}, {Caprini}, {Caputo}, {Castelli}, {Chen}, {Chen}, {Chua}, {Davies}, {Derdzinski}, {Domcke}, {Doneva}, {Dvorkin}, {Mar{\'\i}a Ezquiaga}, {Gair}, {Haiman}, {Harry}, {Hartwig}, {Hees}, {Heffernan}, {Husa}, {Izquierdo-Villalba}, {Karnesis}, {Klein}, {Korol}, {Korsakova}, {Kupfer}, {Laghi}, {Lamberts}, {Larson}, {Le Jeune}, {Lewicki}, {Littenberg}, {Madge}, {Mangiagli}, {Marsat}, {Vilchez}, {Maselli}, {Mathews}, {van de Meent}, {Muratore}, {Nardini},
  {Pani}, {Peloso}, {Pieroni}, {Pound}, {Quelquejay-Leclere}, {Ricciardone}, {Rossi}, {Sartirana}, {Savalle}, {Sberna}, {Sesana}, {Shoemaker}, {Slutsky}, {Sotiriou}, {Speri}, {Staab}, {Steer}, {Tamanini}, {Tasinato}, {Torrado}, {Torres-Orjuela}, {Toubiana}, {Vallisneri}, {Vecchio}, {Volonteri}, {Yagi}, \& {Zwick}}]{Colpi2024arxiv}
{Colpi}, M., {Danzmann}, K., {Hewitson}, M., {et~al.} 2024, arXiv e-prints, arXiv:2402.07571

\bibitem[{{Dav{\'e}} {et~al.}(2019){Dav{\'e}}, {Angl{\'e}s-Alc{\'a}zar}, {Narayanan}, {Li}, {Rafieferantsoa}, \& {Appleby}}]{Dave19}
{Dav{\'e}}, R., {Angl{\'e}s-Alc{\'a}zar}, D., {Narayanan}, D., {et~al.} 2019, \mnras, 486, 2827

\bibitem[{{Davies} {et~al.}(2023){Davies}, {Absil}, {Agapito}, {Agudo Berbel}, {Baruffolo}, {Biliotti}, {Black}, {Bonaglia}, {Bonse}, {Briguglio}, {Campana}, {Cao}, {Carbonaro}, {Cortes}, {Cresci}, {Dallilar}, {Dannert}, {De Rosa}, {Deysenroth}, {Di Antonio}, {Di Cianno}, {Di Rico}, {Doelman}, {Dolci}, {Dorn}, {Eisenhauer}, {Esposito}, {Fantinel}, {Ferruzzi}, {Feuchtgruber}, {Finger}, {F{\"o}rster Schreiber}, {Gao}, {Gemperlein}, {Genzel}, {Gillessen}, {Ginski}, {Glauser}, {Glindemann}, {Grani}, {Hartl}, {Hayoz}, {Heida}, {Henry}, {Hofmann}, {Huber}, {Kasper}, {Keller}, {Kenworthy}, {Kravchenko}, {Kuntschner}, {Lacour}, {Lightfoot}, {Lunney}, {Lutz}, {Macintosh}, {Mannucci}, {Marsset}, {Modigliani}, {Neeser}, {Orban de Xivry}, {Ott}, {Pallanca}, {Patapis}, {Pearson}, {Pe{\~n}a}, {Percheron}, {Puglisi}, {Quanz}, {Rabien}, {Rau}, {Riccardi}, {Salasnich}, {Schmid}, {Schubert}, {Serra}, {Shimizu}, {Snik}, {Sturm}, {Tacconi}, {Taylor}, {Valentini}, {Waring}, {Wiezorrek}, \& {Xompero}}]{Davies2023_eris}
{Davies}, R., {Absil}, O., {Agapito}, G., {et~al.} 2023, \aap, 674, A207

\bibitem[{{Davis} {et~al.}(1985){Davis}, {Efstathiou}, {Frenk}, \& {White}}]{Davis85}
{Davis}, M., {Efstathiou}, G., {Frenk}, C.~S., \& {White}, S.~D.~M. 1985, \apj, 292, 371

\bibitem[{{De Rosa} {et~al.}(2019){De Rosa}, {Vignali}, {Bogdanovi{\'c}}, {Capelo}, {Charisi}, {Dotti}, {Husemann}, {Lusso}, {Mayer}, {Paragi}, {Runnoe}, {Sesana}, {Steinborn}, {Bianchi}, {Colpi}, {del Valle}, {Frey}, {Gab{\'a}nyi}, {Giustini}, {Guainazzi}, {Haiman}, {Herrera Ruiz}, {Herrero-Illana}, {Iwasawa}, {Komossa}, {Lena}, {Loiseau}, {Perez-Torres}, {Piconcelli}, \& {Volonteri}}]{derosa20}
{De Rosa}, A., {Vignali}, C., {Bogdanovi{\'c}}, T., {et~al.} 2019, \nar, 86, 101525

\bibitem[{{De Rosa} {et~al.}(2023){De Rosa}, {Vignali}, {Severgnini}, {Bianchi}, {Bogdanovi{\'c}}, {Charisi}, {Guainazzi}, {Haiman}, {Komossa}, {Paragi}, {Perez-Torres}, {Piconcelli}, {Ducci}, {Parvatikar}, \& {Serafinelli}}]{DeRosa2023}
{De Rosa}, A., {Vignali}, C., {Severgnini}, P., {et~al.} 2023, \mnras, 519, 5149

\bibitem[{{Di Matteo} {et~al.}(2005){Di Matteo}, {Springel}, \& {Hernquist}}]{Dimatteo05}
{Di Matteo}, T., {Springel}, V., \& {Hernquist}, L. 2005, in Growing Black Holes: Accretion in a Cosmological Context, ed. A.~{Merloni}, S.~{Nayakshin}, \& R.~A. {Sunyaev}, 340--345

\bibitem[{{Dolag} {et~al.}(2009){Dolag}, {Borgani}, {Murante}, \& {Springel}}]{Dolag09}
{Dolag}, K., {Borgani}, S., {Murante}, G., \& {Springel}, V. 2009, \mnras, 399, 497

\bibitem[{{Dressler} {et~al.}(2011){Dressler}, {Bigelow}, {Hare}, {Sutin}, {Thompson}, {Burley}, {Epps}, {Oemler}, {Bagish}, {Birk}, {Clardy}, {Gunnels}, {Kelson}, {Shectman}, \& {Osip}}]{IMACS_baade}
{Dressler}, A., {Bigelow}, B., {Hare}, T., {et~al.} 2011, \pasp, 123, 288

\bibitem[{{Du} {et~al.}(2014){Du}, {Hu}, {Lu}, {Wang}, {Qiu}, {Li}, {Bai}, {Kaspi}, {Netzer}, {Wang}, \& {SEAMBH Collaboration}}]{Du14}
{Du}, P., {Hu}, C., {Lu}, K.-X., {et~al.} 2014, \apj, 782, 45

\bibitem[{{Du} {et~al.}(2016){Du}, {Lu}, {Zhang}, {Huang}, {Wang}, {Hu}, {Qiu}, {Li}, {Fan}, {Fang}, {Bai}, {Bian}, {Yuan}, {Ho}, {Wang}, \& {SEAMBH Collaboration}}]{Du16}
{Du}, P., {Lu}, K.-X., {Zhang}, Z.-X., {et~al.} 2016, \apj, 825, 126

\bibitem[{{Du} {et~al.}(2018){Du}, {Zhang}, {Wang}, {Huang}, {Zhang}, {Lu}, {Hu}, {Li}, {Bai}, {Bian}, {Yuan}, {Ho}, {Wang}, \& {SEAMBH Collaboration}}]{Du18}
{Du}, P., {Zhang}, Z.-X., {Wang}, K., {et~al.} 2018, \apj, 856, 6

\bibitem[{{Dubois} {et~al.}(2012){Dubois}, {Devriendt}, {Slyz}, \& {Teyssier}}]{dubois12}
{Dubois}, Y., {Devriendt}, J., {Slyz}, A., \& {Teyssier}, R. 2012, \mnras, 420, 2662

\bibitem[{{Dubois} {et~al.}(2016){Dubois}, {Peirani}, {Pichon}, {Devriendt}, {Gavazzi}, {Welker}, \& {Volonteri}}]{Dubois16}
{Dubois}, Y., {Peirani}, S., {Pichon}, C., {et~al.} 2016, \mnras, 463, 3948

\bibitem[{{Duras} {et~al.}(2020){Duras}, {Bongiorno}, {Ricci}, {Piconcelli}, {Shankar}, {Lusso}, {Bianchi}, {Fiore}, {Maiolino}, {Marconi}, {Onori}, {Sani}, {Schneider}, {Vignali}, \& {La Franca}}]{duras20}
{Duras}, F., {Bongiorno}, A., {Ricci}, F., {et~al.} 2020, \aap, 636, A73

\bibitem[{{EPTA Collaboration} {et~al.}(2023){EPTA Collaboration}, {InPTA Collaboration}, {Antoniadis}, {Arumugam}, {Arumugam}, {Babak}, {Bagchi}, {Bak Nielsen}, {Bassa}, {Bathula}, {Berthereau}, {Bonetti}, {Bortolas}, {Brook}, {Burgay}, {Caballero}, {Chalumeau}, {Champion}, {Chanlaridis}, {Chen}, {Cognard}, {Dandapat}, {Deb}, {Desai}, {Desvignes}, {Dhanda-Batra}, {Dwivedi}, {Falxa}, {Ferdman}, {Franchini}, {Gair}, {Goncharov}, {Gopakumar}, {Graikou}, {Grie{\ss}meier}, {Guillemot}, {Guo}, {Gupta}, {Hisano}, {Hu}, {Iraci}, {Izquierdo-Villalba}, {Jang}, {Jawor}, {Janssen}, {Jessner}, {Joshi}, {Kareem}, {Karuppusamy}, {Keane}, {Keith}, {Kharbanda}, {Kikunaga}, {Kolhe}, {Kramer}, {Krishnakumar}, {Lackeos}, {Lee}, {Liu}, {Liu}, {Lyne}, {McKee}, {Maan}, {Main}, {Mickaliger}, {Ni{\c{t}}u}, {Nobleson}, {Paladi}, {Parthasarathy}, {Perera}, {Perrodin}, {Petiteau}, {Porayko}, {Possenti}, {Prabu}, {Quelquejay Leclere}, {Rana}, {Samajdar}, {Sanidas}, {Sesana}, {Shaifullah}, {Singha}, {Speri}, {Spiewak}, {Srivastava},
  {Stappers}, {Surnis}, {Susarla}, {Susobhanan}, {Takahashi}, {Tarafdar}, {Theureau}, {Tiburzi}, {van der Wateren}, {Vecchio}, {Venkatraman Krishnan}, {Verbiest}, {Wang}, {Wang}, \& {Wu}}]{EPTACollaboration2023}
{EPTA Collaboration}, {InPTA Collaboration}, {Antoniadis}, J., {et~al.} 2023, \aap, 678, A50

\bibitem[{{Feng} {et~al.}(2016){Feng}, {Di-Matteo}, {Croft}, {Bird}, {Battaglia}, \& {Wilkins}}]{Feng16}
{Feng}, Y., {Di-Matteo}, T., {Croft}, R.~A., {et~al.} 2016, \mnras, 455, 2778

\bibitem[{{Ferland} {et~al.}(1998){Ferland}, {Korista}, {Verner}, {Ferguson}, {Kingdon}, \& {Verner}}]{Ferland98}
{Ferland}, G.~J., {Korista}, K.~T., {Verner}, D.~A., {et~al.} 1998, \pasp, 110, 761

\bibitem[{{Ferland} {et~al.}(2013){Ferland}, {Porter}, {van Hoof}, {Williams}, {Abel}, {Lykins}, {Shaw}, {Henney}, \& {Stancil}}]{Ferland13}
{Ferland}, G.~J., {Porter}, R.~L., {van Hoof}, P.~A.~M., {et~al.} 2013, \rmxaa, 49, 137

\bibitem[{{Fu} {et~al.}(2015{\natexlab{a}}){Fu}, {Myers}, {Djorgovski}, {Yan}, {Wrobel}, \& {Stockton}}]{Fu15_a}
{Fu}, H., {Myers}, A.~D., {Djorgovski}, S.~G., {et~al.} 2015{\natexlab{a}}, \apj, 799, 72

\bibitem[{{Fu} {et~al.}(2015{\natexlab{b}}){Fu}, {Wrobel}, {Myers}, {Djorgovski}, \& {Yan}}]{Fu15_b}
{Fu}, H., {Wrobel}, J.~M., {Myers}, A.~D., {Djorgovski}, S.~G., \& {Yan}, L. 2015{\natexlab{b}}, \apjl, 815, L6

\bibitem[{{Gaia Collaboration} {et~al.}(2021){Gaia Collaboration}, {Brown}, {Vallenari}, {Prusti}, {de Bruijne}, {Babusiaux}, {Biermann}, {Creevey}, {Evans}, {Eyer}, {Hutton}, {Jansen}, {Jordi}, {Klioner}, {Lammers}, {Lindegren}, {Luri}, {Mignard}, {Panem}, {Pourbaix}, {Randich}, {Sartoretti}, {Soubiran}, {Walton}, {Arenou}, {Bailer-Jones}, {Bastian}, {Cropper}, {Drimmel}, {Katz}, {Lattanzi}, {van Leeuwen}, {Bakker}, {Cacciari}, {Casta{\~n}eda}, {De Angeli}, {Ducourant}, {Fabricius}, {Fouesneau}, {Fr{\'e}mat}, {Guerra}, {Guerrier}, {Guiraud}, {Jean-Antoine Piccolo}, {Masana}, {Messineo}, {Mowlavi}, {Nicolas}, {Nienartowicz}, {Pailler}, {Panuzzo}, {Riclet}, {Roux}, {Seabroke}, {Sordo}, {Tanga}, {Th{\'e}venin}, {Gracia-Abril}, {Portell}, {Teyssier}, {Altmann}, {Andrae}, {Bellas-Velidis}, {Benson}, {Berthier}, {Blomme}, {Brugaletta}, {Burgess}, {Busso}, {Carry}, {Cellino}, {Cheek}, {Clementini}, {Damerdji}, {Davidson}, {Delchambre}, {Dell'Oro}, {Fern{\'a}ndez-Hern{\'a}ndez}, {Galluccio}, {Garc{\'\i}a-Lario},
  {Garcia-Reinaldos}, {Gonz{\'a}lez-N{\'u}{\~n}ez}, {Gosset}, {Haigron}, {Halbwachs}, {Hambly}, {Harrison}, {Hatzidimitriou}, {Heiter}, {Hern{\'a}ndez}, {Hestroffer}, {Hodgkin}, {Holl}, {Jan{\ss}en}, {Jevardat de Fombelle}, {Jordan}, {Krone-Martins}, {Lanzafame}, {L{\"o}ffler}, {Lorca}, {Manteiga}, {Marchal}, {Marrese}, {Moitinho}, {Mora}, {Muinonen}, {Osborne}, {Pancino}, {Pauwels}, {Petit}, {Recio-Blanco}, {Richards}, {Riello}, {Rimoldini}, {Robin}, {Roegiers}, {Rybizki}, {Sarro}, {Siopis}, {Smith}, {Sozzetti}, {Ulla}, {Utrilla}, {van Leeuwen}, {van Reeven}, {Abbas}, {Abreu Aramburu}, {Accart}, {Aerts}, {Aguado}, {Ajaj}, {Altavilla}, {{\'A}lvarez}, {{\'A}lvarez Cid-Fuentes}, {Alves}, {Anderson}, {Anglada Varela}, {Antoja}, {Audard}, {Baines}, {Baker}, {Balaguer-N{\'u}{\~n}ez}, {Balbinot}, {Balog}, {Barache}, {Barbato}, {Barros}, {Barstow}, {Bartolom{\'e}}, {Bassilana}, {Bauchet}, {Baudesson-Stella}, {Becciani}, {Bellazzini}, {Bernet}, {Bertone}, {Bianchi}, {Blanco-Cuaresma}, {Boch}, {Bombrun}, {Bossini},
  {Bouquillon}, {Bragaglia}, {Bramante}, {Breedt}, {Bressan}, {Brouillet}, {Bucciarelli}, {Burlacu}, {Busonero}, {Butkevich}, {Buzzi}, {Caffau}, {Cancelliere}, {C{\'a}novas}, {Cantat-Gaudin}, {Carballo}, {Carlucci}, {Carnerero}, {Carrasco}, {Casamiquela}, {Castellani}, {Castro-Ginard}, {Castro Sampol}, {Chaoul}, {Charlot}, {Chemin}, {Chiavassa}, {Cioni}, {Comoretto}, {Cooper}, {Cornez}, {Cowell}, {Crifo}, {Crosta}, {Crowley}, {Dafonte}, {Dapergolas}, {David}, {David}, {de Laverny}, {De Luise}, {De March}, {De Ridder}, {de Souza}, {de Teodoro}, {de Torres}, {del Peloso}, {del Pozo}, {Delbo}, {Delgado}, {Delgado}, {Delisle}, {Di Matteo}, {Diakite}, {Diener}, {Distefano}, {Dolding}, {Eappachen}, {Edvardsson}, {Enke}, {Esquej}, {Fabre}, {Fabrizio}, {Faigler}, {Fedorets}, {Fernique}, {Fienga}, {Figueras}, {Fouron}, {Fragkoudi}, {Fraile}, {Franke}, {Gai}, {Garabato}, {Garcia-Gutierrez}, {Garc{\'\i}a-Torres}, {Garofalo}, {Gavras}, {Gerlach}, {Geyer}, {Giacobbe}, {Gilmore}, {Girona}, {Giuffrida}, {Gomel}, {Gomez},
  {Gonzalez-Santamaria}, {Gonz{\'a}lez-Vidal}, {Granvik}, {Guti{\'e}rrez-S{\'a}nchez}, {Guy}, {Hauser}, {Haywood}, {Helmi}, {Hidalgo}, {Hilger}, {H{\l}adczuk}, {Hobbs}, {Holland}, {Huckle}, {Jasniewicz}, {Jonker}, {Juaristi Campillo}, {Julbe}, {Karbevska}, {Kervella}, {Khanna}, {Kochoska}, {Kontizas}, {Kordopatis}, {Korn}, {Kostrzewa-Rutkowska}, {Kruszy{\'n}ska}, {Lambert}, {Lanza}, {Lasne}, {Le Campion}, {Le Fustec}, {Lebreton}, {Lebzelter}, {Leccia}, {Leclerc}, {Lecoeur-Taibi}, {Liao}, {Licata}, {Lindstr{\o}m}, {Lister}, {Livanou}, {Lobel}, {Madrero Pardo}, {Managau}, {Mann}, {Marchant}, {Marconi}, {Marcos Santos}, {Marinoni}, {Marocco}, {Marshall}, {Martin Polo}, {Mart{\'\i}n-Fleitas}, {Masip}, {Massari}, {Mastrobuono-Battisti}, {Mazeh}, {McMillan}, {Messina}, {Michalik}, {Millar}, {Mints}, {Molina}, {Molinaro}, {Moln{\'a}r}, {Montegriffo}, {Mor}, {Morbidelli}, {Morel}, {Morris}, {Mulone}, {Munoz}, {Muraveva}, {Murphy}, {Musella}, {Noval}, {Ord{\'e}novic}, {Orr{\`u}}, {Osinde}, {Pagani}, {Pagano},
  {Palaversa}, {Palicio}, {Panahi}, {Pawlak}, {Pe{\~n}alosa Esteller}, {Penttil{\"a}}, {Piersimoni}, {Pineau}, {Plachy}, {Plum}, {Poggio}, {Poretti}, {Poujoulet}, {Pr{\v{s}}a}, {Pulone}, {Racero}, {Ragaini}, {Rainer}, {Raiteri}, {Rambaux}, {Ramos}, {Ramos-Lerate}, {Re Fiorentin}, {Regibo}, {Reyl{\'e}}, {Ripepi}, {Riva}, {Rixon}, {Robichon}, {Robin}, {Roelens}, {Rohrbasser}, {Romero-G{\'o}mez}, {Rowell}, {Royer}, {Rybicki}, {Sadowski}, {Sagrist{\`a} Sell{\'e}s}, {Sahlmann}, {Salgado}, {Salguero}, {Samaras}, {Sanchez Gimenez}, {Sanna}, {Santove{\~n}a}, {Sarasso}, {Schultheis}, {Sciacca}, {Segol}, {Segovia}, {S{\'e}gransan}, {Semeux}, {Shahaf}, {Siddiqui}, {Siebert}, {Siltala}, {Slezak}, {Smart}, {Solano}, {Solitro}, {Souami}, {Souchay}, {Spagna}, {Spoto}, {Steele}, {Steidelm{\"u}ller}, {Stephenson}, {S{\"u}veges}, {Szabados}, {Szegedi-Elek}, {Taris}, {Tauran}, {Taylor}, {Teixeira}, {Thuillot}, {Tonello}, {Torra}, {Torra}, {Turon}, {Unger}, {Vaillant}, {van Dillen}, {Vanel}, {Vecchiato}, {Viala}, {Vicente},
  {Voutsinas}, {Weiler}, {Wevers}, {Wyrzykowski}, {Yoldas}, {Yvard}, {Zhao}, {Zorec}, {Zucker}, {Zurbach}, \& {Zwitter}}]{Gaia_EDR3_1}
{Gaia Collaboration}, {Brown}, A.~G.~A., {Vallenari}, A., {et~al.} 2021, \aap, 649, A1

\bibitem[{{Ge} {et~al.}(2012){Ge}, {Hu}, {Wang}, {Bai}, \& {Zhang}}]{Ge12_doublepeaked}
{Ge}, J.-Q., {Hu}, C., {Wang}, J.-M., {Bai}, J.-M., \& {Zhang}, S. 2012, \apjs, 201, 31

\bibitem[{{Genel} {et~al.}(2014){Genel}, {Vogelsberger}, {Springel}, {Sijacki}, {Nelson}, {Snyder}, {Rodriguez-Gomez}, {Torrey}, \& {Hernquist}}]{Genel14}
{Genel}, S., {Vogelsberger}, M., {Springel}, V., {et~al.} 2014, \mnras, 445, 175

\bibitem[{{Glikman} {et~al.}(2023){Glikman}, {Langgin}, {Johnstone}, {Yoon}, {Comerford}, {Simmons}, {Stacey}, {Lacy}, \& {O'Meara}}]{Glikman23}
{Glikman}, E., {Langgin}, R., {Johnstone}, M.~A., {et~al.} 2023, \apjl, 951, L18

\bibitem[{{GRAVITY Collaboration} {et~al.}(2024){GRAVITY Collaboration}, {Amorim}, {Bourdarot}, {Brandner}, {Cao}, {Cl{\'e}net}, {Davies}, {de Zeeuw}, {Dexter}, {Drescher}, {Eckart}, {Eisenhauer}, {Fabricius}, {Feuchtgruber}, {F{\"o}rster Schreiber}, {Garcia}, {Genzel}, {Gillessen}, {Gratadour}, {H{\"o}nig}, {Kishimoto}, {Lacour}, {Lutz}, {Millour}, {Netzer}, {Ott}, {Paumard}, {Perraut}, {Perrin}, {Peterson}, {Petrucci}, {Pfuhl}, {Prieto}, {Rabien}, {Rouan}, {Santos}, {Shangguan}, {Shimizu}, {Sternberg}, {Straubmeier}, {Sturm}, {Tacconi}, {Tristram}, {Widmann}, \& {Woillez}}]{gravity_L_R}
{GRAVITY Collaboration}, {Amorim}, A., {Bourdarot}, G., {et~al.} 2024, \aap, 684, A167

\bibitem[{{Green} {et~al.}(2011){Green}, {Myers}, {Barkhouse}, {Aldcroft}, {Trichas}, {Richards}, {Ruiz}, \& {Hopkins}}]{Green11}
{Green}, P.~J., {Myers}, A.~D., {Barkhouse}, W.~A., {et~al.} 2011, \apj, 743, 81

\bibitem[{{Greene} \& {Ho}(2005)}]{GreeneeHo05}
{Greene}, J.~E. \& {Ho}, L.~C. 2005, \apj, 630, 122

\bibitem[{{Gross} {et~al.}(2023){Gross}, {Chen}, {Foord}, {Liu}, {Shen}, {Oguri}, {Goulding}, {Hwang}, {Zakamska}, {Ma}, \& {Nolan}}]{Gross23}
{Gross}, A.~C., {Chen}, Y.-C., {Foord}, A., {et~al.} 2023, \apj, 956, 117

\bibitem[{{Gross} {et~al.}(2025){Gross}, {Chen}, {Oguri}, {Nolan}, {Liu}, {Shen}, {Zhuang}, {Li}, {Zakamska}, {Hwang}, \& {Ishikawa}}]{gross25_vodka}
{Gross}, A.~C., {Chen}, Y.-C., {Oguri}, M., {et~al.} 2025, \apj, 989, 112

\bibitem[{{Gunasekera} {et~al.}(2023){Gunasekera}, {van Hoof}, {Chatzikos}, \& {Ferland}}]{Gunasekera23}
{Gunasekera}, C.~M., {van Hoof}, P. A.~M., {Chatzikos}, M., \& {Ferland}, G.~J. 2023, Research Notes of the American Astronomical Society, 7, 246

\bibitem[{{Haardt} \& {Madau}(1996)}]{HaardtMadau96}
{Haardt}, F. \& {Madau}, P. 1996, \apj, 461, 20

\bibitem[{{Harrison} \& {Ramos Almeida}(2024)}]{Harrison_almeida24}
{Harrison}, C.~M. \& {Ramos Almeida}, C. 2024, Galaxies, 12, 17

\bibitem[{{Henden} {et~al.}(2018){Henden}, {Puchwein}, {Shen}, \& {Sijacki}}]{Henden18_fable}
{Henden}, N.~A., {Puchwein}, E., {Shen}, S., \& {Sijacki}, D. 2018, \mnras, 479, 5385

\bibitem[{{Hennawi} {et~al.}(2010){Hennawi}, {Myers}, {Shen}, {Strauss}, {Djorgovski}, {Fan}, {Glikman}, {Mahabal}, {Martin}, {Richards}, {Schneider}, \& {Shankar}}]{Hennawi10}
{Hennawi}, J.~F., {Myers}, A.~D., {Shen}, Y., {et~al.} 2010, \apj, 719, 1672

\bibitem[{{Hennawi} {et~al.}(2006){Hennawi}, {Strauss}, {Oguri}, {Inada}, {Richards}, {Pindor}, {Schneider}, {Becker}, {Gregg}, {Hall}, {Johnston}, {Fan}, {Burles}, {Schlegel}, {Gunn}, {Lupton}, {Bahcall}, {Brunner}, \& {Brinkmann}}]{Hennawi06}
{Hennawi}, J.~F., {Strauss}, M.~A., {Oguri}, M., {et~al.} 2006, \aj, 131, 1

\bibitem[{{Henriques} {et~al.}(2015){Henriques}, {White}, {Thomas}, {Angulo}, {Guo}, {Lemson}, {Springel}, \& {Overzier}}]{Henriques15}
{Henriques}, B. M.~B., {White}, S. D.~M., {Thomas}, P.~A., {et~al.} 2015, \mnras, 451, 2663

\bibitem[{{Hoffman} \& {Ribak}(1991)}]{1991ApJ...380L...5H}
{Hoffman}, Y. \& {Ribak}, E. 1991, \apjl, 380, L5

\bibitem[{{Hogg}(1999)}]{hogg99}
{Hogg}, D.~W. 1999, arXiv e-prints, astro ph/9905116

\bibitem[{{Hook} {et~al.}(2004){Hook}, {J{\o}rgensen}, {Allington-Smith}, {Davies}, {Metcalfe}, {Murowinski}, \& {Crampton}}]{Gemini_north04}
{Hook}, I.~M., {J{\o}rgensen}, I., {Allington-Smith}, J.~R., {et~al.} 2004, \pasp, 116, 425

\bibitem[{{Husemann} {et~al.}(2018){Husemann}, {Bielby}, {Jahnke}, {Arrigoni-Battaia}, {Worseck}, {Shanks}, {Wardlow}, \& {Scholtz}}]{Husemann18}
{Husemann}, B., {Bielby}, R., {Jahnke}, K., {et~al.} 2018, \aap, 614, L2

\bibitem[{{Hwang} {et~al.}(2020){Hwang}, {Shen}, {Zakamska}, \& {Liu}}]{Hwang20}
{Hwang}, H.-C., {Shen}, Y., {Zakamska}, N., \& {Liu}, X. 2020, \apj, 888, 73

\bibitem[{{Inada} {et~al.}(2008){Inada}, {Oguri}, {Becker}, {Shin}, {Richards}, {Hennawi}, {White}, {Pindor}, {Strauss}, {Kochanek}, {Johnston}, {Gregg}, {Kayo}, {Eisenstein}, {Hall}, {Castander}, {Clocchiatti}, {Anderson}, {Schneider}, {York}, {Lupton}, {Chiu}, {Kawano}, {Scranton}, {Frieman}, {Keeton}, {Morokuma}, {Rix}, {Turner}, {Burles}, {Brunner}, {Sheldon}, {Bahcall}, \& {Masataka}}]{Inada08}
{Inada}, N., {Oguri}, M., {Becker}, R.~H., {et~al.} 2008, \aj, 135, 496

\bibitem[{{Inada} {et~al.}(2010){Inada}, {Oguri}, {Shin}, {Kayo}, {Strauss}, {Hennawi}, {Morokuma}, {Becker}, {White}, {Kochanek}, {Gregg}, {Chiu}, {Johnston}, {Clocchiatti}, {Richards}, {Schneider}, {Frieman}, {Fukugita}, {Gott}, {Hall}, {York}, {Castander}, \& {Bahcall}}]{Inada10}
{Inada}, N., {Oguri}, M., {Shin}, M.-S., {et~al.} 2010, \aj, 140, 403

\bibitem[{{Inada} {et~al.}(2012){Inada}, {Oguri}, {Shin}, {Kayo}, {Strauss}, {Morokuma}, {Rusu}, {Fukugita}, {Kochanek}, {Richards}, {Schneider}, {York}, {Bahcall}, {Frieman}, {Hall}, \& {White}}]{Inada12}
{Inada}, N., {Oguri}, M., {Shin}, M.-S., {et~al.} 2012, \aj, 143, 119

\bibitem[{{Izquierdo-Villalba} {et~al.}(2026){Izquierdo-Villalba}, {Habouzit}, {Bonetti}, {Bonoli}, {Gualandris}, {Volonteri}, {Angeloni}, {Barausse}, {Bhowmick}, {Blecha}, {Bonilla Rivera}, {Bortolas}, {Caliskan}, {Capelo}, {Caramete}, {Caramete}, {Chen}, {Colpi}, {Contini}, {Dav{\'e}}, {Dayal}, {DeGraf}, {Deane}, {Decarli}, {Delpech}, {Di Matteo}, {Dong-P{\'a}ez}, {Graham}, {Haggard}, {Irodotou}, {Johansson}, {Keitaanranta}, {Kelley}, {Khan}, {Langen}, {Li}, {Liao}, {Mangiagli}, {Marsat}, {McCaffrey}, {Ni}, {Pillay}, {Pislan}, {Rawlings}, {Regan}, {Reinoso}, {Roth}, {Ruiz}, {Sergijenko}, {Sesana}, {Shaifullah}, {Singh}, {Spinoso}, {Toubiana}, {Tremmel}, {Trinca}, {Valiante}, {Zhou}, {Dubois}, {Graziani}, {Lovell}, {Peirani}, {Roper}, {Schaye}, {Schneider}, {Trebitsch}, {Vijayan}, {Vogelsberger}, {Wilkins}, \& {Wise}}]{villalba_lisa26}
{Izquierdo-Villalba}, D., {Habouzit}, M., {Bonetti}, M., {et~al.} 2026, arXiv e-prints, arXiv:2605.00092

\bibitem[{{Izquierdo-Villalba} {et~al.}(2022){Izquierdo-Villalba}, {Sesana}, {Bonoli}, \& {Colpi}}]{Izquierdo-Villalba22_Lgalaxies}
{Izquierdo-Villalba}, D., {Sesana}, A., {Bonoli}, S., \& {Colpi}, M. 2022, \mnras, 509, 3488

\bibitem[{{Izquierdo-Villalba} {et~al.}(2023){Izquierdo-Villalba}, {Sesana}, \& {Colpi}}]{2023MNRAS.519.2083I}
{Izquierdo-Villalba}, D., {Sesana}, A., \& {Colpi}, M. 2023, \mnras, 519, 2083

\bibitem[{{Izquierdo-Villalba} {et~al.}(2024){Izquierdo-Villalba}, {Sesana}, {Colpi}, {Spinoso}, {Bonetti}, {Bonoli}, \& {Valiante}}]{Izquierdo-Villalba24}
{Izquierdo-Villalba}, D., {Sesana}, A., {Colpi}, M., {et~al.} 2024, \aap, 686, A183

\bibitem[{{Jing} {et~al.}(2025){Jing}, {Chen}, {Deng}, {Zhu}, {Zou}, \& {Wu}}]{Jing25_desidual}
{Jing}, L., {Chen}, Q., {Deng}, Z., {et~al.} 2025, arXiv e-prints, arXiv:2505.03103

\bibitem[{Jing {et~al.}(2025)Jing, Chen, Deng, Zhu, Zou, \& Wu}]{desi_dual}
Jing, L., Chen, Q., Deng, Z., {et~al.} 2025, A Quasar Pair Catalog Compiled from DESI DR1

\bibitem[{{Johnstone} {et~al.}(2025){Johnstone}, {Treister}, {Bauer}, {Chang}, {Cicone}, {Koss}, {Moral-Castro}, {Muller-Sanchez}, {Privon}, {Ricci}, {Scoville}, {Venturi}, {Barcos-Mu{\~n}oz}, {Armus}, {Blecha}, {Casey}, {Comerford}, {Evans}, {Kawamuro}, {Medling}, {Messias}, {Nagar}, {Rojas}, {Sanders}, {Trakhtenbrot}, {U}, \& {Urry}}]{Johnstone25}
{Johnstone}, M.~A., {Treister}, E., {Bauer}, F.~E., {et~al.} 2025, \apj, 995, 108

\bibitem[{{Junkkarinen} {et~al.}(2001){Junkkarinen}, {Shields}, {Beaver}, {Burbidge}, {Cohen}, {Hamann}, \& {Lyons}}]{Junkkarinen01}
{Junkkarinen}, V., {Shields}, G.~A., {Beaver}, E.~A., {et~al.} 2001, \apjl, 549, L155

\bibitem[{{Kochanek} {et~al.}(2006){Kochanek}, {Mochejska}, {Morgan}, \& {Stanek}}]{Kochanek06}
{Kochanek}, C.~S., {Mochejska}, B., {Morgan}, N.~D., \& {Stanek}, K.~Z. 2006, \apjl, 637, L73

\bibitem[{{Kollatschny} \& {Zetzl}(2013)}]{Kollatschny13}
{Kollatschny}, W. \& {Zetzl}, M. 2013, \aap, 549, A100

\bibitem[{{Koss} {et~al.}(2018){Koss}, {Blecha}, {Bernhard}, {Hung}, {Lu}, {Trakhtenbrot}, {Treister}, {Weigel}, {Sartori}, {Mushotzky}, {Schawinski}, {Ricci}, {Veilleux}, \& {Sanders}}]{Koss18}
{Koss}, M.~J., {Blecha}, L., {Bernhard}, P., {et~al.} 2018, \nat, 563, 214

\bibitem[{{Koss} {et~al.}(2023){Koss}, {Treister}, {Kakkad}, {Casey-Clyde}, {Kawamuro}, {Williams}, {Foord}, {Trakhtenbrot}, {Bauer}, {Privon}, {Ricci}, {Mushotzky}, {Barcos-Munoz}, {Blecha}, {Connor}, {Harrison}, {Liu}, {Magno}, {Mingarelli}, {Muller-Sanchez}, {Oh}, {Shimizu}, {Smith}, {Stern}, {Tello}, \& {Urry}}]{Koss23}
{Koss}, M.~J., {Treister}, E., {Kakkad}, D., {et~al.} 2023, \apjl, 942, L24

\bibitem[{{Kouroumpatzakis} {et~al.}(2026){Kouroumpatzakis}, {Boorman}, {Svoboda}, {Pfeifle}, {Borkar}, {Gupta}, {Stern}, \& {Zezas}}]{Kouroumpatzakis26}
{Kouroumpatzakis}, K., {Boorman}, P.~G., {Svoboda}, J., {et~al.} 2026, arXiv e-prints, arXiv:2607.17729

\bibitem[{{Krone-Martins} {et~al.}(2019){Krone-Martins}, {Graham}, {Stern}, {Djorgovski}, {Delchambre}, {Ducourant}, {Teixeira}, {Drake}, {Scarano}, {Surdej}, {Galluccio}, {Jalan}, {Wertz}, {Kl{\"u}ter}, {Mignard}, {Spindola-Duarte}, {Dobie}, {Slezak}, {Sluse}, {Murphy}, {Boehm}, {Nierenberg}, {Bastian}, {Wambsganss}, \& {LeCampion}}]{KroneMartins19}
{Krone-Martins}, A., {Graham}, M.~J., {Stern}, D., {et~al.} 2019, arXiv e-prints, arXiv:1912.08977

\bibitem[{{Lagos} {et~al.}(2018){Lagos}, {Tobar}, {Robotham}, {Obreschkow}, {Mitchell}, {Power}, \& {Elahi}}]{Lagos18_shark}
{Lagos}, C. d.~P., {Tobar}, R.~J., {Robotham}, A. S.~G., {et~al.} 2018, \mnras, 481, 3573

\bibitem[{{Lamperti} {et~al.}(2026){Lamperti}, {Mannucci}, {Bertola}, {Marconi}, {Cresci}, {Nardini}, {D'Amato}, {Perna}, {Rojas-Lilay{\'u}}, {Bracci}, {Braito}, {Cataldi}, {Ceci}, {Chakraborty}, {Cicone}, {De Rosa}, {Feltre}, {Ginolfi}, {Lusso}, {Marconcini}, {Moreschini}, {Portaluri}, {Rubinur}, {Scialpi}, {Severgnini}, {Tozzi}, {Trindade Falc{\~a}o}, {Ulivi}, {Venturi}, {Vignali}, \& {Zanchettin}}]{Lamperti26}
{Lamperti}, I., {Mannucci}, F., {Bertola}, E., {et~al.} 2026, \aap, 710, A193

\bibitem[{{Larkin} {et~al.}(2006){Larkin}, {Barczys}, {Krabbe}, {Adkins}, {Aliado}, {Amico}, {Brims}, {Campbell}, {Canfield}, {Gasaway}, {Honey}, {Iserlohe}, {Johnson}, {Kress}, {LaFreniere}, {Lyke}, {Magnone}, {Magnone}, {McElwain}, {Moon}, {Quirrenbach}, {Skulason}, {Song}, {Spencer}, {Weiss}, \& {Wright}}]{osiris_Larkin2006}
{Larkin}, J., {Barczys}, M., {Krabbe}, A., {et~al.} 2006, in Society of Photo-Optical Instrumentation Engineers (SPIE) Conference Series, Vol. 6269, Ground-based and Airborne Instrumentation for Astronomy, ed. I.~S. {McLean} \& M.~{Iye}, 62691A

\bibitem[{{Lemon} {et~al.}(2020){Lemon}, {Auger}, {McMahon}, {Anguita}, {Apostolovski}, {Chen}, {Fassnacht}, {Melo}, {Motta}, {Shajib}, {Treu}, {Agnello}, {Buckley-Geer}, {Schechter}, {Birrer}, {Collett}, {Courbin}, {Rusu}, {Abbott}, {Allam}, {Annis}, {Avila}, {Bertin}, {Brooks}, {Burke}, {Carnero Rosell}, {Carrasco Kind}, {Carretero}, {Costanzi}, {da Costa}, {De Vicente}, {Desai}, {Eifler}, {Flaugher}, {Frieman}, {Garc{\'\i}a-Bellido}, {Gaztanaga}, {Gerdes}, {Gruen}, {Gruendl}, {Gschwend}, {Gutierrez}, {Honscheid}, {James}, {Kim}, {Krause}, {Kuehn}, {Kuropatkin}, {Lahav}, {Lima}, {Lin}, {Maia}, {March}, {Marshall}, {Menanteau}, {Miquel}, {Palmese}, {Paz-Chinch{\'o}n}, {Plazas}, {Roodman}, {Sanchez}, {Schubnell}, {Serrano}, {Smith}, {Soares-Santos}, {Suchyta}, {Tarle}, \& {Walker}}]{Lemon20}
{Lemon}, C., {Auger}, M.~W., {McMahon}, R., {et~al.} 2020, \mnras, 494, 3491

\bibitem[{{Lemon} {et~al.}(2019){Lemon}, {Auger}, \& {McMahon}}]{Lemon19}
{Lemon}, C.~A., {Auger}, M.~W., \& {McMahon}, R.~G. 2019, \mnras, 483, 4242

\bibitem[{{Lemon} {et~al.}(2018){Lemon}, {Auger}, {McMahon}, \& {Ostrovski}}]{Lemon18}
{Lemon}, C.~A., {Auger}, M.~W., {McMahon}, R.~G., \& {Ostrovski}, F. 2018, \mnras, 479, 5060

\bibitem[{{Li} {et~al.}(2023){Li}, {Cai}, {Bian}, {Lin}, {Li}, {Wu}, {Sun}, {Zhang}, {Golden-Marx}, {Sun}, {Zou}, {Fan}, {Egami}, {Charlot}, {Bruzual}, \& {Chevallard}}]{Li2023}
{Li}, M., {Cai}, Z., {Bian}, F., {et~al.} 2023, \apjl, 955, L18

\bibitem[{{Li} {et~al.}(2025){Li}, {Conselice}, {Duan}, {Austin}, {Harvey}, {Adams}, {Bendo}, {Westcott}, {Rusakov}, {Cai}, {Ning}, \& {Zhang}}]{Li+25}
{Li}, Q., {Conselice}, C.~J., {Duan}, Q., {et~al.} 2025, \mnras, 544, 4160

\bibitem[{{Liu} {et~al.}(2011){Liu}, {Shen}, {Strauss}, \& {Hao}}]{Liu11_sdsspair}
{Liu}, X., {Shen}, Y., {Strauss}, M.~A., \& {Hao}, L. 2011, \apj, 737, 101

\bibitem[{{Lupi} {et~al.}(2024){Lupi}, {Trinca}, {Volonteri}, {Dotti}, \& {Mazzucchelli}}]{Lupi24}
{Lupi}, A., {Trinca}, A., {Volonteri}, M., {Dotti}, M., \& {Mazzucchelli}, C. 2024, \aap, 689, A128

\bibitem[{{Lupi} {et~al.}(2019){Lupi}, {Volonteri}, {Decarli}, {Bovino}, {Silk}, \& {Bergeron}}]{2019MNRAS.488.4004L}
{Lupi}, A., {Volonteri}, M., {Decarli}, R., {et~al.} 2019, \mnras, 488, 4004

\bibitem[{{Lyke} {et~al.}(2020){Lyke}, {Higley}, {McLane}, {Schurhammer}, {Myers}, {Ross}, {Dawson}, {Chabanier}, {Martini}, {Busca}, {Mas des Bourboux}, {Salvato}, {Streblyanska}, {Zarrouk}, {Burtin}, {Anderson}, {Bautista}, {Bizyaev}, {Brandt}, {Brinkmann}, {Brownstein}, {Comparat}, {Green}, {de la Macorra}, {Mu{\~n}oz Guti{\'e}rrez}, {Hou}, {Newman}, {Palanque-Delabrouille}, {P{\^a}ris}, {Percival}, {Petitjean}, {Rich}, {Rossi}, {Schneider}, {Smith}, {Vivek}, \& {Weaver}}]{Lyke20}
{Lyke}, B.~W., {Higley}, A.~N., {McLane}, J.~N., {et~al.} 2020, \apjs, 250, 8

\bibitem[{Mahadevan(1997)}]{Mahadevan1997}
Mahadevan, R. 1997, \apj, 477, 585

\bibitem[{{Maiolino} {et~al.}(2024){Maiolino}, {Scholtz}, {Curtis-Lake}, {Carniani}, {Baker}, {de Graaff}, {Tacchella}, {{\"U}bler}, {D'Eugenio}, {Witstok}, {Curti}, {Arribas}, {Bunker}, {Charlot}, {Chevallard}, {Eisenstein}, {Egami}, {Ji}, {Jones}, {Lyu}, {Rawle}, {Robertson}, {Rujopakarn}, {Perna}, {Sun}, {Venturi}, {Williams}, \& {Willott}}]{Maiolino2024dual}
{Maiolino}, R., {Scholtz}, J., {Curtis-Lake}, E., {et~al.} 2024, \aap, 691, A145

\bibitem[{{Mannucci} {et~al.}(2022){Mannucci}, {Pancino}, {Belfiore}, {Cicone}, {Ciurlo}, {Cresci}, {Lusso}, {Marasco}, {Marconi}, {Nardini}, {Pinna}, {Severgnini}, {Saracco}, {Tozzi}, \& {Yeh}}]{Mannucci22}
{Mannucci}, F., {Pancino}, E., {Belfiore}, F., {et~al.} 2022, Nature Astronomy, 6, 1185

\bibitem[{{Mannucci} {et~al.}(2023){Mannucci}, {Scialpi}, {Ciurlo}, {Yeh}, {Marconcini}, {Tozzi}, {Cresci}, {Marconi}, {Amiri}, {Belfiore}, {Carniani}, {Cicone}, {Nardini}, {Pancino}, {Rubinur}, {Severgnini}, {Ulivi}, {Venturi}, {Vignali}, {Volonteri}, {Pinna}, {Rossi}, {Puglisi}, {Agapito}, {Plantet}, {Ghose}, {Carbonaro}, {Xompero}, {Grani}, {Esposito}, {Power}, {Guerra Ramon}, {Lefebvre}, {Cavallaro}, {Davies}, {Riccardi}, {Macintosh}, {Taylor}, {Dolci}, {Baruffolo}, {Feuchtgruber}, {Kravchenko}, {Rau}, {Sturm}, {Wiezorrek}, {Dallilar}, \& {Kenworthy}}]{Mannucci23}
{Mannucci}, F., {Scialpi}, M., {Ciurlo}, A., {et~al.} 2023, \aap, 680, A53

\bibitem[{{Marziani} {et~al.}(2022){Marziani}, {Olmo}, {Negrete}, {Dultzin}, {Piconcelli}, {Vietri}, {Mart{\'\i}nez-Aldama}, {D'Onofrio}, {Bon}, {Bon}, {Deconto Machado}, {Stirpe}, \& {Buendia Rios}}]{Marziani22}
{Marziani}, P., {Olmo}, A.~d., {Negrete}, C.~A., {et~al.} 2022, \apjs, 261, 30

\bibitem[{{Marziani} \& {Sulentic}(2012)}]{Marziani12}
{Marziani}, P. \& {Sulentic}, J.~W. 2012, \nar, 56, 49

\bibitem[{{Massey}(1951)}]{Massey_KS}
{Massey}, F. J.~J. 1951, Journal of the American Statistical Association, 46, 68

\bibitem[{{Mathews} \& {Ferland}(1987)}]{Mathews_Ferland87}
{Mathews}, W.~G. \& {Ferland}, G.~J. 1987, \apj, 323, 456

\bibitem[{{Matsuoka} {et~al.}(2024){Matsuoka}, {Izumi}, {Onoue}, {Strauss}, {Iwasawa}, {Kashikawa}, {Akiyama}, {Aoki}, {Arita}, {Imanishi}, {Ishimoto}, {Kawaguchi}, {Kohno}, {Lee}, {Nagao}, {Silverman}, \& {Toba}}]{Matsuoka2024}
{Matsuoka}, Y., {Izumi}, T., {Onoue}, M., {et~al.} 2024, \apjl, 965, L4

\bibitem[{{Mej{\'\i}a-Restrepo} {et~al.}(2016){Mej{\'\i}a-Restrepo}, {Trakhtenbrot}, {Lira}, {Netzer}, \& {Capellupo}}]{Mejia-Restrepo16}
{Mej{\'\i}a-Restrepo}, J.~E., {Trakhtenbrot}, B., {Lira}, P., {Netzer}, H., \& {Capellupo}, D.~M. 2016, \mnras, 460, 187

\bibitem[{{Mezcua} {et~al.}(2026){Mezcua}, {Laloux}, {Scialpi}, {Siudek}, {Er{\'o}stegui}, {Ricci}, {Matamoro Zatarain}, {Visser}, {Rottgering}, {Gutierrez}, {Feltre}, {Bisigello}, \& {Saulder}}]{mezcua26}
{Mezcua}, M., {Laloux}, B., {Scialpi}, M., {et~al.} 2026, arXiv e-prints, arXiv:2604.13170

\bibitem[{{Monaco} {et~al.}(2002){Monaco}, {Theuns}, {Taffoni}, {Governato}, {Quinn}, \& {Stadel}}]{Monaco02}
{Monaco}, P., {Theuns}, T., {Taffoni}, G., {et~al.} 2002, \apj, 564, 8

\bibitem[{{Morgan} {et~al.}(2000){Morgan}, {Burley}, {Costa}, {Maza}, {Persson}, {Ruiz}, {Schechter}, {Thompson}, \& {Winn}}]{Morgan00}
{Morgan}, N.~D., {Burley}, G., {Costa}, E., {et~al.} 2000, \aj, 119, 1083

\bibitem[{{Munari} {et~al.}(2017){Munari}, {Monaco}, {Sefusatti}, {Castorina}, {Mohammad}, {Anselmi}, \& {Borgani}}]{Munari17}
{Munari}, E., {Monaco}, P., {Sefusatti}, E., {et~al.} 2017, \mnras, 465, 4658

\bibitem[{{Nelson} {et~al.}(2019){Nelson}, {Springel}, {Pillepich}, {Rodriguez-Gomez}, {Torrey}, {Genel}, {Vogelsberger}, {Pakmor}, {Marinacci}, {Weinberger}, {Kelley}, {Lovell}, {Diemer}, \& {Hernquist}}]{Nelson19}
{Nelson}, D., {Springel}, V., {Pillepich}, A., {et~al.} 2019, Computational Astrophysics and Cosmology, 6, 2

\bibitem[{{Netzer} {et~al.}(2007){Netzer}, {Lira}, {Trakhtenbrot}, {Shemmer}, \& {Cury}}]{Netzer07}
{Netzer}, H., {Lira}, P., {Trakhtenbrot}, B., {Shemmer}, O., \& {Cury}, I. 2007, \apj, 671, 1256

\bibitem[{{Newville} {et~al.}(2016){Newville}, {Stensitzki}, {Allen}, {Rawlik}, {Ingargiola}, \& {Nelson}}]{lmfit}
{Newville}, M., {Stensitzki}, T., {Allen}, D.~B., {et~al.} 2016, {Lmfit: Non-Linear Least-Square Minimization and Curve-Fitting for Python}, Astrophysics Source Code Library, record ascl:1606.014

\bibitem[{{Ni} {et~al.}(2025){Ni}, {Chen}, {Zhou}, {Park}, {Yang}, {Di Matteo}, {Bird}, \& {Croft}}]{Ni25}
{Ni}, Y., {Chen}, N., {Zhou}, Y., {et~al.} 2025, \apj, 990, 120

\bibitem[{{Ni} {et~al.}(2022){Ni}, {Di Matteo}, {Bird}, {Croft}, {Feng}, {Chen}, {Tremmel}, {DeGraf}, \& {Li}}]{Ni22_astrid}
{Ni}, Y., {Di Matteo}, T., {Bird}, S., {et~al.} 2022, \mnras, 513, 670

\bibitem[{{Onorato} {et~al.}(2026){Onorato}, {Farina}, {Decarli}, {Protu{\v{s}}ov{\'a}}, {Pizzati}, {Pelliccia}, {Chu}, {Kimura}, {Lozi}, {Peavy}, {Arrigoni-Battaia}, {Ba{\~n}ados}, {Barth}, {Belladitta}, {Bischetti}, {Bosman}, {Choi}, {Costa}, {Eilers}, {Fan}, {Gallerani}, {Geballe}, {Gloudemans}, {Kim}, {Lemaux}, {Liu}, {Matsuoka}, {Mazzucchelli}, {Nitta}, {Schindler}, {Stephens}, {Suh}, {Walter}, {Wang}, \& {Yang}}]{Onorato26}
{Onorato}, S., {Farina}, E.~P., {Decarli}, R., {et~al.} 2026, arXiv e-prints, arXiv:2609.31835

\bibitem[{{Pati{\~n}o-{\'A}lvarez} {et~al.}(2026){Pati{\~n}o-{\'A}lvarez}, {Guerrero-Gonz{\'a}lez}, {Chavushyan}, {Monjardin-Ward}, {Arshakian}, \& {Cruz-Gonz{\'a}lez}}]{PatinoAlvarez26}
{Pati{\~n}o-{\'A}lvarez}, V.~M., {Guerrero-Gonz{\'a}lez}, J.~U., {Chavushyan}, V., {et~al.} 2026, \aap, 705, A83

\bibitem[{{Patton} {et~al.}(2024){Patton}, {Faria}, {Hani}, {Torrey}, {Ellison}, {Thakur}, \& {Westlake}}]{Patton24_illustristng}
{Patton}, D.~R., {Faria}, L., {Hani}, M.~H., {et~al.} 2024, \mnras, 529, 1493

\bibitem[{{Perna} {et~al.}(2025){Perna}, {Arribas}, {Lamperti}, {Circosta}, {Bertola}, {P{\'e}rez-Gonz{\'a}lez}, {D'Eugenio}, {{\"U}bler}, {Cresci}, {Volonteri}, {Mannucci}, {Maiolino}, {Rodr{\'\i}guez Del Pino}, {B{\"o}ker}, {Bunker}, {Charlot}, {Willott}, {Carniani}, {Curti}, {Jones}, {Kumari}, {Marshall}, {Venturi}, {Saxena}, {Scholtz}, \& {Witstok}}]{Perna2025dual}
{Perna}, M., {Arribas}, S., {Lamperti}, I., {et~al.} 2025, \aap, 696, A59

\bibitem[{{Pfeifle} {et~al.}(2019){Pfeifle}, {Satyapal}, {Manzano-King}, {Cann}, {Sexton}, {Rothberg}, {Canalizo}, {Ricci}, {Blecha}, {Ellison}, {Gliozzi}, {Secrest}, {Constantin}, \& {Harvey}}]{Pfeifle19_triple}
{Pfeifle}, R.~W., {Satyapal}, S., {Manzano-King}, C., {et~al.} 2019, \apj, 883, 167

\bibitem[{{Pfeifle} {et~al.}(2023){Pfeifle}, {Weaver}, {Satyapal}, {Ricci}, {Secrest}, {Gliozzi}, {Blecha}, \& {Rothberg}}]{pfeifle23_dual}
{Pfeifle}, R.~W., {Weaver}, K., {Satyapal}, S., {et~al.} 2023, \apj, 954, 116

\bibitem[{{Pfeifle} {et~al.}(2025){Pfeifle}, {Weaver}, {Secrest}, {Rothberg}, \& {Patton}}]{Bigmac25}
{Pfeifle}, R.~W., {Weaver}, K.~A., {Secrest}, N.~J., {Rothberg}, B., \& {Patton}, D.~R. 2025, \apjs, 281, 25

\bibitem[{{Pillepich} {et~al.}(2018){Pillepich}, {Springel}, {Nelson}, {Genel}, {Naiman}, {Pakmor}, {Hernquist}, {Torrey}, {Vogelsberger}, {Weinberger}, \& {Marinacci}}]{Pillepich18}
{Pillepich}, A., {Springel}, V., {Nelson}, D., {et~al.} 2018, \mnras, 473, 4077

\bibitem[{{Planck Collaboration} {et~al.}(2016){Planck Collaboration}, {Ade}, {Aghanim}, {Arnaud}, {Ashdown}, {Aumont}, {Baccigalupi}, {Banday}, {Barreiro}, {Bartlett}, {Bartolo}, {Battaner}, {Battye}, {Benabed}, {Beno{\^\i}t}, {Benoit-L{\'e}vy}, {Bernard}, {Bersanelli}, {Bielewicz}, {Bock}, {Bonaldi}, {Bonavera}, {Bond}, {Borrill}, {Bouchet}, {Boulanger}, {Bucher}, {Burigana}, {Butler}, {Calabrese}, {Cardoso}, {Catalano}, {Challinor}, {Chamballu}, {Chary}, {Chiang}, {Chluba}, {Christensen}, {Church}, {Clements}, {Colombi}, {Colombo}, {Combet}, {Coulais}, {Crill}, {Curto}, {Cuttaia}, {Danese}, {Davies}, {Davis}, {de Bernardis}, {de Rosa}, {de Zotti}, {Delabrouille}, {D{\'e}sert}, {Di Valentino}, {Dickinson}, {Diego}, {Dolag}, {Dole}, {Donzelli}, {Dor{\'e}}, {Douspis}, {Ducout}, {Dunkley}, {Dupac}, {Efstathiou}, {Elsner}, {En{\ss}lin}, {Eriksen}, {Farhang}, {Fergusson}, {Finelli}, {Forni}, {Frailis}, {Fraisse}, {Franceschi}, {Frejsel}, {Galeotta}, {Galli}, {Ganga}, {Gauthier}, {Gerbino}, {Ghosh}, {Giard},
  {Giraud-H{\'e}raud}, {Giusarma}, {Gjerl{\o}w}, {Gonz{\'a}lez-Nuevo}, {G{\'o}rski}, {Gratton}, {Gregorio}, {Gruppuso}, {Gudmundsson}, {Hamann}, {Hansen}, {Hanson}, {Harrison}, {Helou}, {Henrot-Versill{\'e}}, {Hern{\'a}ndez-Monteagudo}, {Herranz}, {Hildebrandt}, {Hivon}, {Hobson}, {Holmes}, {Hornstrup}, {Hovest}, {Huang}, {Huffenberger}, {Hurier}, {Jaffe}, {Jaffe}, {Jones}, {Juvela}, {Keih{\"a}nen}, {Keskitalo}, {Kisner}, {Kneissl}, {Knoche}, {Knox}, {Kunz}, {Kurki-Suonio}, {Lagache}, {L{\"a}hteenm{\"a}ki}, {Lamarre}, {Lasenby}, {Lattanzi}, {Lawrence}, {Leahy}, {Leonardi}, {Lesgourgues}, {Levrier}, {Lewis}, {Liguori}, {Lilje}, {Linden-V{\o}rnle}, {L{\'o}pez-Caniego}, {Lubin}, {Mac{\'\i}as-P{\'e}rez}, {Maggio}, {Maino}, {Mandolesi}, {Mangilli}, {Marchini}, {Maris}, {Martin}, {Martinelli}, {Mart{\'\i}nez-Gonz{\'a}lez}, {Masi}, {Matarrese}, {McGehee}, {Meinhold}, {Melchiorri}, {Melin}, {Mendes}, {Mennella}, {Migliaccio}, {Millea}, {Mitra}, {Miville-Desch{\^e}nes}, {Moneti}, {Montier}, {Morgante}, {Mortlock},
  {Moss}, {Munshi}, {Murphy}, {Naselsky}, {Nati}, {Natoli}, {Netterfield}, {N{\o}rgaard-Nielsen}, {Noviello}, {Novikov}, {Novikov}, {Oxborrow}, {Paci}, {Pagano}, {Pajot}, {Paladini}, {Paoletti}, {Partridge}, {Pasian}, {Patanchon}, {Pearson}, {Perdereau}, {Perotto}, {Perrotta}, {Pettorino}, {Piacentini}, {Piat}, {Pierpaoli}, {Pietrobon}, {Plaszczynski}, {Pointecouteau}, {Polenta}, {Popa}, {Pratt}, \& {Pr{\'e}zeau}}]{Planck16}
{Planck Collaboration}, {Ade}, P.~A.~R., {Aghanim}, N., {et~al.} 2016, \aap, 594, A13

\bibitem[{{Planck Collaboration} {et~al.}(2020){Planck Collaboration}, {Aghanim}, {Akrami}, {Ashdown}, {Aumont}, {Baccigalupi}, {Ballardini}, {Banday}, {Barreiro}, {Bartolo}, {Basak}, {Battye}, {Benabed}, {Bernard}, {Bersanelli}, {Bielewicz}, {Bock}, {Bond}, {Borrill}, {Bouchet}, {Boulanger}, {Bucher}, {Burigana}, {Butler}, {Calabrese}, {Cardoso}, {Carron}, {Challinor}, {Chiang}, {Chluba}, {Colombo}, {Combet}, {Contreras}, {Crill}, {Cuttaia}, {de Bernardis}, {de Zotti}, {Delabrouille}, {Delouis}, {Di Valentino}, {Diego}, {Dor{\'e}}, {Douspis}, {Ducout}, {Dupac}, {Dusini}, {Efstathiou}, {Elsner}, {En{\ss}lin}, {Eriksen}, {Fantaye}, {Farhang}, {Fergusson}, {Fernandez-Cobos}, {Finelli}, {Forastieri}, {Frailis}, {Fraisse}, {Franceschi}, {Frolov}, {Galeotta}, {Galli}, {Ganga}, {G{\'e}nova-Santos}, {Gerbino}, {Ghosh}, {Gonz{\'a}lez-Nuevo}, {G{\'o}rski}, {Gratton}, {Gruppuso}, {Gudmundsson}, {Hamann}, {Handley}, {Hansen}, {Herranz}, {Hildebrandt}, {Hivon}, {Huang}, {Jaffe}, {Jones}, {Karakci}, {Keih{\"a}nen},
  {Keskitalo}, {Kiiveri}, {Kim}, {Kisner}, {Knox}, {Krachmalnicoff}, {Kunz}, {Kurki-Suonio}, {Lagache}, {Lamarre}, {Lasenby}, {Lattanzi}, {Lawrence}, {Le Jeune}, {Lemos}, {Lesgourgues}, {Levrier}, {Lewis}, {Liguori}, {Lilje}, {Lilley}, {Lindholm}, {L{\'o}pez-Caniego}, {Lubin}, {Ma}, {Mac{\'\i}as-P{\'e}rez}, {Maggio}, {Maino}, {Mandolesi}, {Mangilli}, {Marcos-Caballero}, {Maris}, {Martin}, {Martinelli}, {Mart{\'\i}nez-Gonz{\'a}lez}, {Matarrese}, {Mauri}, {McEwen}, {Meinhold}, {Melchiorri}, {Mennella}, {Migliaccio}, {Millea}, {Mitra}, {Miville-Desch{\^e}nes}, {Molinari}, {Montier}, {Morgante}, {Moss}, {Natoli}, {N{\o}rgaard-Nielsen}, {Pagano}, {Paoletti}, {Partridge}, {Patanchon}, {Peiris}, {Perrotta}, {Pettorino}, {Piacentini}, {Polastri}, {Polenta}, {Puget}, {Rachen}, {Reinecke}, {Remazeilles}, {Renzi}, {Rocha}, {Rosset}, {Roudier}, {Rubi{\~n}o-Mart{\'\i}n}, {Ruiz-Granados}, {Salvati}, {Sandri}, {Savelainen}, {Scott}, {Shellard}, {Sirignano}, {Sirri}, {Spencer}, {Sunyaev}, {Suur-Uski}, {Tauber}, {Tavagnacco},
  {Tenti}, {Toffolatti}, {Tomasi}, {Trombetti}, {Valenziano}, {Valiviita}, {Van Tent}, {Vibert}, {Vielva}, {Villa}, {Vittorio}, {Wandelt}, {Wehus}, {White}, {White}, {Zacchei}, \& {Zonca}}]{Planck2020}
{Planck Collaboration}, {Aghanim}, N., {Akrami}, Y., {et~al.} 2020, \aap, 641, A6

\bibitem[{{Puerto-S{\'a}nchez} {et~al.}(2025){Puerto-S{\'a}nchez}, {Habouzit}, {Volonteri}, {Ni}, {Foord}, {Angl{\'e}s-Alc{\'a}zar}, {Chen}, {Guetzoyan}, {Dav{\'e}}, {Di Matteo}, {Dubois}, {Koss}, \& {Rosas-Guevara}}]{puertosanchez24}
{Puerto-S{\'a}nchez}, C., {Habouzit}, M., {Volonteri}, M., {et~al.} 2025, \mnras, 536, 3016

\bibitem[{{Reardon} {et~al.}(2023){Reardon}, {Zic}, {Shannon}, {Hobbs}, {Bailes}, {Di Marco}, {Kapur}, {Rogers}, {Thrane}, {Askew}, {Bhat}, {Cameron}, {Cury{\l}o}, {Coles}, {Dai}, {Goncharov}, {Kerr}, {Kulkarni}, {Levin}, {Lower}, {Manchester}, {Mandow}, {Miles}, {Nathan}, {Os{\l}owski}, {Russell}, {Spiewak}, {Zhang}, \& {Zhu}}]{2023ApJ...951L...6R}
{Reardon}, D.~J., {Zic}, A., {Shannon}, R.~M., {et~al.} 2023, \apjl, 951, L6

\bibitem[{{Ricci} {et~al.}(2017){Ricci}, {Bauer}, {Treister}, {Schawinski}, {Privon}, {Blecha}, {Arevalo}, {Armus}, {Harrison}, {Ho}, {Iwasawa}, {Sanders}, \& {Stern}}]{Ricci17}
{Ricci}, C., {Bauer}, F.~E., {Treister}, E., {et~al.} 2017, \mnras, 468, 1273

\bibitem[{{Richards} {et~al.}(2006){Richards}, {Lacy}, {Storrie-Lombardi}, {Hall}, {Gallagher}, {Hines}, {Fan}, {Papovich}, {Vanden Berk}, {Trammell}, {Schneider}, {Vestergaard}, {York}, {Jester}, {Anderson}, {Budav{\'a}ri}, \& {Szalay}}]{Richards06}
{Richards}, G.~T., {Lacy}, M., {Storrie-Lombardi}, L.~J., {et~al.} 2006, \apjs, 166, 470

\bibitem[{{Rodriguez-Gomez} {et~al.}(2015){Rodriguez-Gomez}, {Genel}, {Vogelsberger}, {Sijacki}, {Pillepich}, {Sales}, {Torrey}, {Snyder}, {Nelson}, {Springel}, {Ma}, \& {Hernquist}}]{Rodriguez-Gomez15}
{Rodriguez-Gomez}, V., {Genel}, S., {Vogelsberger}, M., {et~al.} 2015, \mnras, 449, 49

\bibitem[{{Rosas-Guevara} {et~al.}(2019){Rosas-Guevara}, {Bower}, {McAlpine}, {Bonoli}, \& {Tissera}}]{Rosas-Guevara19}
{Rosas-Guevara}, Y.~M., {Bower}, R.~G., {McAlpine}, S., {Bonoli}, S., \& {Tissera}, P.~B. 2019, \mnras, 483, 2712

\bibitem[{{Runnoe} {et~al.}(2012){Runnoe}, {Brotherton}, \& {Shang}}]{Runnoe12}
{Runnoe}, J.~C., {Brotherton}, M.~S., \& {Shang}, Z. 2012, \mnras, 427, 1800

\bibitem[{{Runnoe} {et~al.}(2013){Runnoe}, {Brotherton}, {Shang}, {Wills}, \& {DiPompeo}}]{Runnoe13}
{Runnoe}, J.~C., {Brotherton}, M.~S., {Shang}, Z., {Wills}, B.~J., \& {DiPompeo}, M.~A. 2013, \mnras, 429, 135

\bibitem[{{Saeedzadeh} {et~al.}(2024){Saeedzadeh}, {Babul}, {Mukherjee}, {Tremmel}, {Quinn}, \& {Mayer}}]{Saeedzadeh24}
{Saeedzadeh}, V., {Babul}, A., {Mukherjee}, S., {et~al.} 2024, \apj, 975, 265

\bibitem[{Satyapal {et~al.}(2017)Satyapal, Secrest, Ricci, Ellison, Rothberg, Blecha, Constantin, Gliozzi, McNulty, \& Ferguson}]{Satyapal17}
Satyapal, S., Secrest, N.~J., Ricci, C., {et~al.} 2017, The Astrophysical Journal, 848, 126

\bibitem[{{Schaye} {et~al.}(2015){Schaye}, {Crain}, {Bower}, {Furlong}, {Schaller}, {Theuns}, {Dalla Vecchia}, {Frenk}, {McCarthy}, {Helly}, {Jenkins}, {Rosas-Guevara}, {White}, {Baes}, {Booth}, {Camps}, {Navarro}, {Qu}, {Rahmati}, {Sawala}, {Thomas}, \& {Trayford}}]{Schaye15}
{Schaye}, J., {Crain}, R.~A., {Bower}, R.~G., {et~al.} 2015, \mnras, 446, 521

\bibitem[{{Schechter} {et~al.}(2017){Schechter}, {Morgan}, {Chehade}, {Metcalfe}, {Shanks}, \& {McDonald}}]{Scheckter17}
{Schechter}, P.~L., {Morgan}, N.~D., {Chehade}, B., {et~al.} 2017, \aj, 153, 219

\bibitem[{{Scialpi} {et~al.}(2026){Scialpi}, {Mannucci}, {D'Amato}, {Marconcini}, {Cresci}, {Marconi}, {Ulivi}, {Fumagalli}, {Rosati}, {Tozzi}, {Zanchettin}, {Battistini}, {Bertola}, {Bracci}, {Carniani}, {Cataldi}, {Ceci}, {Chakraborty}, {Cicone}, {Ciurlo}, {De Rosa}, {Di Rosa}, {Feltre}, {Ginolfi}, {Lamperti}, {Lusso}, {Moreschini}, {Nardini}, {Parvatikar}, {Perna}, {Rubinur}, {Severgnini}, {Singh}, {Spingola}, {Venturi}, {Vignali}, \& {Volonteri}}]{Scialpi+25_arxive}
{Scialpi}, M., {Mannucci}, F., {D'Amato}, Q., {et~al.} 2026, \aap, 709, A236

\bibitem[{{Scialpi} {et~al.}(2024){Scialpi}, {Mannucci}, {Marconcini}, {Venturi}, {Pancino}, {Marconi}, {Cresci}, {Belfiore}, {Amiri}, {Bertola}, {Carniani}, {Cicone}, {Ciurlo}, {D'Amato}, {Ginolfi}, {Lusso}, {Marasco}, {Nardini}, {Rubinur}, {Severgnini}, {Tozzi}, {Ulivi}, {Vignali}, \& {Volonteri}}]{Scialpi24}
{Scialpi}, M., {Mannucci}, F., {Marconcini}, C., {et~al.} 2024, \aap, 690, A57

\bibitem[{{Sesana} {et~al.}(2008){Sesana}, {Vecchio}, \& {Colacino}}]{Sesana08}
{Sesana}, A., {Vecchio}, A., \& {Colacino}, C.~N. 2008, \mnras, 390, 192

\bibitem[{{Shakura} \& {Sunyaev}(1973)}]{ShakuraSunyaev73}
{Shakura}, N.~I. \& {Sunyaev}, R.~A. 1973, \aap, 24, 337

\bibitem[{{Shectman} \& {Johns}(2003)}]{Magellan_telescope03}
{Shectman}, S.~A. \& {Johns}, M. 2003, in Society of Photo-Optical Instrumentation Engineers (SPIE) Conference Series, Vol. 4837, Large Ground-based Telescopes, ed. J.~M. {Oschmann} \& L.~M. {Stepp}, 910--918

\bibitem[{{Shen} {et~al.}(2023){Shen}, {Hwang}, {Oguri}, {Chen}, {Di Matteo}, {Ni}, {Bird}, {Zakamska}, {Liu}, {Chen}, \& {Kratter}}]{Shen23b}
{Shen}, Y., {Hwang}, H.-C., {Oguri}, M., {et~al.} 2023, \apj, 943, 38

\bibitem[{{Shen} {et~al.}(2019){Shen}, {Hwang}, {Zakamska}, \& {Liu}}]{Shen19}
{Shen}, Y., {Hwang}, H.-C., {Zakamska}, N., \& {Liu}, X. 2019, \apjl, 885, L4

\bibitem[{{Shen} {et~al.}(2011){Shen}, {Richards}, {Strauss}, {Hall}, {Schneider}, {Snedden}, {Bizyaev}, {Brewington}, {Malanushenko}, {Malanushenko}, {Oravetz}, {Pan}, \& {Simmons}}]{Shen+11}
{Shen}, Y., {Richards}, G.~T., {Strauss}, M.~A., {et~al.} 2011, \apjs, 194, 45

\bibitem[{{Silverman} {et~al.}(2020){Silverman}, {Tang}, {Lee}, {Hartwig}, {Goulding}, {Strauss}, {Schramm}, {Ding}, {Riffel}, {Fujimoto}, {Hikage}, {Imanishi}, {Iwasawa}, {Jahnke}, {Kayo}, {Kashikawa}, {Kawaguchi}, {Kohno}, {Luo}, {Matsuoka}, {Matsuda}, {Nagao}, {Oguri}, {Ono}, {Onoue}, {Ouchi}, {Shimasaku}, {Suh}, {Suzuki}, {Taniguchi}, {Toba}, {Ueda}, \& {Yasuda}}]{Silverman20}
{Silverman}, J.~D., {Tang}, S., {Lee}, K.-G., {et~al.} 2020, \apj, 899, 154

\bibitem[{{Singh} {et~al.}(2026){Singh}, {Severgnini}, {Cammelli}, {De Rosa}, {Vignali}, {Rigamonti}, {Valiante}, {Monaco}, {Tan}, {Battistini}, {Della Ceca}, {Harms}, \& {Parvatikar}}]{singh26}
{Singh}, J., {Severgnini}, P., {Cammelli}, V., {et~al.} 2026, \aap, 706, A196

\bibitem[{{Smyth} {et~al.}(2019){Smyth}, {Ballance}, \& {Ramsbottom}}]{Smyth19}
{Smyth}, R.~T., {Ballance}, C.~P., \& {Ramsbottom}, C.~A. 2019, \apj, 874, 144

\bibitem[{{Spinoso} {et~al.}(2023){Spinoso}, {Bonoli}, {Valiante}, {Schneider}, \& {Izquierdo-Villalba}}]{Spinoso23}
{Spinoso}, D., {Bonoli}, S., {Valiante}, R., {Schneider}, R., \& {Izquierdo-Villalba}, D. 2023, \mnras, 518, 4672

\bibitem[{{Springel}(2010)}]{springel10}
{Springel}, V. 2010, \mnras, 401, 791

\bibitem[{{Steinborn} {et~al.}(2016){Steinborn}, {Dolag}, {Comerford}, {Hirschmann}, {Remus}, \& {Teklu}}]{Steinborn2016}
{Steinborn}, L.~K., {Dolag}, K., {Comerford}, J.~M., {et~al.} 2016, \mnras, 458, 1013

\bibitem[{{Suh} {et~al.}(2026){Suh}, {Decarli}, {Farina}, {Scharw{\"a}chter}, {Mezcua}, {Lanzuisi}, {Loiacono}, {Lemaux}, {Yi}, {Marchesi}, {Volonteri}, {Hasinger}, {Civano}, {Gloudemans}, {Choi}, {Nam}, {Foord}, \& {Onorato}}]{suh26}
{Suh}, H., {Decarli}, R., {Farina}, E.~P., {et~al.} 2026, arXiv e-prints, arXiv:2607.19491

\bibitem[{{Sutherland} \& {Dopita}(1993)}]{Sutherland93}
{Sutherland}, R.~S. \& {Dopita}, M.~A. 1993, \apjs, 88, 253

\bibitem[{{Tang} {et~al.}(2026){Tang}, {Silverman}, {Prochaska}, {Banerji}, {Ding}, {Onoue}, \& {Jahnke}}]{Tang+26}
{Tang}, S., {Silverman}, J., {Prochaska}, X., {et~al.} 2026, \mnras, 547, stag411

\bibitem[{{Temple} {et~al.}(2021){Temple}, {Hewett}, \& {Banerji}}]{Temple21}
{Temple}, M.~J., {Hewett}, P.~C., \& {Banerji}, M. 2021, \mnras, 508, 737

\bibitem[{{Teyssier}(2002)}]{Teyssier02_ramses}
{Teyssier}, R. 2002, \aap, 385, 337

\bibitem[{{Thomas} {et~al.}(2019){Thomas}, {Dav{\'e}}, {Angl{\'e}s-Alc{\'a}zar}, \& {Jarvis}}]{Thomas19}
{Thomas}, N., {Dav{\'e}}, R., {Angl{\'e}s-Alc{\'a}zar}, D., \& {Jarvis}, M. 2019, \mnras, 487, 5764

\bibitem[{{Treister} {et~al.}(2020){Treister}, {Messias}, {Privon}, {Nagar}, {Medling}, {U}, {Bauer}, {Cicone}, {Mu{\~n}oz}, {Evans}, {Muller-Sanchez}, {Comerford}, {Armus}, {Chang}, {Koss}, {Venturi}, {Schawinski}, {Casey}, {Urry}, {Sanders}, {Scoville}, \& {Sheth}}]{Treister20}
{Treister}, E., {Messias}, H., {Privon}, G.~C., {et~al.} 2020, \apj, 890, 149

\bibitem[{{Treister} {et~al.}(2018){Treister}, {Privon}, {Sartori}, {Nagar}, {Bauer}, {Schawinski}, {Messias}, {Ricci}, {U}, {Casey}, {Comerford}, {Muller-Sanchez}, {Evans}, {Finlez}, {Koss}, {Sanders}, \& {Urry}}]{Treister18}
{Treister}, E., {Privon}, G.~C., {Sartori}, L.~F., {et~al.} 2018, \apj, 854, 83

\bibitem[{{Tremmel} {et~al.}(2017){Tremmel}, {Karcher}, {Governato}, {Volonteri}, {Quinn}, {Pontzen}, {Anderson}, \& {Bellovary}}]{Tremmel17_ROMULUS}
{Tremmel}, M., {Karcher}, M., {Governato}, F., {et~al.} 2017, \mnras, 470, 1121

\bibitem[{{Trevese} {et~al.}(2014){Trevese}, {Perna}, {Vagnetti}, {Saturni}, \& {Dadina}}]{Trevese14}
{Trevese}, D., {Perna}, M., {Vagnetti}, F., {Saturni}, F.~G., \& {Dadina}, M. 2014, \apj, 795, 164

\bibitem[{{Troncoso} {et~al.}(2025){Troncoso}, {Treister}, {Rojas}, {Boquien}, {Bauer}, {Koss}, {Assef}, {Parra Tello}, {del Moral-Castro}, {Ricci}, {Dai}, {Oh}, {Ricci}, {Peca}, {Urry}, {Gupta}, {Venturi}, {Signorini}, {Mushotzky}, \& {Sanders}}]{Troncoso25}
{Troncoso}, M., {Treister}, E., {Rojas}, A., {et~al.} 2025, \apj, 994, 181

\bibitem[{{Truant} {et~al.}(2026){Truant}, {Izquierdo-Villalba}, {Sesana}, {Mohiuddin Shaifullah}, {Bonetti}, {Spinoso}, \& {Bonoli}}]{Truant26}
{Truant}, R.~J., {Izquierdo-Villalba}, D., {Sesana}, A., {et~al.} 2026, \aap, 706, A115

\bibitem[{{Tweed} {et~al.}(2009){Tweed}, {Devriendt}, {Blaizot}, {Colombi}, \& {Slyz}}]{Tweed09}
{Tweed}, D., {Devriendt}, J., {Blaizot}, J., {Colombi}, S., \& {Slyz}, A. 2009, \aap, 506, 647

\bibitem[{{{\"U}bler} {et~al.}(2024){{\"U}bler}, {Maiolino}, {P{\'e}rez-Gonz{\'a}lez}, {D'Eugenio}, {Perna}, {Curti}, {Arribas}, {Bunker}, {Carniani}, {Charlot}, {Rodr{\'\i}guez Del Pino}, {Baker}, {B{\"o}ker}, {Cresci}, {Dunlop}, {Grogin}, {Jones}, {Kumari}, {Lamperti}, {Laporte}, {Marshall}, {Mazzolari}, {Parlanti}, {Rawle}, {Scholtz}, {Venturi}, \& {Witstok}}]{Ubler2024dual}
{{\"U}bler}, H., {Maiolino}, R., {P{\'e}rez-Gonz{\'a}lez}, P.~G., {et~al.} 2024, \mnras, 531, 355

\bibitem[{{{\"U}bler} {et~al.}(2025){{\"U}bler}, {Mazzolari}, {Maiolino}, {D'Eugenio}, {Davari}, {Juod{\v{z}}balis}, {Schneider}, {Valiante}, {Arribas}, {Bertola}, {Bunker}, {Bromm}, {Carniani}, {Charlot}, {Cresci}, {Curti}, {Davies}, {Eisenhauer}, {Fabian}, {F{\"o}rster Schreiber}, {Genzel}, {Inayoshi}, {Ivey}, {Jones}, {Liu}, {Lutz}, {Mackenzie}, {Matthee}, {Parlanti}, {Perna}, {Robertson}, {Rodr{\'\i}guez del Pino}, {Shimizu}, {Sijacki}, {Sturm}, {Tacchella}, {Tacconi}, {Tozzi}, {Trinca}, {Venturi}, {Volonteri}, {Willot}, \& {Zhang}}]{Ubler2025triple}
{{\"U}bler}, H., {Mazzolari}, G., {Maiolino}, R., {et~al.} 2025, arXiv e-prints, arXiv:2509.21575

\bibitem[{{Ulivi} {et~al.}(2025){Ulivi}, {Mannucci}, {Scialpi}, {Marconcini}, {Cresci}, {Marconi}, {Feltre}, \& {Ginolfi}}]{Ulivi2025ArXiv}
{Ulivi}, L., {Mannucci}, F., {Scialpi}, M., {et~al.} 2025, arXiv e-prints, arXiv:2508.19494

\bibitem[{{Vestergaard} \& {Peterson}(2006)}]{VestergaardPeterson06}
{Vestergaard}, M. \& {Peterson}, B.~M. 2006, \apj, 641, 689

\bibitem[{{Vogelsberger} {et~al.}(2014){Vogelsberger}, {Genel}, {Springel}, {Torrey}, {Sijacki}, {Xu}, {Snyder}, {Nelson}, \& {Hernquist}}]{Vogelsberger14}
{Vogelsberger}, M., {Genel}, S., {Springel}, V., {et~al.} 2014, \mnras, 444, 1518

\bibitem[{{Volonteri} {et~al.}(2016){Volonteri}, {Dubois}, {Pichon}, \& {Devriendt}}]{Volonteri16}
{Volonteri}, M., {Dubois}, Y., {Pichon}, C., \& {Devriendt}, J. 2016, \mnras, 460, 2979

\bibitem[{{Volonteri} {et~al.}(2022){Volonteri}, {Pfister}, {Beckmann}, {Dotti}, {Dubois}, {Massonneau}, {Musoke}, \& {Tremmel}}]{Volonteri22}
{Volonteri}, M., {Pfister}, H., {Beckmann}, R., {et~al.} 2022, \mnras, 514, 640

\bibitem[{{Wang} {et~al.}(2014){Wang}, {Du}, {Hu}, {Netzer}, {Bai}, {Lu}, {Kaspi}, {Qiu}, {Li}, {Wang}, \& {SEAMBH Collaboration}}]{Wang14}
{Wang}, J.-M., {Du}, P., {Hu}, C., {et~al.} 2014, \apj, 793, 108

\bibitem[{{Wang} {et~al.}(2026){Wang}, {Ding}, {Silverman}, {Prochaska}, {Treu}, {Yesuf}, {Goulding}, {Imanishi}, {Kashikawa}, {Kayo}, {Kohno}, {Liao}, {Matsuoka}, {Strauss}, \& {Tang}}]{wang+26}
{Wang}, Q., {Ding}, X., {Silverman}, J., {et~al.} 2026, \apj, 997, 269

\bibitem[{{White} \& {Frenk}(1991)}]{White+91}
{White}, S. D.~M. \& {Frenk}, C.~S. 1991, \apj, 379, 52

\bibitem[{{Woodgate} {et~al.}(1998){Woodgate}, {Kimble}, {Bowers}, {Kraemer}, {Kaiser}, {Danks}, {Grady}, {Loiacono}, {Brumfield}, {Feinberg}, {Gull}, {Heap}, {Maran}, {Lindler}, {Hood}, {Meyer}, {Vanhouten}, {Argabright}, {Franka}, {Bybee}, {Dorn}, {Bottema}, {Woodruff}, {Michika}, {Sullivan}, {Hetlinger}, {Ludtke}, {Stocker}, {Delamere}, {Rose}, {Becker}, {Garner}, {Timothy}, {Blouke}, {Joseph}, {Hartig}, {Green}, {Jenkins}, {Linsky}, {Hutchings}, {Moos}, {Boggess}, {Roesler}, \& {Weistrop}}]{STIS98}
{Woodgate}, B.~E., {Kimble}, R.~A., {Bowers}, C.~W., {et~al.} 1998, \pasp, 110, 1183

\bibitem[{{Wu} \& {Shen}(2022)}]{WuShen22}
{Wu}, Q. \& {Shen}, Y. 2022, \apjs, 263, 42

\bibitem[{{Xu} {et~al.}(2023){Xu}, {Chen}, {Guo}, {Jiang}, {Wang}, {Xu}, {Xue}, {Caballero}, {Yuan}, {Xu}, {Wang}, {Hao}, {Luo}, {Lee}, {Han}, {Jiang}, {Shen}, {Wang}, {Wang}, {Xu}, {Wu}, {Manchester}, {Qian}, {Guan}, {Huang}, {Sun}, \& {Zhu}}]{2023RAA....23g5024X}
{Xu}, H., {Chen}, S., {Guo}, Y., {et~al.} 2023, Research in Astronomy and Astrophysics, 23, 075024

\bibitem[{{Yue} {et~al.}(2021){Yue}, {Fan}, {Yang}, \& {Wang}}]{yue+21}
{Yue}, M., {Fan}, X., {Yang}, J., \& {Wang}, F. 2021, \apjl, 921, L27

\bibitem[{{Yue} {et~al.}(2023){Yue}, {Fan}, {Yang}, \& {Wang}}]{yue+23}
{Yue}, M., {Fan}, X., {Yang}, J., \& {Wang}, F. 2023, \aj, 165, 191

\bibitem[{{Zamora} {et~al.}(2025){Zamora}, {Venturi}, {Carniani}, {Bertola}, {Parlanti}, {Perna}, {Arribas}, {B{\"o}ker}, {Bunker}, {Charlot}, {D'Eugenio}, {Maiolino}, {Del Pino}, {{\"U}bler}, {Cresci}, {Jones}, \& {Lamperti}}]{Zamora25}
{Zamora}, S., {Venturi}, G., {Carniani}, S., {et~al.} 2025, \aap, 702, A102

\end{thebibliography}


\appendix

\section{AGN spectral fitting}
\label{app:fit}

In this appendix, we present representative examples of the spectral decomposition obtained with the fitting pipeline described in Section~\ref{sec:specfit}. The model simultaneously optimizes the continuum, broad- and narrow-line emission, Fe,\textsc{ii} emission complexes, and absorption features within a unified $\chi^2$ minimization framework.

The data reduction and source classification are described in \cite{Junkkarinen01,Scheckter17,Mannucci22,Mannucci23,Ciurlo23,Scialpi24,Chen23a,Scialpi+25_arxive}, as well as in Zanchettin, Ceci et al. (in prep.), Tozzi et al. (in prep.), and Zanchettin et al. (subm.). The spectral fitting analysis presented in this work is applied to these reduced datasets. Before performing the spectral fits, all spectra were corrected for Galactic extinction.

\begin{figure}[h!]
    \centering
    \includegraphics[width=\linewidth]{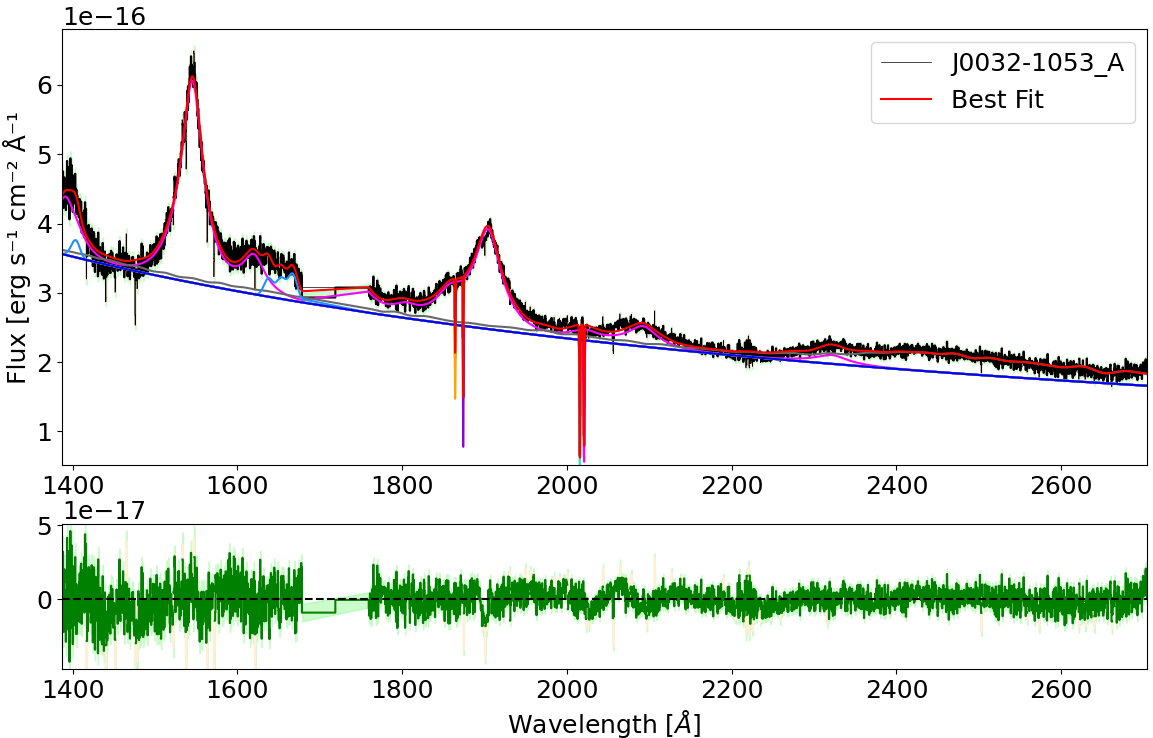}
    \includegraphics[width=\linewidth]{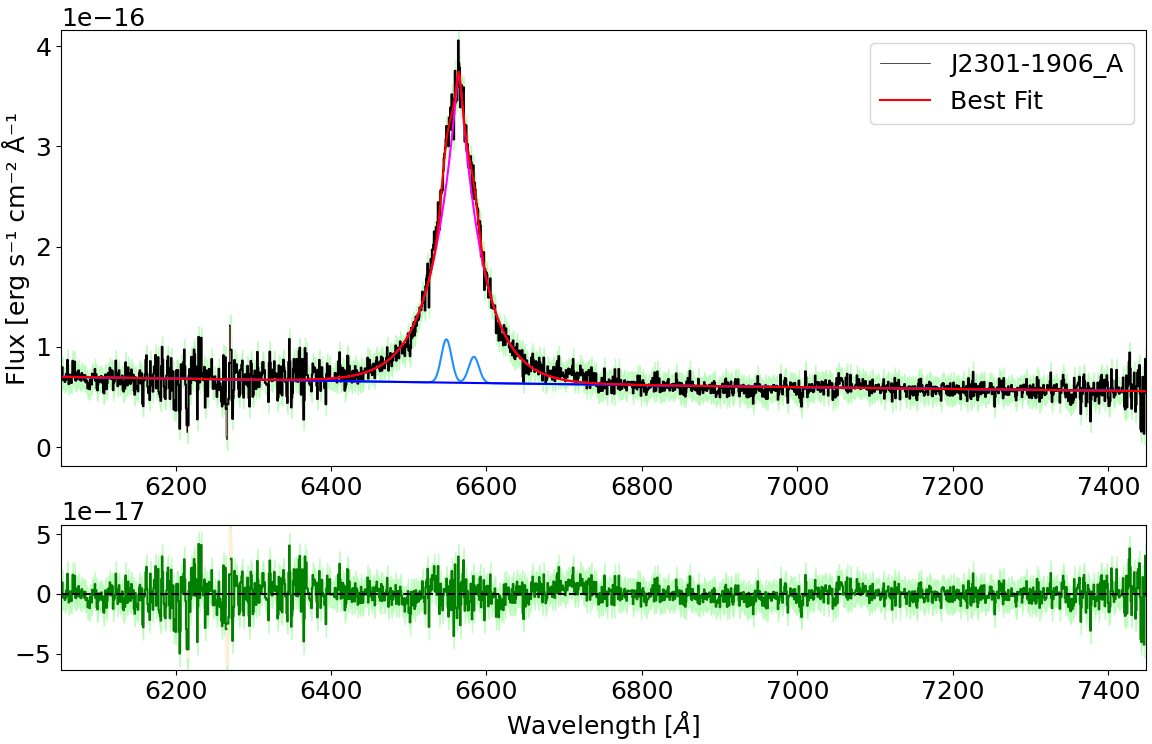}
    \caption{
    Representative examples of spectral decomposition for Component A in two dual AGN systems observed with different instruments. J0032$-$1053 
    is taken from \cite{Scialpi+25_arxive} (MUSE observations), J2301$-$1906 from ERIS observations (Zanchettin in prep.). 
    The observed spectrum is shown in black, while the best-fitting model is shown in red. Individual model components include the power-law continuum (blue), BLR emission (pink), NLR emission (light blue), Fe,\textsc{ii} emission complexes (grey), and absorption features (orange, violet, and cyan). Residuals (data$-$model) are shown in green in the lower panels.
    }
    \label{fig:fits_fig}
\end{figure}

Each spectrum is decomposed into a power-law continuum, broad-line region (BLR) emission, narrow-line region (NLR) emission, Fe,\textsc{ii} templates, and, where required, absorption features. Figure~\ref{fig:fits_fig} shows two representative examples of one system fitted using rest-frame UV emission lines (Mg\,II and C\,IV) and one based on the H$\alpha$ emission line.

\section{Bolometric luminosity calibration}
\label{app:lbol}

\subsection{Continuum-based bolometric luminosities}
In this appendix, we test the consistency of bolometric luminosities derived at three rest-frame wavelengths (3000\ \AA, 4400\ \AA, and 5100\ \AA), with the goal of validating the adoption of 4400\ \AA\ as the reference wavelength for constructing a homogeneous luminosity scale used in the main analysis (Sect.~\ref{sec:properties}).

For each object, we compute $L_{\mathrm{bol}}$ independently at 3000\ \AA, 4400\ \AA, and 5100\ \AA\ using the corresponding continuum measurements. If a wavelength falls outside the spectral coverage, we estimate the continuum via extrapolation of the observed spectrum.
We then fit pairwise linear relations of the form:
$$
\log L_{\mathrm{bol},y} = a \, \log L_{\mathrm{bol},x} + b,
$$
as reported in Table~\ref{tab:lbol_fit}. These fits are performed on the full combined sample, including both directly measured and extrapolated values. The pairwise relations are used as a diagnostic of consistency between wavelength regimes, while the final homogenization is always performed with respect to 4400\ \AA.

\begin{figure}[h!]
    \centering
    \includegraphics[width=0.97\linewidth]{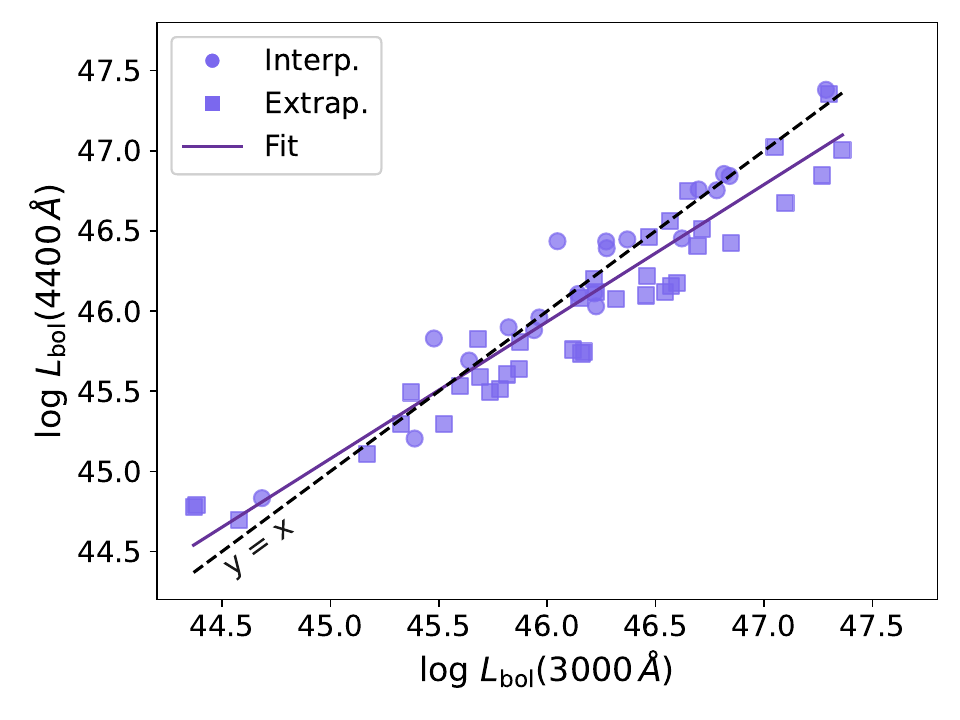}
        \includegraphics[width=0.97\linewidth]{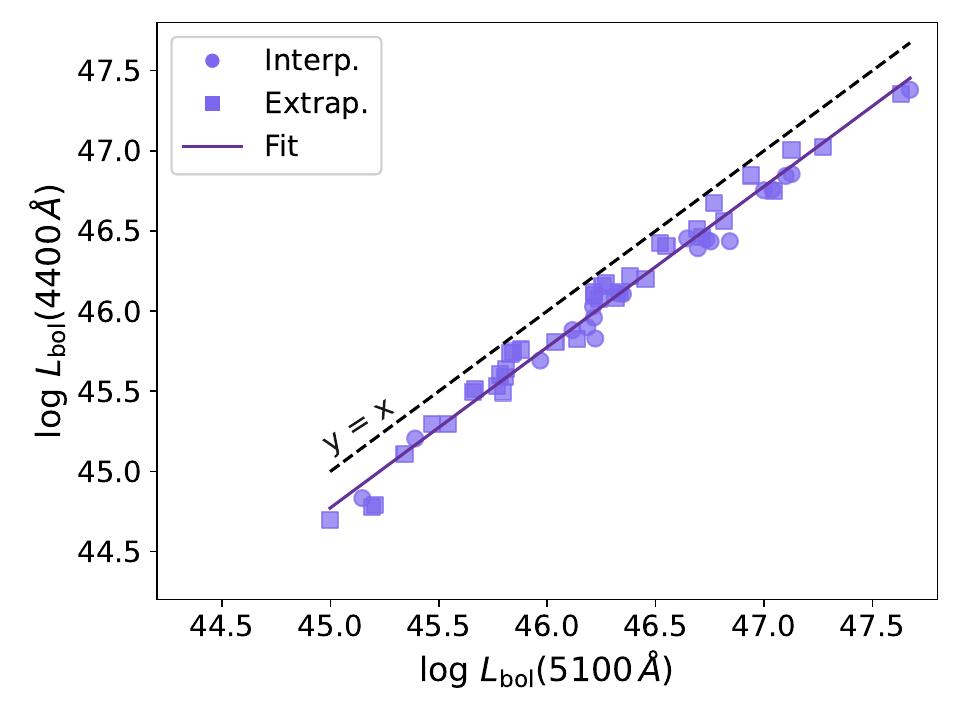}
                \includegraphics[width=0.97\linewidth]{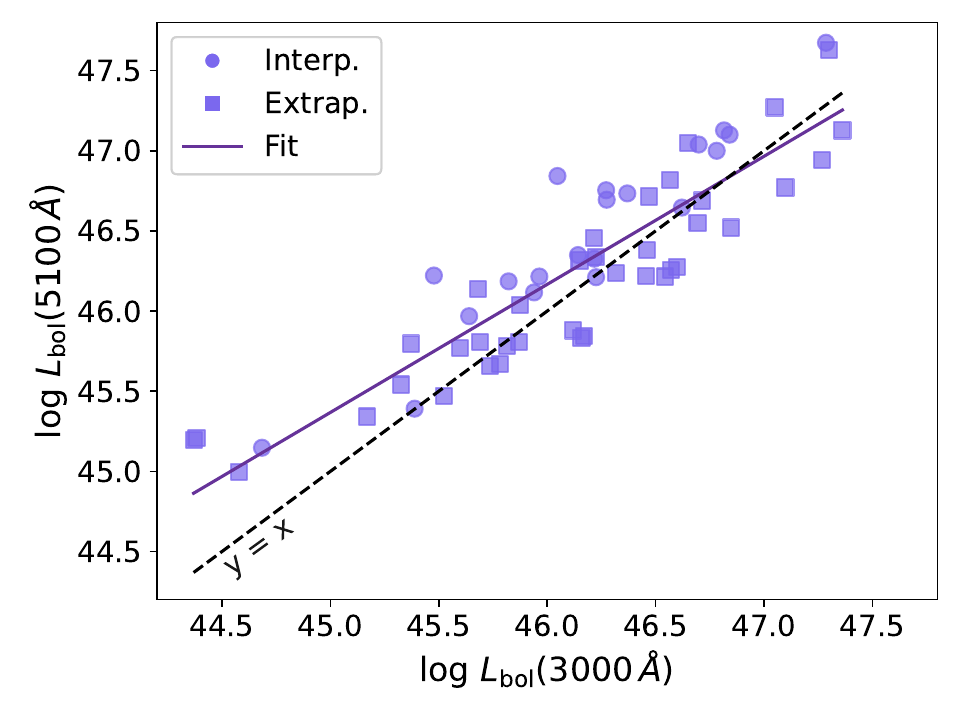}
    \caption{Comparison between bolometric luminosities derived at different wavelengths. The three panels show the relations between 3000\ \AA, 4400\ \AA, and 5100\ \AA\ luminosities. The solid line shows the best-fit relation obtained from the full combined sample. Dots and squares indicate interpolated and extrapolated measurements, respectively. The black dashed line indicates the one-to-one relation.}
    \label{fig:lbol}
\end{figure}

We also verified that fitting separately the two regimes yields statistically consistent results within the uncertainties, indicating that the derived relations are not driven by a specific subset of the data. In Fig.~\ref{fig:lbol}, the two populations are shown separately for diagnostic purposes only, while the solid line corresponds to the global fit of the full dataset.

The tightest correlation is found between 4400\ \AA\ and 5100\ \AA, whereas those involving 3000\ \AA\ show larger scatter and systematic deviations from linearity (see Table~\ref{tab:lbol_fit}). The bootstrap analysis indicates no statistically significant differences between directly measured and extrapolated regimes, with all $p$-values $\gg 0.05$, suggesting that the extrapolation does not introduce systematic biases.

\begin{table}[h!]
\centering
\caption{Best-fit parameters for bolometric luminosity relations.}
\label{tab:lbol_fit}

\setlength{\tabcolsep}{10pt} 
\setlength{\tabcolsep}{10pt}
\renewcommand{\arraystretch}{1.2}

\begin{tabular}{lccc}
\hline
Relation & $a$ & $b$ & $\sigma$ \\
\hline
4400--3000\ \AA & 0.870 & 5.922 & 0.201 \\
4400--5100\ \AA & 0.994 & 0.037 & 0.085 \\
5100--3000\ \AA & 0.820 & 8.468 & 0.279 \\
\hline
\end{tabular}
\end{table}

Finally, for each object we compute $L_{\mathrm{bol}}$ independently at 3000\ \AA, 4400\ \AA, and 5100\ \AA. If the continuum at a given wavelength is not covered by the spectrum, we estimate it via extrapolation. 
The resulting luminosities are then converted to a common scale at 4400\ \AA\ using the empirical relations reported in Table~\ref{tab:lbol_fit}:
$$
\log L_{\mathrm{bol}}^{4400} = a \, \log L_{\mathrm{bol}}^{\lambda} + b,
$$
ensuring a homogeneous definition of $L_{\mathrm{bol}}^{4400}$ across the full sample. Uncertainties are propagated through the linear transformations assuming standard error propagation in log-space.

\subsection{Line-based bolometric luminosities}
As a further validation of the adopted luminosity scale, we compared the bolometric luminosities derived from the continuum at 4400 $\AA$ with independent estimates obtained from broad emission-line luminosities.
For objects with H$\alpha$ measurements, we first derived the monochromatic continuum luminosity at 5100\ \AA\ using the empirical calibration of \citet{GreeneeHo05}. The resulting $L_{5100}$ values were then converted into bolometric luminosities using the bolometric correction of \cite{Runnoe12}.
For sources with Mg\,{\sc ii} measurements, we estimated the monochromatic luminosity at 3000\ \AA\ using the empirical relation presented in \cite{PatinoAlvarez26}, and subsequently converted it into bolometric luminosity adopting the same prescription of \citet{Runnoe12}.

\begin{figure}[h!]
\centering
\includegraphics[width=0.98\columnwidth]{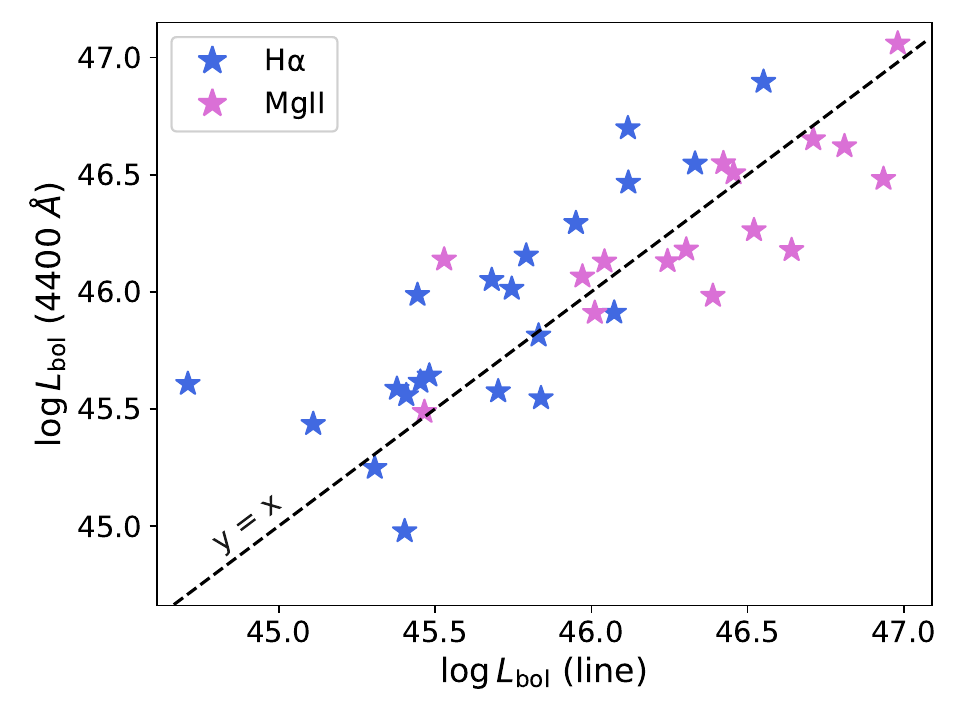}
\caption{Comparison of bolometric luminosities from broad emission lines (colors indicate H$\alpha$ or Mg {\sc ii}) and the 4400\ \AA\ continuum. The grey solid line marks the one-to-one relation.}
\label{fig:lbol_lines}
\end{figure}

Figure~\ref{fig:lbol_lines} compares the continuum- and line-based bolometric luminosities. Both H$\alpha$- and Mg\,{\sc ii}-based estimates show a strong correlation with the continuum luminosities (Pearson $r=0.79$ and $0.81$, respectively; $p<2\times10^{-4}$ in both cases), confirming the internal consistency of the adopted calibrations.

For the H$\alpha$ sample, we measure a mean offset of $0.20$ dex between the continuum- and line-based bolometric luminosities, indicating that the H$\alpha$ estimates are systematically lower than those derived from the continuum by a factor of $\sim1.6$. This offset is statistically significant ($p=0.007$). The scatter around the relation is $0.30$ dex, comparable to the intrinsic dispersion expected for these empirical calibrations. In contrast, the Mg\,{\sc ii}-based luminosities show a much smaller mean offset ($-0.07$ dex), which is not statistically significant ($p=0.32$). Considering the full sample, the average offset decreases to only $0.08$ dex and is also consistent with zero ($p=0.12$), indicating no significant systematic bias overall.

The small systematic difference observed for H$\alpha$ may arise from a combination of host-galaxy contamination affecting the optical continuum around 5100\ \AA, whereas the continuum at 3000\ \AA\ is expected to be much less affected, and the more limited spectral coverage available for the H$\alpha$ sample. 
Indeed, the H$\alpha$ measurements are based on ERIS spectra, whose relatively narrow wavelength range requires a larger extrapolation of the continuum to estimate the luminosity at 4400\ \AA, increasing the associated uncertainties. In contrast, the Mg\,{\sc ii} sample is entirely observed with MUSE, whose broader wavelength coverage allows both the 3000\ \AA\ and 4400\ \AA\ continuum luminosities to be measured more directly, reducing the impact of extrapolation. 
Overall, the close agreement between continuum- and line-based bolometric luminosities indicates that our adopted luminosity scale is robust.
\vspace{-4mm}


\vspace{-2mm}
\section{Observed dual AGN properties}
Properties of the 30 dual AGN analyzed in this work. Projected separations and redshifts are taken from the discovery papers, while continuum luminosities and line measurements are obtained following the procedures described in Sect. \ref{sec:specfit}. Black hole masses, bolometric luminosities, and Eddington ratios are then derived as described in Section \ref{sec:properties}.

\newpage
\renewcommand{\arraystretch}{1.8}
\begin{table*}[h!]
\centering

\begin{tabular}{|@{\hspace{0.35cm}}c@{\hspace{0.35cm}}|C{1cm}C{0.8cm}cc@{\hspace{0.35cm}}|c@{\hspace{0.35cm}}cc@{\hspace{0.35cm}}|c@{\hspace{0.35cm}}cc@{\hspace{0.35cm}}|}\hline
Target & & &  & 
& \multicolumn{3}{c|}{Brighter} 
& \multicolumn{3}{c|}{Fainter} \\
 & z & $r_{\rm p}$ & $\Delta v$  & line 
& $\log(M_{BH})$ & $\log(L_{Bol})$ & $\lambda_{Edd}$ 
& $\log(M_{BH})$ & $\log(L_{Bol})$ & $\lambda_{Edd}$ \\

       &     & $kpc$ & $km\ s^{-1}$  &   
       & $M_\odot$ & $erg\ s^{-1}$  &
       & $M_\odot$ & $erg\ s^{-1}$  & \\
\hline
J0032--1053 & 2.439 & 2.1 & < 440 &  CIV & 9.5$\pm$ 0.5 & 47.0$\pm$ 0.3 & 0.24 & 9.5$\pm$ 0.5& 46.5$\pm$ 0.3&0.08\\

J0032+1624 & 2.356 & 2.3&190 $\pm$ 250 &  H$\alpha$ & 10.0$\pm$ 0.4 & 46.5$\pm$ 0.3& 0.03& 9.2$\pm$ 0.4 & 46.5$\pm$ 0.3&0.17\\

J0052--3045 & 1.455 & 3.7 &1200 $\pm$ 250 &  MgII & 8.9$\pm$ 0.4 &46.2$\pm$ 0.3 & 0.17& 8.8$\pm$ 0.4& 46.2$\pm$ 0.3&0.19\\

J0103--2753 & 0.834 & 2.3 & 3900 $\pm$ 1500&  CIV & - &- & -&9.3$\pm$ 0.5 &46.8$\pm$0.3 &0.25 \\

J0109--3529 & 1.570 & 6.0 & 290 $\pm$ 160&  H$\alpha$ &  8.5$\pm$ 0.5 &46.1 $\pm$ 0.3 & 0.35& 7.8$\pm$ 0.5& 45.4$\pm$ 0.3&0.30\\

J0144--5745 & 0.793 & 3.4 & < 90 &  H$\beta$ & 8.9$\pm$ 0.4& 45.7$\pm$ 0.3&0.05 & 8.3$\pm$ 0.4& 44.8$\pm$ 0.3&0.02\\

J0216+1430 & 0.973 &4.5 & 800 $\pm$ 150&  MgII & 8.9$\pm$ 0.4& 45.6$\pm$ 0.3& 0.04&8.4$\pm$ 0.4 &45.0$\pm$ 0.3&0.03\\

J0300--0532 &0.679 &3.6 &950 $\pm$ 250&  H$\alpha$ & 8.9$\pm$ 0.4 & 45.3$\pm$ 0.3&0.02 &8.5$\pm$ 0.4 &44.8$\pm$ 0.3&0.01\\

J0400+2754 & 1.349&3.1& 450 $\pm$ 80& MgII &  8.5$\pm$ 0.4& 46.0$\pm$ 0.3&0.24 & 8.9$\pm$ 0.4& 45.0$\pm$ 0.3&0.01\\

J0402--3237 & 1.280 & 5.6 & 130 $\pm$ 70 &  H$\alpha$ &  9.0$\pm$ 0.4& 46.3$\pm$ 0.3& 0.16& 8.8$\pm$ 0.4&46.0$\pm$ 0.3& 0.12\\

J0533--2303 & 1.373& 2.8&435 $\pm$ 90 &  H$\alpha$ &  9.1$\pm$ 0.4& 46.9$\pm$ 0.3&0.47 & 8.7$\pm$ 0.4&45.5$\pm$ 0.3& 0.05\\

J0602--3118 & 1.305& 2.2& 400 $\pm$ 70&  H$\alpha$ & 9.2$\pm$ 0.4 &45.8$\pm$ 0.3 & 0.03& 8.6$\pm$ 0.4& 45.6$\pm$ 0.3&0.08\\

J0637--4137 & 2.743 & 3.3 & < 300 & CIII&10.1$\pm$ 0.4 &47.2$\pm$ 0.3 & 0.09&  9.4$\pm$ 0.4&46.4$\pm$ 0.3&0.07 \\

J0749+2255 & 2.170 & 3.9 & 300 $\pm$ 200 & MgII & 9.6$\pm$ 0.4& 46.5$\pm$ 0.3&0.07 &9.0$\pm$ 0.4 & 46.1$\pm$ 0.3&0.10\\

J0840--0529 & 1.386 & 3.3 & 250 $\pm$ 170 &  MgII &9.2$\pm$ 0.4 & 46.7$\pm$ 0.3& 0.25&8.6$\pm$ 0.4 &46.1$\pm$ 0.3& 0.23\\

J0841+4825 & 2.950 & 3.6 & < 450 & CIV &8.7$\pm$0.5 &46.6 $\pm$0.3 &0.66 & 9.0$\pm$0.5&46.4 $\pm$0.3&0.22\\

J1048+4541 & 1.445 &2.8 & 580 $ \pm$ 120 & H$\alpha$ & 8.5$\pm$ 0.5&45.6$\pm$ 0.3 &0.11 &9.3$\pm$ 0.4 & 46.0$\pm$ 0.3&0.04\\

J1103+2348 &1.441 &4.5 & 580 $ \pm$ 110 & H$\alpha$ & 8.9$\pm$ 0.4& 46.3$\pm$ 0.3&0.21 & 8.4$\pm$ 0.5&45.6$\pm$ 0.3& 0.12\\

J1423+1544 & 1.327& 6.9& < 370 & H$\alpha$ &9.4$\pm$ 0.4 & 46.0$\pm$ 0.3&0.03 & 9.7$\pm$ 0.4&46.0$\pm$ 0.3&0.02 \\

J1456+5742 & 1.371 & 4.4 &< 450 & MgII& 8.6$\pm$ 0.4& 46.1$\pm$ 0.3&0.27 &  7.9$\pm$ 0.4& 45.4$\pm$ 0.3&0.23\\

J1613+1708 & 1.545 & 6.6& 160 $\pm$ 300 & H$\alpha$ &8.9$\pm$ 0.4 &46.2$\pm$ 0.3 &0.17 &9.5$\pm$ 0.4 & 45.9$\pm$ 0.3&0.02\\

J1837--6711 & 1.124 & 3.5 & < 440 &- &- & 46.8 $\pm$ 0.3& -&- &45.7 $\pm$ 0.3&- \\

J2100+0612 & 0.785 & 2.9 & < 90  & MgII &9.3 $\pm$ 0.4 &45.9$\pm$ 0.3 &0.03 &7.9$\pm$ 0.4 & 45.2$\pm$ 0.3&0.15\\

J2204--6530 & 1.880 & 3.6 & 1236 $\pm$ 450 & MgII & 9.4$\pm$ 0.4& 46.7$\pm$ 0.3& 0.15& 8.9$\pm$ 0.4&45.9$\pm$ 0.3& 0.09\\

J2215--3056 & 1.347 & 6.1 & 330 $\pm$ 30  & MgII &8.9$\pm$ 0.4 &46.6$\pm$ 0.3 & 0.39& 8.7$\pm$ 0.4& 46.1$\pm$ 0.3&0.22\\

J2228--3047 & 1.949 & 3.3 & 200 $\pm$ 170 & MgII &9.2$\pm$ 0.4 &47.2$\pm$ 0.3 & 0.87& 8.8$\pm$ 0.4& 46.6$\pm$ 0.3&0.45\\

J2301--1906 & 0.811& 4.0& 590 $\pm$ 170 & H$\alpha$ &8.5$\pm$ 0.4 &46.1$\pm$ 0.3 &0.28 & 7.4$\pm$ 0.4& 45.1$\pm$ 0.3&0.39\\

J2332+2243 &0.886 & 3.6& < 350 & H$\alpha$ & 8.3$\pm$ 0.5& 45.6$\pm$ 0.3&0.16 &8.3$\pm$ 0.5 & 45.4$\pm$ 0.3&0.10\\

J2334+2950 &2.282 & 1.9&815 $\pm$ 290 & H$\alpha$ & 9.0$\pm$ 0.4& 46.7$\pm$ 0.3& 0.43& 8.8$\pm$ 0.4 & 45.7$\pm$ 0.3&0.07\\
J2335+3201 &0.903 &4.3 & 450 $\pm$ 180 & H$\alpha$ & 8.5 $\pm$ 0.4&45.6$\pm$ 0.3 & 0.11&8.5$\pm$ 0.5 & 45.2$\pm$ 0.3&0.04\\
\hline
\end{tabular}
\vspace{4mm}
\caption{Redshift, projected separation, velocity offset, and emission-line properties of all targets. For the brighter and fainter components, we report the SE black hole mass, bolometric luminosity and Eddington ratio.}

\label{tab:line_measurements}.
\end{table*}
\clearpage
\newpage

\section{The simulated AGN photometry}
In this appendix, we describe the model adopted to compute the Gaia G-band magnitudes of our simulated AGN sample. We first outline the calculation of the spectral energy distribution (SED) and its conversion into an observed G-band magnitude. We then present the dust attenuation model used to account for the absorption of radiation by dust along the line of sight between the AGN and the observer.
\subsection{\textsc{G magnitudes}}
\label{app:gmag}
The Gaia $G$-band apparent magnitude ($G_{\mathrm{SIM}}$) for each simulated AGN is computed following the procedure of \cite{Truant26}. This ensures a consistent application of the same magnitude selection criteria to both simulations and observations.

The AB magnitude is given by:
\begin{equation}\label{eq:magnitude}
    G_{SIM} = -2.5 \log_{10} \left( 
    \frac{\int_{\lambda_{\rm min}}^{\lambda_{\rm max}} d\lambda \, S_g(\lambda)\,\lambda^{-1}\,L_{\nu}(\lambda)}
    {4 \pi d_L^2 \int_{\lambda_{\rm min}}^{\lambda_{\rm max}} d\lambda \, S_g(\lambda)\,\lambda^{-1}} 
    \right) - 48.6 \,, 
\end{equation}
where $L_{\nu}(\lambda)$ corresponds to the observed spectral energy distribution (SED) of the accreting SMBH, expressed per unit frequency, and $\lambda$ is the observed wavelength. The quantity $d_L$ is the luminosity distance to the source, while $S_g(\lambda)$ is the transmission function of the Gaia $g$-band filter. 

To derive $L_{\nu}(\lambda)$ we model SED using either a geometrically thin, optically thick accretion disc \citep{ShakuraSunyaev73} or an advection-dominated accretion flow \citep[ADAF,][]{Mahadevan1997}. The thin disc model is adopted when the Eddington ratio of the accreting SMBH, defined as $f_{\rm EDD} = \dot{M}/\dot{M}_{\rm Edd}$, exceeds 0.01 \citep{Bonoli25_Lgalaxies}, in which case the SED of the AGN is computed following \cite{ShakuraSunyaev73}:
\begin{equation}\label{eq:EmissionSS}
    L_{\nu} = \frac{2 \pi h r_{\rm Schw}}{c^{2}}\nu^{3}
        \ \int_{x_{\rm in}}^{x_{\rm out}}
        \ \frac{4 \pi x }{{\rm exp}\left(\frac{h \nu}{k_{\rm B} \, T(x)}\right)-1} dx, 
\end{equation}
where $k_{\text{\tiny B}}$ is the Boltzmann constant, $h$ is the Planck constant, $c$ is the speed of light, and $\nu$ is the rest-frame frequency. The variable $x$ is defined as $x \, {\equiv}\, r/r_{\rm Schw}$, with $r$ being the radial distance to the SMBH and $r_{\rm Schw}$ the Schwarzschild radius. The temperature of the circumbinary disk at a given distance $x$, $T(x)$, is determined by:
\begin{equation} \label{eq:TemperatureSS}
    T(x) \,{=}\, \left[ \frac{3\dot{M}_{\rm BH}c^6}{64\pi \sigma_{\rm B} G^2 M^{2}_{\rm BH}} \right]^{\frac{1}{4}} \left[ \frac{1}{x^{3}} \left(1 - \sqrt{\frac{3}{x}}\right)\right]^{\frac{1}{4}},
\end{equation}
where $G$ is the gravitational constant and $\sigma_{\rm B}$ is the Stefan-Boltzmann constant. In the case where the SMBH accretes in the ADAF regime ($f_{\rm EDD} < 0.03$), we follow the SED model of \citet{Mahadevan1997}, in which the total emission is characterised by synchrotron radiation (\(L_{\nu}^{\rm Sync}\)), inverse Compton scattering (\(L_{\nu}^{\rm Comp}\)), and bremsstrahlung radiation (\(L_{\nu}^{\rm Brem}\)):
\begin{equation} \label{eq:ADAF_model}
        L_{\nu} \,{=}\, L_{\nu}^{\rm  Sync} + L_{\nu}^{\rm Comp} + L_{\nu}^{\rm Brem}.
\end{equation}
where:
\begin{equation*}
        L_{\nu}^{ \rm Sync}\,{=}\, (s_1 \ s_2)^{8/5} \  s_3 \ \mathrm{M_{BH}}^{6/5} \ \dot{m}^{4/5} \ T_e^{21/5}\nu^{2/5},
\end{equation*}
\begin{equation*}
        L_{\nu}^{\rm Comp}\,{=}\, L_{\nu}^{\rm Sync} \left(\frac{\nu}{\nu_p}\right)^{\alpha_c},
\end{equation*}
\begin{equation*}
\begin{split}
    L_{\nu}^{\rm Brem} = \mathcal{A}\, \alpha^{-2} c_1^{-2} \ln \left(\frac{r_{\rm  max}}{r_{\rm min}}\right)   \frac{\mathcal{F}(\theta_e)}{T_e} \mathrm{M_{BH}} \, \dot{m}^2 \exp\left(-\frac{h \nu}{ k_{\rm  B} T_e}\right),
\end{split}
\end{equation*}
being $m$ the accretion rate of the SMBH in units of Eddington accretion, $\alpha \,{=}\,0.3$, $\beta\,{=}\,0.5$, $\delta\,{=}\,2000^{-1}$ and $\nu_p$ is the frequency at which the synchrotron radiation is maximum. $T_e$ is the electron temperature, $r_{\rm min}\,{=}\,3\,r_{\rm Schw}$, $r_{\rm max}\,{=}\,10^3\,r_{\rm Schw}$ and $s_1$, $s_2$, $s_3$, $\mathcal{A}$ are constants. For the sake of brevity, we refer to \cite{Mahadevan1997} for the specific values and computations of the parameters. 

\subsection{Dust correction}
\label{sec:dust}
The impact of dust obscuration on the observed luminosities of a given AGN is modelled following the diffuse ISM attenuation prescription adopted by \citet{Henriques15}. In this framework, the attenuation is produced by dust distributed within the host galaxy and depends on the amount of cold gas, its chemical enrichment and the orientation of the galaxy relative to the observer. 
The optical depth is therefore expressed as:
\begin{equation} \label{eq:optical_Depth}
\tau_{\lambda}^{\rm ISM} = (1+z)^{-1}
\left(\frac{A_{\lambda}}{A_V}\right)_{Z_\odot}
\left(\frac{Z_{\rm gas}}{Z_\odot}\right)^s
\left(
\frac{\langle N_{\rm H}\rangle}
{2.1\times10^{21}\ {\rm atoms\ cm^{-2}}}
\right),
\end{equation}
where the first term accounts for the redshift evolution of the dust content, $(A_{\lambda}/A_V)_{Z_\odot}$ describes the wavelength dependence of the extinction curve at solar metallicity, $Z_{\rm gas}/Z_\odot$ accounts for the gas metallicity, and $\langle N_{\rm H}\rangle$ is the mean hydrogen column density.

The metallicity dependence is set $s$, which accounts for the variation of the dust extinction efficiency with wavelength: 
\begin{equation}
s=
\begin{cases}
1.35, & \lambda < 2000~{\rm \AA},\\
1.60, & \lambda \geq 2000~{\rm \AA}.
\end{cases}
\end{equation}
On the other hand, the extinction curve ($(A_{\lambda}/A_V)_{Z_\odot}$) is evaluated using the Galactic extinction law of \citet{Cardelli1989}, while the gas metallicity is expressed as:
\begin{equation}
\frac{Z_{\rm gas}}{Z_\odot} =
\frac{M_{{\rm Z,cold}}}{M_{\rm cold}}
\frac{1}{0.02},
\end{equation}
where $Z_\odot=0.02$, and $M_{{\rm Z,cold}}$ and $M_{\rm cold}$ are the total metal and cold-gas masses of the galaxy disc, respectively. The amount of intervening material is quantified through the mean hydrogen column density, which is derived from the cold-gas distribution 
as:

\begin{equation}
\langle N_{\rm H}\rangle =
\frac{M_{\rm cold}}
{1.4\,\pi\,(aR_{\rm gas})^2},
\end{equation}
where $R_{\rm gas}$ is the exponential gas-disc scale length of the galaxy cold gas disc, the factor of 1.4 accounts for the contribution of helium to the total gas mass, and $a$ is set to $1.68$ for $\langle N_{\rm H} \rangle$ to represent the mass-weighted average column density of an exponential disc.

To convert the optical depth (Eq.~\ref{eq:optical_Depth}) into an observable attenuation, we adopt the slab geometry assumed in \citet{Henriques15}. For a galaxy observed at an inclination angle $\theta$ with respect to the line of sight, where the orientations are assumed to be randomly distributed, the attenuation in magnitudes is given by:
\begin{equation}
A_{\lambda}^{\rm ISM} = -2.5\log_{10} \left[
\frac{1-\exp\left(-\tau_{\lambda}^{\rm ISM}\sec\theta\right)}
{\tau_{\lambda}^{\rm ISM}\sec\theta}
\right],
\end{equation}
The final attenuated magnitude is then obtained by applying the following correction to the intrinsic luminosity:
\begin{equation}
M_{\rm att}=M_{\rm int}+A_{\lambda}^{\rm ISM},
\end{equation}
where $M_{\rm int}$ is the intrinsic (unattenuated) magnitude of the source.

\subsection{SDSS application}
\begin{figure}
    \centering
    \includegraphics[width=1\linewidth]{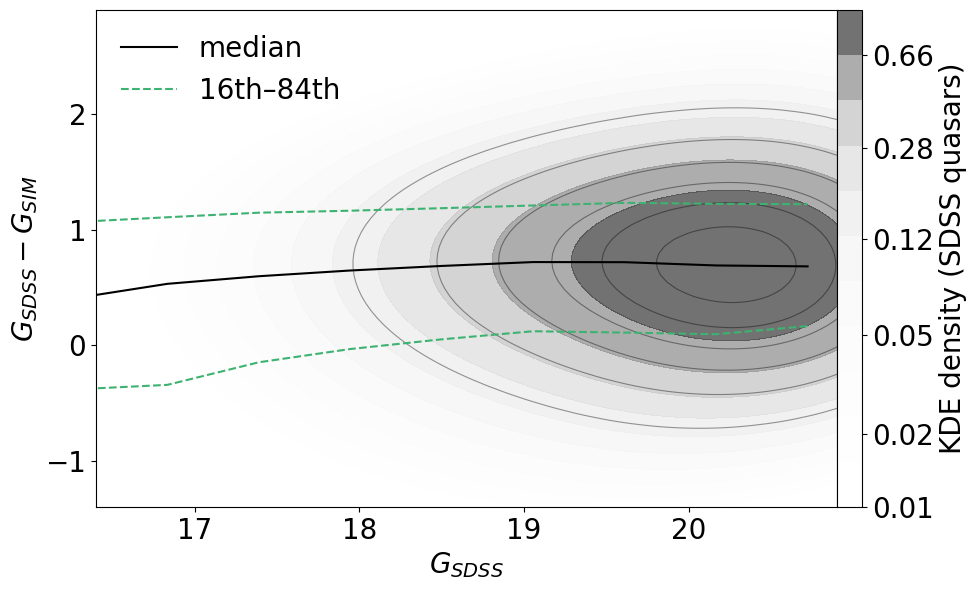}
    \caption{KDE distribution of SDSS QSOs in the $G_{\rm SDSS}-G_{\rm sim}$ vs. $G_{\rm SDSS}$ plane. Black and green lines show the median and the 16th/84th percentiles, respectively.}
    \label{fig:gmag}
\end{figure}

Figure~\ref{fig:gmag} shows the comparison between the $G$ magnitudes estimated from SDSS photometry ($G_{\rm SDSS}$, \cite{Lyke20}) and those computed with the method described above ($G_{\rm SIM}$) as a function of $G_{\rm SDSS}$. 
We find a systematic offset of $\sim0.5$ mag, consistent within the $1\sigma$ uncertainties and without any significant dependence on magnitude. 
This corresponds to $\Delta \log L \simeq 0.2$ dex and does not affect our analysis: applying the offset to the full statistical sample leaves the selection boundaries unchanged and does not alter any of the results presented in this work within the uncertainties. Although dust attenuation cannot be included consistently because the SDSS catalogue lacks some of the quantities required by our model (Sect. \ref{sec:dust}), this comparison provides a useful independent check of our photometric modelling. The positive offset, $G_{\rm SDSS}-G_{\rm SIM}>0$, is consistent with the expected effect of attenuation, while the agreement within the uncertainties supports the validity of our approach.

\section{Comparison with models}
\label{app:simulations}

\renewcommand{\arraystretch}{1.3}

\begin{table*}[t]
\centering
{\small
\setlength{\tabcolsep}{5.4pt}
\begin{tabular}{|l|c|c|c|c|c|c|c|c|c|}
\hline
Model & Type & Geometry & Size & $N_\mathrm{particle}$  & Softening & Ref. &$N_{\rm obj}$& Density\\
      &      &          & cMpc$^3$ &  & [kpc] & &\small{GMP-like}& $\mathrm{cMpc}^{-3}$\\
\hline

\textsc{L-GalaxiesBH} 
& Semi-analytical 
& Lightcone 
& $7.3 \times 10^{9}$
& -- 
& -- 
& \cite{Bonoli25_Lgalaxies} 
& 185 
& $2.5\times10^{-8}$\\

\textsc{Astrid} 
& Hydro-dynamical 
& Box 
& 250
& $2\times5500^3$ 
& 1.5 
& \cite{Ni25} 
& 31 
& $2.0\times10^{-6}$\\
\textsc{Fable} 
& Hydro-dynamical 
& Box 
& 100
& $2\times1280^3$ 
& 2.39 
& \cite{Buttigieg25_fable} 
& 4 
& $3.5\times10^{-7}$\\

\textsc{Horizon-AGN} 
& Hydro-dynamical 
& Box 
& 142
& $\sim1024^3$ 
& 2 
& \cite{Volonteri22} 
& 1 
& $4.0\times10^{-6}$\\
\textsc{Romulus} 
& Hydro-dynamical 
& Box 
& 25
& -- 
& 0.25 
& \cite{Saeedzadeh24} 
& 0 
& 0\\
\hline
\end{tabular}
}
\vspace{4mm}
\caption{Overview of the models used in this work. The columns report: (i) the simulation name; (ii) the model type (semi-analytical or hydrodynamical); (iii) the geometry of the simulation (lightcone or box); (iv) the comoving size of the volume ; (v) the number of particles used in hydrodynamical simulations,
(vi) the gravitational softening length; (vii) the reference to the dual AGN simulation paper; (viii) $N_{\mathrm{obj}}$, the number of GMP-like systems, and (ix) their density.
\label{tab:simulations}}
\end{table*}

In this appendix, we present a full comparison of all simulated samples considered in this work, including those not directly used in the main analysis due to selection limitations. We compare the physical properties of the primary and secondary components, defined as the most and least massive black holes in dual AGN systems, respectively.
Across the different simulations, dual AGN systems are identified by selecting pairs of active SMBH separated by less than $30$ physical kpc, and hierarchical procedure is applied to avoid overcounting systems belonging to higher-order multiples. We then apply the observational selection criteria described in Sect.~\ref{sec:models_gmplike} to identify the GMP-like subsample used for the comparison with observations.

As discussed in Section~\ref{sec:models}, comparing observations with cosmological simulations presents a fundamental challenge: simultaneously achieving sufficient spatial resolution to resolve dual AGN at $\sim$kpc separations and sufficiently large cosmological volumes to include the rare, luminous systems currently accessible to observations. This is particularly relevant for the GMP sample, which is limited to relatively bright sources ($G < 21$).
In this context, the spatial resolution sets the minimum resolvable separation for dual AGN systems, while the simulation volume primarily determines the ability to statistically sample the bright end of the AGN population, including massive black holes and high-luminosity systems at $z \sim 1$--3.

\begin{figure*}
        \centering
    \includegraphics[width=0.495\linewidth]{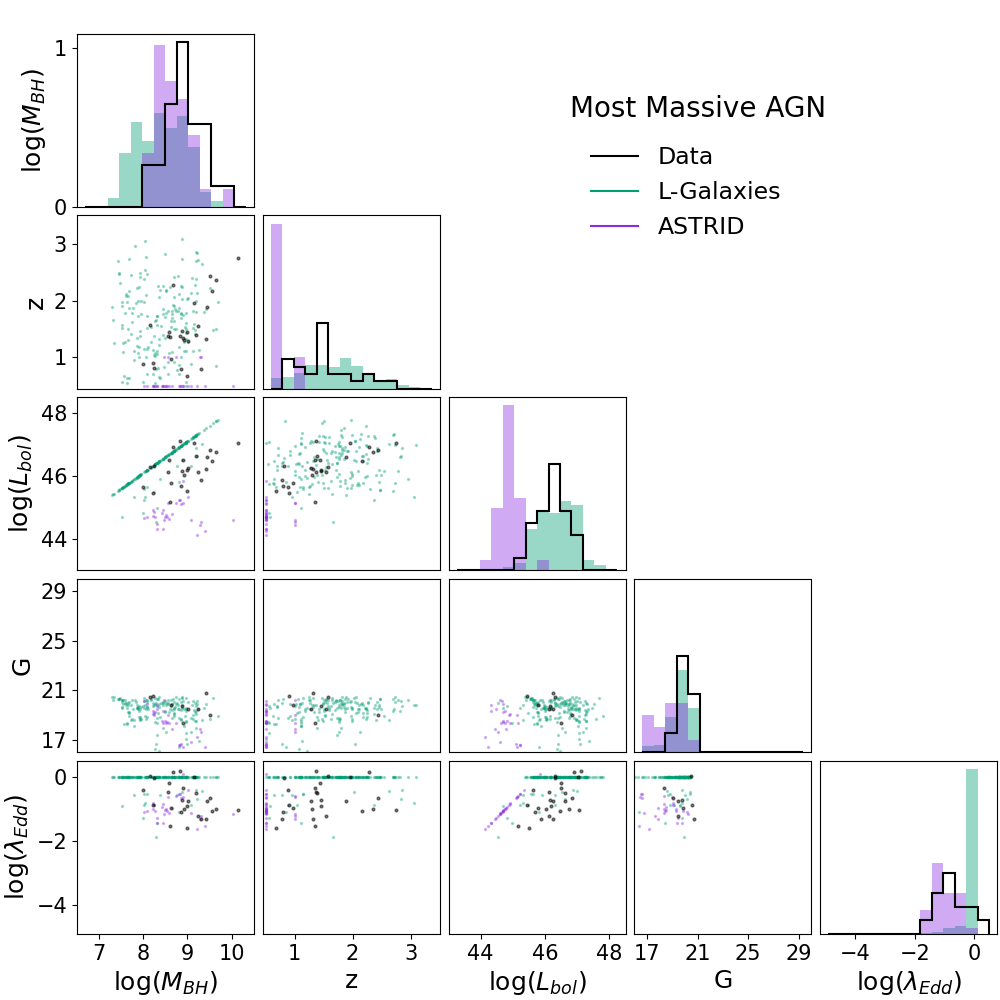}
        \includegraphics[width=0.495\linewidth]{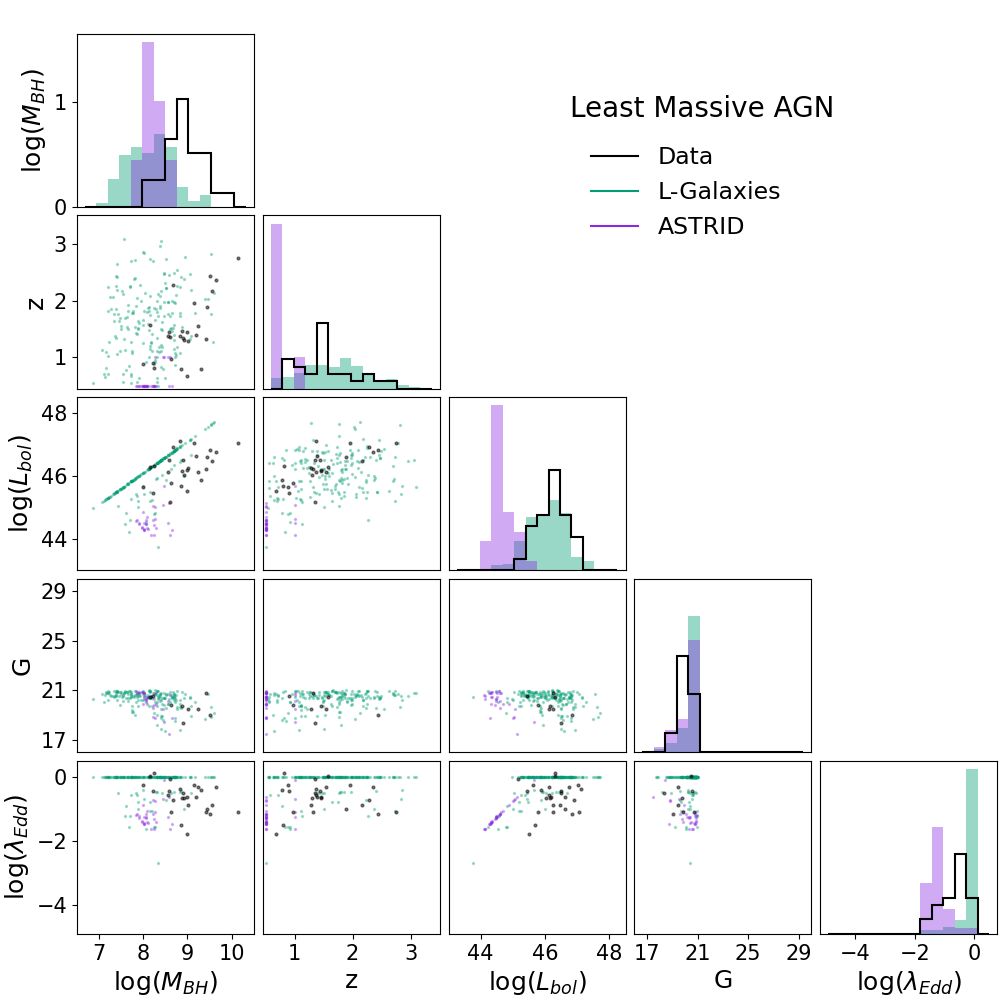}\\
        \vspace{5mm}
            \includegraphics[width=0.495\linewidth]{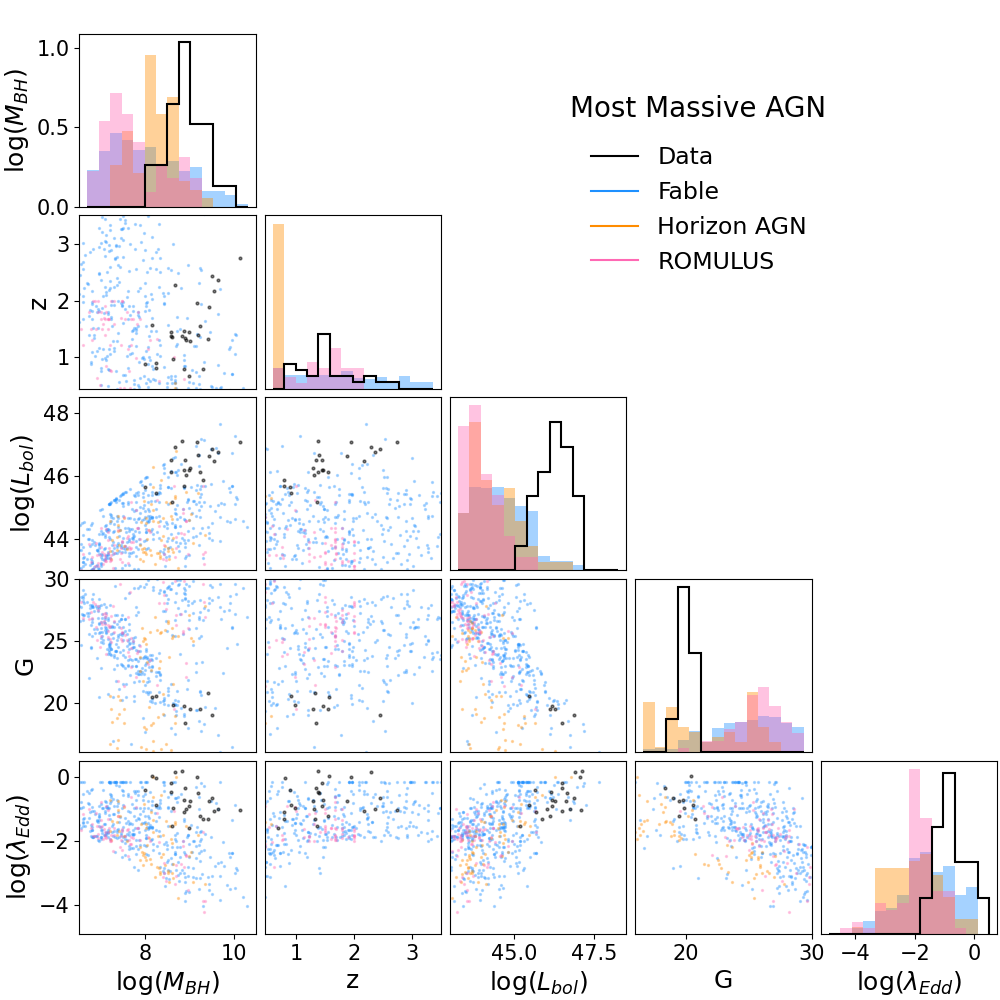}
    \includegraphics[width=0.495\linewidth]{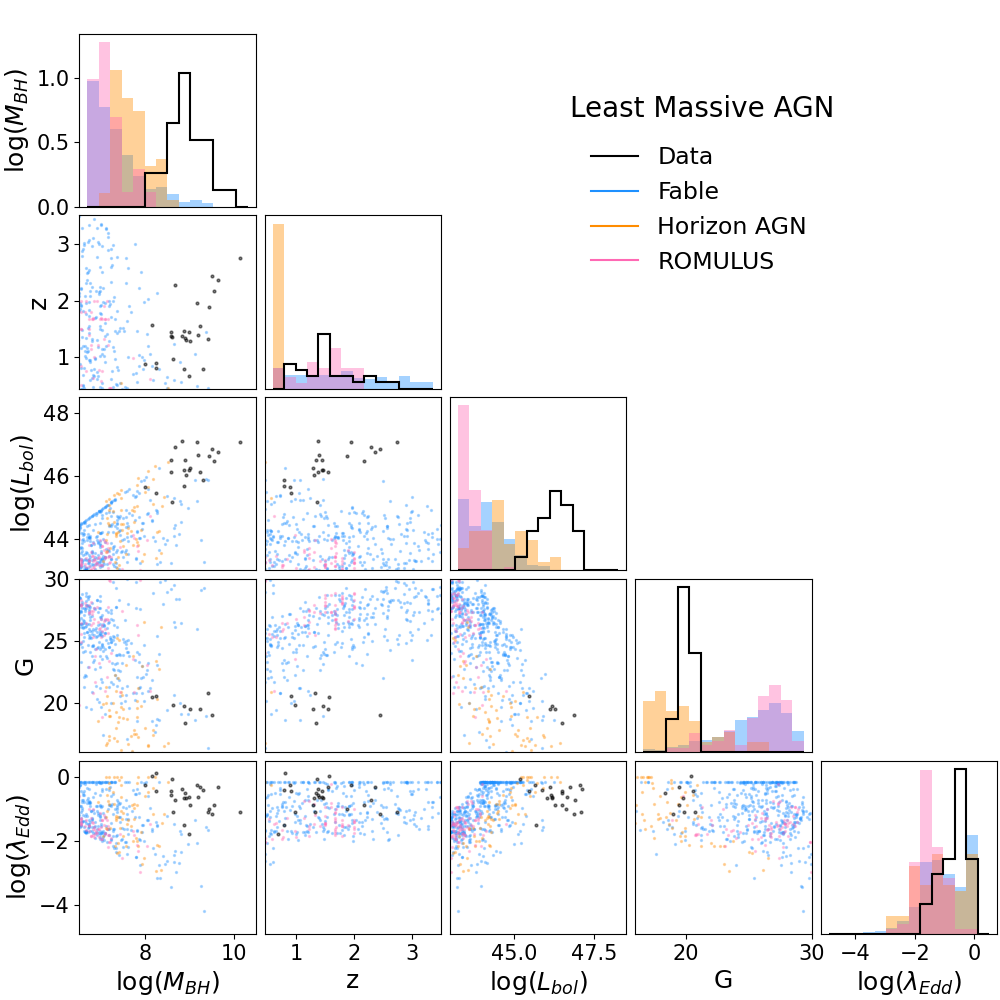}
\vspace{3mm}
\caption{Comparison of the distributions of $M_{BH}$, redshift ($z$), $L_{\rm bol}$, G magnitude, and $\lambda_{\rm Edd}$ for the most massive (left panels) and least massive (right panels) AGN in dual AGN systems, as predicted by the \textsc{L-GalaxiesBH} (teal) and \textsc{Astrid} (purple) simulations in the top panels, and by the \textsc{Fable} (blue), \textsc{Horizon-AGN} (orange), and \textsc{Romulus} (pink) simulations in the bottom panels, compared with observational data (black). 
For \textsc{L-GalaxiesBH} and \textsc{Astrid}, the distributions are shown after applying the GMP-like selection cuts, while for \textsc{Fable}, \textsc{Horizon-AGN}, and \textsc{Romulus} we consider objects with $G<30$. The diagonal panels show normalized one-dimensional histograms, while the off-diagonal panels display the corresponding two-dimensional scatter distributions. }
\label{fig:G_Z_distirb}
\end{figure*}

Figure~\ref{fig:G_Z_distirb} presents the distributions of black hole mass, redshift, bolometric luminosity, $G$ magnitude, and Eddington ratio for the simulated and observed GMP dual AGN populations, highlighting systematic differences among the simulations while considering all simulated objects satisfying the minimal selection criterion $G < 30$. Although all simulations reproduce part of the observed parameter space, only large-volume runs, such as \textsc{L-GalaxiesBH} and \textsc{ASTRID}, are able to host luminous and massive objects at cosmic noon comparable to the GMP sources. 
In fact, applying the observational cut described in Sect.~\ref{sec:models}, the resulting number of dual AGN systems surviving the full selection is strongly model-dependent: we find 185 systems for \textsc{L-GalaxiesBH}, 31 for \textsc{Astrid}, 4 for \textsc{Fable}, 1 for \textsc{Horizon-AGN}, and none for \textsc{Romulus}. These correspond to number densities of $2.5\times10^{-8}$, $2.0\times10^{-6}$, $3.5\times10^{-7}$, $4.0\times10^{-6}$, and $0~\mathrm{Mpc}^{-3}$, respectively. The large variation in number density reflects the combined effects of simulation volume, resolution, and redshift- and luminosity-dependent selection, rather than simply the total number of systems produced by each model. In particular, the much lower density predicted by \textsc{L-GalaxiesBH} despite its large volume highlights the rarity of GMP-like systems in the selected parameter space, while the higher densities in \textsc{Astrid} and \textsc{Horizon-AGN} are based on substantially smaller volumes and therefore suffer from larger statistical uncertainties. These differences highlight the limited overlap between several hydrodynamical simulations and the observational parameter space.

The redshift cut is one of the limiting factors for several simulations. In particular, \textsc{Horizon-AGN}, \textsc{Fable}, and \textsc{Romulus} lose a significant fraction of systems when the cut $z\geq0.5$ is applied. \textsc{Astrid} is also strongly affected, with most remaining objects concentrated at the discrete snapshot $z=0.5$. This may bias the comparison, as the observed GMP sample spans a broader redshift range and, at fixed $G$-band magnitude, probes more luminous AGN at higher redshifts. This effect is also reflected in the black hole mass distribution.
\textsc{Fable} and \textsc{Romulus} preferentially populate the low-luminosity regime, resulting in a deficit of massive black holes compared to the observed GMP sample. Objects that satisfy the $G$-magnitude cut tend to be either low-mass and/or at low redshift, whereas those that fail this criterion generally have black hole masses comparable to those in the observed sample. By contrast, \textsc{L-GalaxiesBH} and \textsc{Astrid} better reproduce the high-mass tails of the observed distributions.

Regarding the Eddington ratios in \textsc{L-GalaxiesBH}, they are not fully emergent quantities as in the other simulations, but are instead set by an explicit modelling assumption following \citet{Bonoli25_Lgalaxies} (Section 2.5.2). In this framework, whenever a merger or a disc instability channels gas towards the galaxy centre, the black hole is assumed to accrete at the Eddington limit for a finite time until the black hole has consumed 70\% of the gas mass transferred to the SMBH during the event, regardless of whether it is triggered by a merger or a disc instability. As a consequence, the resulting Eddington ratios are partially model-driven rather than purely emergent, and may not be directly comparable to those obtained in the other simulations.

\subsection{\textsc{Romulus}}
We make use of the \textsc{Romulus25} cosmological hydrodynamical simulation \citep{Tremmel17_ROMULUS}, which follows galaxy formation and supermassive black hole (SMBH) evolution within a $(25~\mathrm{Mpc})^3$ comoving volume. The simulation is performed with the Tree+SPH code \textsc{ChaNGa} and adopts a Plummer-equivalent gravitational softening of $\sim250$~pc, reaching a maximum spatial resolution of $\sim70$~pc. The mass resolution is $M_{\mathrm{DM}} = 3.39 \times 10^5~M_\odot$ and $M_{\mathrm{gas}} = 2.12 \times 10^5~M_\odot$, with cosmological parameters consistent with \cite{Planck16}.

A key feature of \textsc{Romulus25} is the physically motivated treatment of SMBH seeding, growth, and dynamics. In contrast to many simulations where black holes are seeded based on halo mass thresholds, SMBH formation depends on local gas properties. Seeds of mass $10^6~M_\odot$ form in dense, low-metallicity gas ($Z < 3 \times 10^{-4}$) with temperatures $\sim10^4$~K, allowing black holes to appear in low-mass halos and at high redshift ($z \gtrsim 5$), often resulting in multiple SMBHs within a single halo.

The dynamical evolution of SMBHs is followed self-consistently without artificial repositioning. A subgrid model accounts for unresolved dynamical friction from stars and DM, enabling accurate tracking of SMBH orbits down to sub-kiloparsec scales. SMBH mergers are triggered when pairs are gravitationally bound, have sufficiently low relative velocities, and reach separations below twice the Plummer-equivalent gravitational softening length of 0.35 kpc (i.e., $\sim0.7kpc$). This approach ensures that mergers are physically motivated while avoiding spurious coalescences at unresolved scales.

SMBH growth proceeds via gas accretion and mergers. Gas accretion is modeled using a modified Bondi--Hoyle--Lyttleton prescription that includes corrections for gas angular momentum and bulk motion, as well as a density-dependent boost factor to account for unresolved multiphase structure. Feedback from accretion is implemented as thermal energy injection into the surrounding gas, assuming a radiative efficiency of 10\% and a coupling efficiency of 2\%.
Halos and subhalos are identified using the \textsc{AHF} halo finder and their evolution is tracked with \textsc{TANGOS}.

\subsection{\textsc{Horizon-AGN}}
We make use of the \textsc{Horizon-AGN} cosmological hydrodynamical simulation \citep{Dubois16,Volonteri22}, which follows galaxy formation and MBH evolution within a $(142~\mathrm{Mpc})^3$ comoving volume. The simulation is performed with the adaptive mesh refinement code \textsc{RAMSES} \citep{Teyssier02_ramses} and reaches a maximum spatial resolution of $\Delta x = 1$~kpc. The DM particle mass is $8 \times 10^7~M_\odot$, while stellar particles have masses of $2 \times 10^6~M_\odot$. Gas cooling is modeled following \cite{Sutherland93} down to temperatures of $10^4$~K, and a uniform UV background is included after reionization at $z_{\mathrm{reion}} = 10$ \citep{HaardtMadau96}. Star formation follows a Schmidt law with an efficiency of 2\% in regions exceeding a hydrogen number density threshold of $n_0 = 0.1~\mathrm{H\,cm^{-3}}$.

A key feature of \textsc{Horizon-AGN} is the physically motivated implementation of MBH formation and evolution. In contrast to simulations that seed black holes based on halo mass thresholds, MBH seeds are generated according to local gas properties. Seeds of mass $10^5~M_\odot$ form in dense gas cells with densities above $n_0$ and velocity dispersions exceeding $100~\mathrm{km\,s^{-1}}$, until $z = 1.5$, after which seeding is halted. To avoid multiple seeds forming within the same galaxy, an exclusion radius of $50$ comoving kpc is imposed.
The dynamical evolution of MBHs is followed self-consistently, including an explicit correction for unresolved gas dynamical friction \citep{dubois12}. Gas accretion is modeled using a modified Bondi--Hoyle--Lyttleton prescription, including a density-dependent boost factor, $\alpha = (n/n_0)^2$ for $n > n_0$, to account for unresolved multiphase gas structure. The radiative efficiency is fixed at 10\%, and accretion rates are capped at the Eddington limit.

AGN feedback is implemented in two modes depending on the accretion state. Above 1\% of the Eddington rate, 15\% of the emitted luminosity is isotropically coupled to the surrounding gas as thermal energy within $4\Delta x$. Below this threshold, feedback switches to a kinetic mode, in which energy is injected into bipolar outflows with velocities of $10^4~\mathrm{km\,s^{-1}}$. MBHs are merged when they become gravitationally bound and their separation falls below $4$~kpc, corresponding to $4\Delta x$.
DM halos and subhalos are identified with \textsc{HaloMaker}, based on the \textsc{AdaptaHOP} structure finder \citep{Aubert04,Tweed09}. 

\subsection{\textsc{Fable}}

We make use of the \textsc{Fable} cosmological hydrodynamical simulation suite \citep{Henden18_fable}, performed with the moving-mesh code \textsc{Arepo} \citep{springel10}. We employ the larger \textsc{Fable} box, which follows structure formation within a $(100~h^{-1},\mathrm{cMpc})^3$ volume using $1280^3$ DM particles and an equal number of gas resolution elements. The DM particle and target gas-cell masses are $m_{\mathrm{DM}} = 3.4 \times 10^7~h^{-1}M_\odot$ and $m_{\mathrm{gas}} = 6.4 \times 10^6~h^{-1}M_\odot$, respectively. The Plummer-equivalent gravitational softening length is $2.393~h^{-1}$~kpc in physical units below $z=5$ and fixed in comoving coordinates at higher redshift. The simulation adopts a cosmology consistent with \cite{Planck16}.

SMBHs are modeled as massive collisionless sink particles, seeded with masses of $10^5~h^{-1}M_\odot$ in DM halos exceeding $5 \times 10^{10}~h^{-1}M_\odot$ that do not already host a black hole. SMBHs are repositioned to the minimum of the host halo gravitational potential to prevent spurious displacements and ensure efficient central growth. Their growth proceeds through gas accretion and mergers, with gas accretion following an Eddington-limited Bondi--Hoyle--Lyttleton prescription with a boost factor of $\alpha=100$ to account for unresolved multiphase gas structure. A radiative efficiency of 10\% is assumed.

AGN feedback operates in two modes depending on the Eddington ratio. Above 1\% of the Eddington limit, feedback is implemented through isotropic thermal energy injection in quasar mode, while below this threshold, radio-mode feedback injects hot bubbles into the circumgalactic medium. SMBHs are merged when they approach within one smoothing length, typically $\sim1$~kpc, with the remnant assigned the summed mass of the progenitors.
DM halos and subhalos are identified using the \textsc{FoF} and \textsc{Subfind} algorithms \citep{Davis85,Dolag09}, while merger trees are constructed with \textsc{SubLink} \citep{Rodriguez-Gomez15}.

\end{document}